\documentclass[reprint,superscriptaddress,amsmath,amssymb,aps]{revtex4-1}
\usepackage[utf8]{inputenc}
\usepackage[T1]{fontenc} 
\usepackage{graphicx}
\usepackage{rotating}
\usepackage{makeidx}
\usepackage{threeparttable}
\usepackage{float}
\usepackage{amsmath}
\usepackage{rotating}
\usepackage{hyperref}
\usepackage{relsize}
\usepackage{etoolbox}
\usepackage[most]{tcolorbox}
\usepackage{hyperref}
\usepackage{dcolumn}
\usepackage{bm}
\usepackage{wrapfig}
\usepackage{float}  
\usepackage[toc,page]{appendix}
\usepackage{epstopdf}
\usepackage{subfiles}
\usepackage{comment} 
\usepackage[autostyle=true]{csquotes} 
\usepackage{makecell}
\usepackage{siunitx}
\usepackage[all]{hypcap} 
\usepackage[most]{tcolorbox}
\usepackage{hyperref}
\usepackage{xcolor}
\usepackage{nicefrac}
\usepackage{chemformula}
\usepackage{ulem}  
\definecolor{gainsboro}{rgb}{0.86, 0.86, 0.86}
\usepackage[backgroundcolor=gainsboro]{todonotes}
\hypersetup{
  colorlinks   = true, 
  urlcolor     = red, 
  linkcolor    = red, 
  citecolor   = red 
}

\newcommand{\eg}{$\rm{e_g}$}
\newcommand{\ttg}{$\rm{t_{2g}}$}
\newcommand{\e}[1]{{\rm e}^{#1}}
\newcommand{\tab}[1]{Tab.~\ref{#1}}
\newcommand{\fig}[1]{Fig.~\ref{#1}}
\newcommand{\pcmohalf}{{$\rm{Pr_{\nicefrac{1}{2}}Ca_{\nicefrac{1}{2}}MnO_3}$}}
\newcommand{\pcmox}{{$\rm{Pr_{x}Ca_{1-x}MnO_3}$}}
\newcommand{\weg}[1]{}

\begin{document}

\title{Evolution of the magnetic and polaronic order of
  $\rm{Pr_{1/2}Ca_{1/2}MnO_3}$ following an ultrashort
  light pulse.}

\author{Sangeeta Rajpurohit}
\affiliation{Institute for Theoretical physics, 
Clausthal University of Technology, Germany}

\author{Christian Jooss}
\affiliation{Institute for Material Physics, 
Georg-August-Universit{\"a}t G{\"o}ttingen, Germany}

\author{Peter E. Bl{\"o}chl} 
\email{peter.bloechl@tu-clausthal.de.}
\affiliation{Institute for Theoretical
  physics, Clausthal University of Technology, Germany}
\affiliation{Institute for Theoretical Physics, 
Georg-August-Universit{\"a}t G{\"o}ttingen, Germany}
\date{\today}
\pacs{}

\begin{abstract}
The dynamics of electrons, spins and phonons induced by optical
femtosecond pulses has been simulated for the polaronic crystal
$\rm{Pr_{1/2}Ca_{1/2}MnO_3}$.  The model used for the
simulation has been derived from \textit{first-principles}
calculations.  The simulations reproduce the experimentally observed
melting of charge/orbital order with increasing fluence. The loss of
charge order in the high-fluence regime induces a transition to a
ferromagnetic metal. At low fluence, the dynamics is deterministic and
coherent phonons are created by the repopulation of electronic
orbitals, which are strongly coupled to the phonon degrees of freedom.
In contrast to the low-fluence regime, the magnetic transitions
occurring at higher fluence can be attributed to a quasi-thermal
transition of a cold-plasma-like state with hot electrons and cold
phonons and spins.  The findings can be rationalized in a more
complete picture of the electronic structure that goes beyond the
simple ionic picture of charge order.
\end{abstract}

\maketitle
\section{Introduction}
Unlike traditional spectroscopy, ultrafast pump-probe spectroscopy is
a powerful tool to dynamically track the interactions in a strongly
correlated system.  Specifically, manganites are a suitable model
system, because different types of correlations between electrons,
spins and phonons have similar strength.  A number of different
electronic ground states can be realized just by changing temperature
or doping.  In these materials, high-resolution ultrafast pump-probe
spectroscopy experiments have unraveled interesting physical effects,
such as photo-induced phase transitions
\cite{ren08_prb78_014408,matsubara07_prl99_207401,ogasawara03_prb68_180407} and transient "hidden"
phases \cite{ichikawa11_naturematerials10_101}.
 
In manganite perovskites, a growing number of ultrafast experiments
provide access to the dynamics on different time scales from femto- to
nanoseconds.  The dynamics depends on the phase of the selected
manganite as well as on the photon energy and intensity of the pump
pulse.  In $\rm{GdSrMnO_3}$ close to half doping, a photo-induced
transition from a charge-order phase to a ferromagnetic metallic phase
within 200~fs has been observed\cite{matsubara07_prl99_207401}.  An
ultrafast metal-insulator transition has been induced in
$\rm{Pr_{0.7}Ca_{0.3}MnO_3}$ by selectively exciting phonon modes at
625 cm$^{-1}$ \cite{rini07_nature449_72}.  On shorter timescales,
coherent oscillations in the sample resistance in the
insulator-to-metal dynamics with 30~THz are interpreted as orbital
waves.\cite{polli07_naturematerials6_643} Several ultrafast
optical-pump terahertz-probe studies in manganites focused on probing
the nature of the quasi particles and their dynamics within a given
phase \cite{polli07_naturematerials6_643,rini07_nature449_72,raiser17_aenm7_1602174}.

Optical pump-probe experiments suggest a two-component relaxation
process in
$\rm{Nd_{0.5}Sr_{0.5}MnO_3}$\cite{prasankumar07_prb76_020402} and
$\rm{La_{0.7}Ca_{0.3}MnO_3}$
\cite{averitt01_prl87_017401,wu09_jap105_043901,bielecki10_prb81_064434}. In
the paramagnetic insulating phase of $\rm{La_{0.7}Ca_{0.3}MnO_3}$, the
fast component ${<}1$~ps involves thermalization of electronic
subsystem and its energy redistribution with the lattice
subsystem. The slower component on the 20-200~ps timescale, with a
$\rm{T_c}$-dependent lifetime, is attributed to the spin-lattice
relaxation
\cite{ogasawara03_prb68_180407,wu09_jap105_043901,bielecki10_prb81_064434}.

Recently, long-lived polaron-type optical excitations with lifetime of
1-2~ns are observed in the charge-ordered phase of
$\rm{Pr_{0.7}Ca_{0.3}MnO_3}$ \cite{raiser17_aenm7_1602174}.  Another
long-lived intermediate state, with a mixture of ferromagnetic
metallic and charge-ordered nanoscale domains, is observed in
$\rm{La_{0.325}Pr_{0.3}Ca_{0.375}MnO_3}$ \cite{lin18_prl120_267202}.

Depending on the phase and excitation intensity, coherent acoustic
phonons as well as oscillating strain waves are observed on
time scales of several 10~ps
\cite{thomsen84_prl53_989,zeiger92_prb45_768}; for manganites see
\cite{jang10_prb81_214416}.

The goal of this work is to augment previous experimental studies with
a detailed description of the microscopic processes occurring during
the first few picoseconds. For this purpose, we perform simulations,
that are verified by comparing our findings with experimental
observations. The work presented here is parameter free in the sense
that the model parameters have been extracted from first-principles
calculations\cite{sotoudeh17_prb95_235150}.

We simulate the photo-excitation of {\pcmohalf} by a
femto-second light pulse and the subsequent relaxation of the magnetic
and polaronic microstructure for the first few picoseconds.
Ehrenfest dynamics\cite{tully98_faradaydiscuss110_407} is adopted to
propagate wave functions, spins and atoms.  Peierls
substitution has been employed to incorporate the external light pulse. A
systematic study is performed to investigate the relaxation process by
varying the intensity of the light pulse.

The paper is organized as follows: The methods of the paper are
covered in section~\ref{sec:methods}: The tight-binding model used for
the electron, spin and phonon degrees of freedom is described first,
followed by the dynamical equations of motion, the treatment of the
optical excitation and quantities used in the analysis. In
section~\ref{sec:equilibrium}, we revisit the ground state of
{\pcmohalf}, which experiences the optical excitation. Thereafter, we
describe in section~\ref{sec:results} the results of our simulations
and discuss the underlying mechanisms. Finally we summarize the
findings in section~\ref{sec:summary}.

\section{Methods}
\label{sec:methods}
\subsection{Tight-Binding model}
\label{secs:model}
To investigate the electronic, atomic and magnetic microstructure of
manganites, we employ a tight-binding
model\cite{dagotto01_physrep344_1,hotta02_arxiv0212_466,hotta06_rpp69_2061}.
The selection of energy terms and the parameter values have been
extracted from the first-principles
calculations\cite{sotoudeh17_prb95_235150}.

The model describes the correlations between electron, spin and phonon
degrees of freedom.  The explicit electronic degrees of freedom
describe the Mn-d-electrons with {\eg}-character.  The spin degrees of
freedom describe the three majority-spin d-orbitals of each Mn
ion with {\ttg}~character. The phonon degrees of freedom are two
Jahn-Teller active vibration modes and one breathing mode of each
MnO$_6$ octahedron.  In addition, we allow for a global expansion of
the lattice.

The potential-energy functional of the system is 
 \begin{eqnarray}
E_{pot}=E_{e} +E_{S}+E_{ph}+E_{e-ph}+E_{e-S}\;,
\label{eq:1_1}
 \end{eqnarray} 
where the $E_{e}$, $E_{S}$ and $E_{ph}$ are the energies of the
isolated electronic, spin and phonon subsystems, respectively.
$E_{e-ph}$ is the electron-phonon coupling of Mn-{\eg}~electrons with
the Jahn-Teller active modes as well as with  the
breathing mode of the MnO$_6$ octahedra.  $E_{e-S}$ is Hund's coupling
between the {\eg}~electrons and the spins of the Mn-{\ttg}~electrons.

Our model avoids the common infinite-Hund's-coupling
limit,\cite{muellerhartmann96_prb54_R6819} and uses a more realistic
description with explicit minority-spin electrons and a fully
non-collinear treatment of the electron spin. Furthermore, we include
the strong cooperativity of Jahn-Teller distortions and octahedral
breathing modes by expressing them in terms of the explicit oxygen
positions, which are shared each by two adjacent MnO$_6$ octahedra.

 The Mn {\eg}~electrons are described by a Slater determinant formed
 by the one-particle wave functions $|\psi_n\rangle$. The latter are
 expressed as superposition of local orbitals
 $|\chi_{\sigma,\alpha,R}\rangle$ with complex-valued
   coefficients $\psi_{\sigma,\alpha,R,n}$ 
 \begin{eqnarray}
 |\psi_n\rangle=\sum\limits_{\sigma,\alpha,R}|\chi_{\sigma,\alpha,R}
\rangle\psi_{\sigma,\alpha,R,n}\;.
\label{eq:1}
 \end{eqnarray}
The local orbital $|\chi_{\sigma,\alpha,R}\rangle$ is a spin-orbital
with {\eg}~character at the Mn site $R$. It is a spin
eigenstate with spin $\sigma\in\{\uparrow,\downarrow\}$ and
spatial orbital character $\alpha$ $\in$ $\{a,b\}$ (where
$a{=}d_{x^2-y^2}$ and $b{=}d_{3z^2-r^2}$) \cite{sotoudeh17_prb95_235150}.  The
wavefunctions are Pauli spinor wavefunctions that account for the
non-collinear nature of the magnetization.

The three majority-spin {\ttg}~electrons of a Mn site with site index
$R$ are described by a spin vector $\vec{S}_R$
\cite{sotoudeh17_prb95_235150}.  The two Jahn-Teller active
  phonon modes $Q_{2,R}$ and $Q_{3,R}$ for each $\rm{MnO_6}$
  octahedron as well as the breathing mode $Q_{1,R}$ are expressed by
  the displacements of oxygen ions along the Mn-O-Mn bridge using
  Eq.~22 of \cite{sotoudeh17_prb95_235150}.

\subsection{Dynamics}
\label{secs:method}
The dynamics of the optical excitation and its relaxation processes
are described by Ehrenfest
dynamics\cite{mclachlan64_molphys8_3,tully98_faradaydiscuss110_407}:
That is, the electrons evolve under the time-dependent Schr\"odinger
equation, while the atoms are treated as classical particles and obey
Newton's equations of motion.
\begin{eqnarray}
i\hbar\partial_t \psi_{\sigma,\alpha,R,n}&=&
\frac{\partial E_{pot}}{\partial\psi^*_{\sigma,\alpha,R,n}}
\\
\partial_t S_{j,R}&=&\frac{2m_S}{\hbar}\vec{B}_R\times\vec{S}_R
\\
M_O\partial^2_t R_j&=& -\frac{\partial E_{pot}}{\partial R_j}
\end{eqnarray}
$R_j$ are the structural degrees of freedom of the oxygen ions and
$M_O$ is their mass. $m_S$ is the absolute value of the magnetic
moment of the \ttg-spins and the quasi-magnetic field $\vec{B}_R$ is
\begin{eqnarray}
B_{j,R}=\frac{3\hbar}{2m_S}\frac{\partial E}{\partial S_{j,R}}
\end{eqnarray}

Further details are given in
appendices~\ref{secs:apen1} and \ref{sec:spindynamics}.  We also allowed for a
strain dynamics, which does not have notable consequences on the
results presented here. For the sake of completeness, it is described
in appendix~\ref{sec:straindynamics}.

\subsection{Light Pulse} 
\label{sec:lightpulse}
The light pulse is described by a spatially homogeneous, but
time-dependent electromagnetic field (see appendix~\ref{secs:apen2})
\begin{eqnarray}
\vec{E}(t)=\vec{e}_A\omega
{\rm Im}\left(A_0^*\e{i\omega t)}\right)g(t)\;,
\label{eq:lightpulse}
\end{eqnarray}
where $A_0$ is the amplitude of the vector potential and $\hbar\omega$
is the photon energy. $c$ is the speed of light.  The polarization of
the electric field and of the vector potential is the unit vector
$\vec{e}_A$.

The temporal profile of the laser pulse is described by a Gaussian
\begin{eqnarray}
g(t)=\frac{1}{\sqrt[4]{\pi c_w^2}}\e{-\frac{t^2}{2c_w^2}}
\label{eq:pulsesshapefunction}
\end{eqnarray}
The intensity, which is proportional to $|g(t)|^2$, has the
full-width-at-half-maximum (FWHM) of $2c_w\sqrt{\ln2}$.

The light pulse is implemented with the Peierls-substitution method
\cite{peierls33_zp80_763,hofstadter76_prb14_2239}. Details are given in
Appendix~\ref{secs:apen2}.

\subsection{Parameters of the simulation}
\tab{tab:t1} summarizes the relevant parameters
used in this paper.
\begin{table}[!htb]
\centering 
\caption{Simulation parameters. For explanation, see text.}
\label{tab:t1}
\begin{tabular}{|l|l|}
\hline\hline
k-grid        & $1{\times}1{\times}1$ \\
supercell     & $N_x{\times}N_y{\times}N_z{=}8{\times}8{\times}4$\\ 
Mn sites per unit cell  &$N_{Mn}=256$\\
O sites per unit cell  &$N_{O}=768$\\
Mn-Mn spacing & $d_{Mn-Mn}=3.84$~\AA\\ 
time step     & $\Delta_t{=}0.040 (4\pi\epsilon_0)^2\hbar^3/(m_ee^4)$\\
               & $=0.97\times 10^{-18}$~s\\
oxygen mass   &  $M_{O}=15.998$~u \\
fictitious cell mass &  $M_{g}{=}8.0{\times}10^{10}$~m$_e$  \\
photon energy &$\hbar\omega=1.17$~eV\\
pulse length (FWHM) & $2\sqrt{\ln2}c_w=100~{\rm fs}$  \\
polarization  &  $\vec{e}_A=(\vec{e}_x+\vec{e}_y)/\sqrt{2}$   \\
initial tetragonal distortion& $g_x=g_y=1.0388$, $g_z=1.0077$\\
\hline
\hline
\end{tabular}
\end{table}  
We use a Cartesian coordinate system with the coordinate axes
$\vec{e}_x$, $\vec{e}_y$ and $\vec{e}_z$ pointing along the Mn-Mn
nearest-neighbor distances. The vectors $\vec{e}_j$ are defined with
length 1.

The Mn-Mn nearest-neighbor vectors are $g_xd_{Mn-Mn}\vec{e}_x$,
  $g_yd_{Mn-Mn}\vec{e}_y$, and $g_zd_{Mn-Mn}\vec{e}_z$ with
  $d_{Mn-Mn}$ from \tab{tab:t1} and scale factors $g_x$, $g_y$
  and $g_z$.  With $N_x$, $N_y$ and $N_z$, we denote the number of Mn
  sites in each of the three spatial directions in the supercell used
  in the calculation.

To describe perovskites, one usually refers to the lattice vectors of
Pbnm unit cell, an non-standard setting of space group 62.  The
lattice vectors $\vec{a}$, $\vec{b}$ and $\vec{c}$ of the Pbnm unit
cell are
\begin{eqnarray}
\vec{a}&=&\left(g_x\vec{e}_x-g_y\vec{e}_y\right)d_{Mn-Mn}
\nonumber\\
\vec{b}&=&\left(g_x\vec{e}_x+g_y\vec{e}_y\right)d_{Mn-Mn}
\nonumber\\
\vec{c}&=&2g_z\vec{e}_zd_{Mn-Mn}
\end{eqnarray}

The pulse length has been chosen consistent with the laser
pulses used in ultrafast pump-probe experiments\cite{raiser17_aenm7_1602174}.

\subsection{Diffraction patterns}
In order to link our results with diffraction experiments, we inspect
the intensities of diffraction for charges, orbitals and spins.  

The intensity of diffraction of an observable $\hat{X}$, such as
number of electrons, orbital occupations, or spins, with density
$x(\vec{r})$ is\cite{feil77_israeljchem16_103}
\begin{eqnarray}
I_X(\vec{q}) :=I_{X,0}(\vec{q})\left|\int_V d^3r\;
x(\vec{r})\e{-i\vec{q}\vec{r}}\right|^2
\end{eqnarray}
where the integration is performed over the illuminated sample volume
$V$.  $I_{X,0}(\vec{q})$ is the intensity of diffraction of a point
scatterer $x(\vec{r})=X\delta(\vec{r})$.

For a periodic lattice of atom-centered distributions
$x(\vec{r})=\sum_{R,\vec{t}} x_{R}(\vec{r}-\vec{R}_{R}-\vec{t})$, the
intensity of diffraction is
\begin{eqnarray}
I_X(\vec{q})
&=&
I_{X,0}(\vec{q})
N_{\in V}
\Omega_{G}\sum_G\delta(\vec{q}-\vec{G})
 C_X(\vec{G})
\end{eqnarray}
with the correlation function\cite{hotta00_prb61_R11879}
\begin{eqnarray}
C_X(\vec{G})=\frac{1}{N}
\biggl|\sum_{R=1}^N\e{-i\vec{G}\vec{R}_R}X_R(\vec{G})\biggr|^2
\end{eqnarray}
The distributions $x_R(\vec{r})$ are placed at the lattice sites
$\vec{R}_R+\vec{t}$, where $\vec{R}_R$ is the position of an atom
inside the first unit cell and $\vec{t}$ is the lattice translation
vector of a specific unit cell.  $N$ is the number of atoms in the
unit cell, $\Omega_T$ is the unit-cell volume,
$\Omega_G=(2\pi)^3/\Omega_T$ is the unit-cell volume of the reciprocal
lattice and $\vec{G}$ are the corresponding general reciprocal-lattice
vectors. $N_{\in V}$ is the number of sites in the illuminated region.
$X_R(\vec{q}):=\int d^3r\; x_{R}(\vec{r})\e{-i\vec{q}\vec{r}}$ is the
form factor of $x_R$.  Note that the correlation function is
meaningful only at the reciprocal lattice vectors $\vec{G}$.

Specifically, we address the following diffraction patterns:
\begin{itemize}
\item The charge-order correlation function\cite{hotta00_prb61_R11879} 
\begin{eqnarray}
C_{Q}(\vec{G})&:=&
\frac{1}{N}\biggl|\sum_{R=1}^N
\e{i\vec{G}\vec{R}_R}
\Big(n_{R}-\langle{n}\rangle\Big)\biggr|^2
\end{eqnarray}
 probes the deviation of the electron density from its mean value, i.e
 $X_R=n_R-\langle{n}\rangle$, where
 $n_R=\sum_{\alpha,\sigma}\rho_{\sigma,\alpha,R,\sigma,\alpha,R}$ is
 the number of {\eg}-electrons on Mn-site $R$ and
 $\langle{n}\rangle=1-x$ with the doping $x=\frac{1}{2}$ is the
 average number of electrons on Mn sites.
The one-particle-reduced density matrix 
\begin{eqnarray}
\rho_{\sigma,\alpha,R,\sigma',\alpha',R'}
=:\sum_n f_n \psi_{\sigma,\alpha,R,n} \psi^*_{\sigma',\alpha',R',n}
\end{eqnarray}
is given by the wave-function coefficients $\psi_{\sigma,\alpha,R,n}$
and the occupations $f_n$.
\item The orbital-order correlation function
\begin{eqnarray}
C_{O}(\vec{G})&{:=}&
\frac{1}{N}\biggl|\sum_{R=1}^N
\e{i\vec{G}\vec{R}_R}
\left(n_{x,R}-n_{y,R}\right)\biggr|^2
\label{eq:19}
\end{eqnarray}
probes difference between the orbital occupations  
  $X_{j,R}=n_{x,R}-n_{y,R}$,
where 
\begin{eqnarray}
n_{j,R}{=}\sum_{\alpha,\beta}\langle\theta_j|\chi_{\alpha}\rangle
\biggl(\sum_{\sigma}
\rho_{\sigma,\alpha,R,\sigma,\beta,R}\biggr)
\langle\chi_{\beta}|\theta_{j}\rangle
\end{eqnarray}
are calculated for the set of orthonormal orbitals $|\theta_j\rangle$
with $j\in\{x,y\}$, which are nearly axial in the $x$, respectively
the $y$ direction.  These orbitals are defined in terms of the
original basis states $|d_{x^2-y^2}\rangle$ and $|d_{3z^2-r^2}\rangle$
as
\begin{eqnarray}
|\theta_x\rangle&:=&
\left(|d_{3z^2-r^2}\rangle -|d_{x^2-y^2}\rangle\right)\frac{1}{\sqrt{2}}
\label{eq:19_2}
\end{eqnarray}
and 
\begin{eqnarray}
|\theta_y\rangle&:=&
-\left(|d_{3z^2-r^2}\rangle+|d_{x^2-y^2}\rangle\right)\frac{1}{\sqrt{2}}
\;.
\label{eq:19_3}
\end{eqnarray}

We skip spin and site indices of the Wannier-like orbitals for scalar
products, where they are identical.
\item The spin-correlation function\cite{hotta00_prb61_R11879} 
  \begin{eqnarray} 
  C_{S}(\vec{G})&:=&\frac{1}{N}\left|
  \sum_{R=1}^N{\rm{e}}^{i\vec{G}\vec{R}_R}
  (\vec{S}_R+\vec{s}_R)\right|^2
  \end{eqnarray}
  probes the total spin $\vec{X}_R=\vec{S}_R+\vec{s}_R$ of the
  Mn-sites, where $\vec{S}_R$ is the spin of the {\ttg}~electrons
  and $\vec{s}_R$ is the spin of the {\eg}~electrons at Mn-site $R$.
\end{itemize}

In order to account for the blurring of the diffraction peaks due to
fluctuations, we included for the time-dependent spin correlation
functions, in Figs~\ref{fig:region2mag}, \ref{fig:region3mag} and
\ref{fig:region4mag},  the contribution from a
$(3\times3\times3)$ set of reciprocal lattice vectors of the super
cell centered at the specified reciprocal lattice vector.

\section{Equilibrium}
\label{sec:equilibrium}
Before investigating the optically-induced dynamics, let us remind of
the salient features of the spin, charge, orbital, and lattice order
of {\pcmohalf} in equilibrium. The low-temperature phase serves as
initial state for the excitation and it determines the time-evolution
of the system.

{\pcmox} has a perovskite lattice formed by a network of
corner-sharing MnO$_6$ octahedra. Large cations such as Ca$^{2+}$ and
Pr$^{3+}$ fill the voids in between the octahedra.  The octahedral
network distorts to optimize the Coulomb energy between the ions,
which results in a characteristic pattern of the octahedral tilts.
This tilt pattern fits into the orthorhombic Pbnm 
unit cell, which holds four octahedra.

The Mn-ions occur in the formal Mn$^{4+}$ and Mn$^{3+}$ oxidation
states, with spin-aligned d-electrons on each Mn site. The additional
electron of Mn$^{3+}$ produces a Jahn-Teller distortion, which is
highly cooperative.

\begin{figure}[hbtp]
\begin{center}
\includegraphics[width=0.45\linewidth]{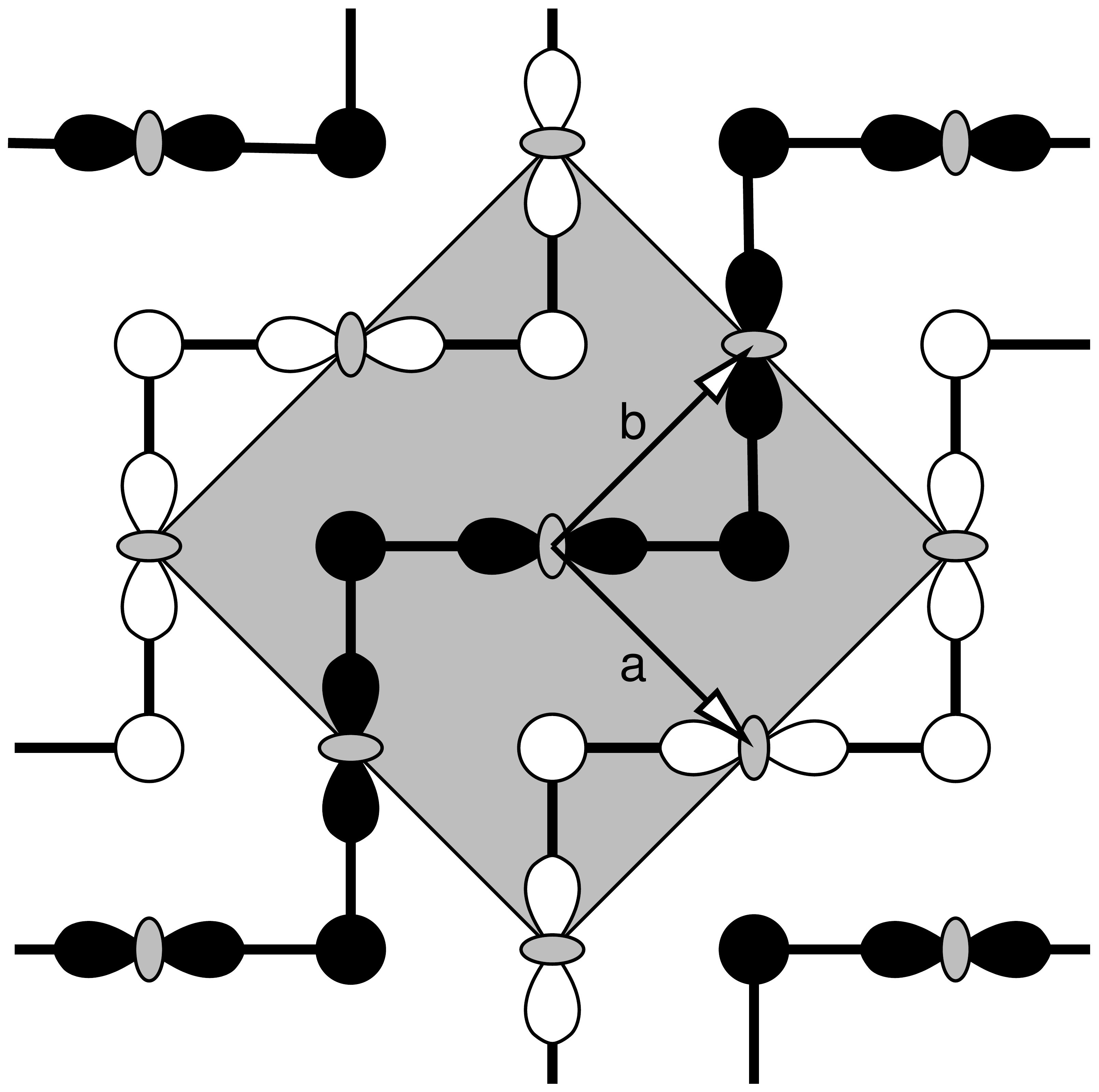}
\hfill
\includegraphics[width=0.45\linewidth]{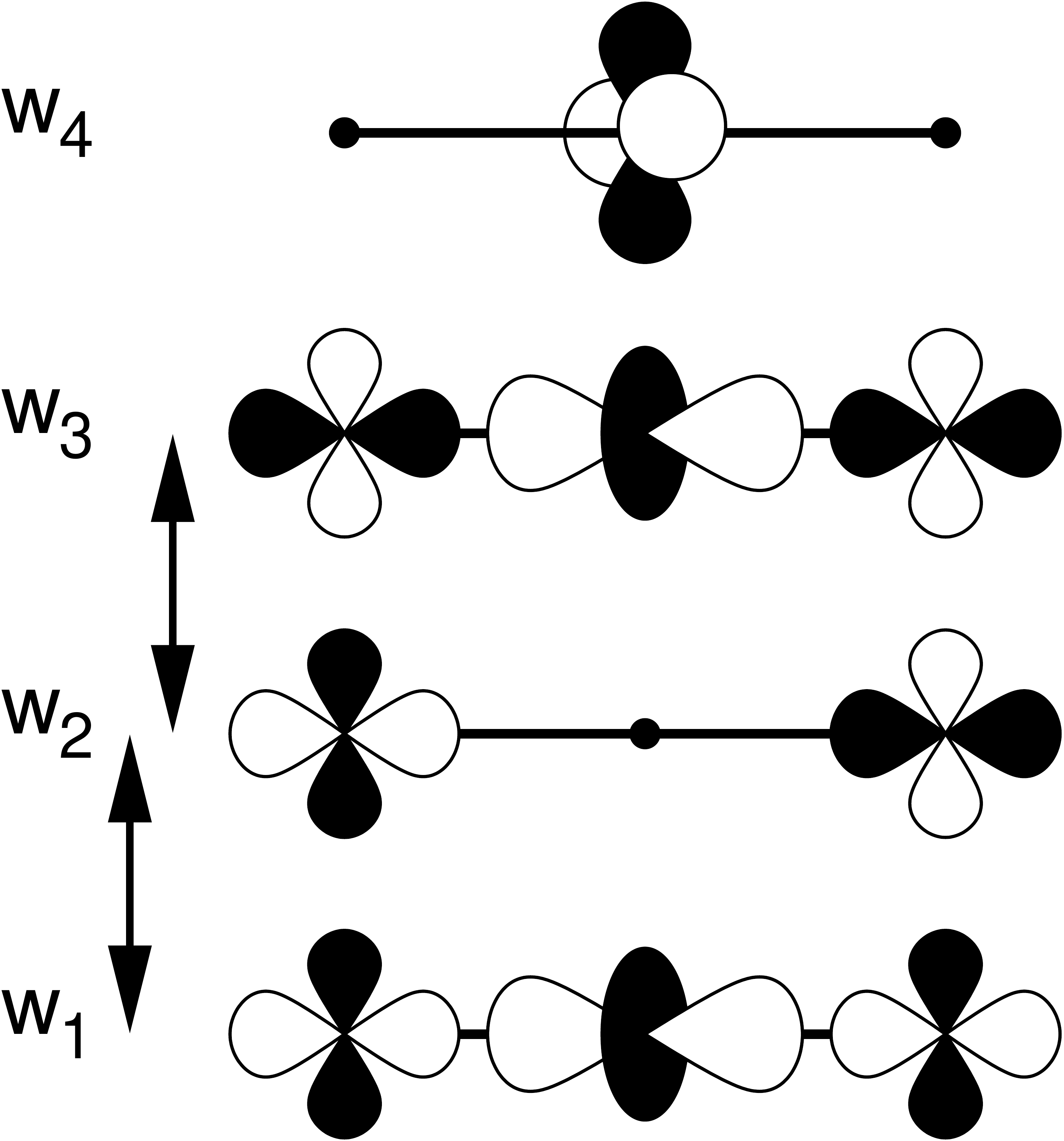}
\end{center}
\caption{\label{fig:wannierlikeorb} Left: CE-type magnetic order and
  orbital order in the ab-plane of the ground state of {\pcmohalf}.
  Black and white symbols indicate up- and down-spins. The
  orbital-polarized central orbitals are indicated by a $d_{3z^2-r^2}$
  orbital symbol in the corresponding direction. The corner sites have
  no orbital polarization and are indicated by circles. The gray
  square indicates the magnetic unit cell of the material in the
  ab-plane.  Also shown are the lattice vectors $a$ and $b$ of the
  orthorhombic Pbnm unit cell.  Right: Wannier-like orbitals along a
  trimer of the zig-zag chains of the CE-type magnetic structure. The
  sign of the orbital-lobes are indicated by black and white. The
  Wannier-like orbitals are orthogonal within and between trimers.
  The arrows connect orbitals with dipole-allowed transitions.
  Transitions between all other orbital pairs within and between
  trimers are dipole forbidden. The orbital $|w_1\rangle$ in the
  majority-spin direction is filled. The optical excitation lifts
  electrons from $|w_1\rangle$ to $|w_2\rangle$ in the majority-spin
  direction.}
\end{figure}

At half doping, {\pcmox} has a low-temperature phase, which is
described as charge and orbital
ordered.\cite{wollan55_pr100_545,goodenough55_pr100_564,jirak85_jmmm53_153}
The low-temperature phase of {\pcmohalf} is shown schematically in
\fig{fig:wannierlikeorb}.  It has a CE-type antiferromagnetic order
exhibiting ferromagnetic zig-zag chains in the ab-plane, which proceed
along the b-direction.  These zig-zag chains are antiferromagnetically
coupled among each other. Along the zig-zag chain, we can distinguish
alternating central and corner sites.  The central sites are
described formally as Mn$^{3+}$ ions and exhibit Jahn-Teller
distortion, while the corner sites are formally Mn$^{4+}$ ions with a
negligible Jahn-Teller distortion.

\begin{table}[!htb]
\caption{\label{tab:diffracpatt}Typical diffraction patterns. The
  diffraction spots are shown for various spin orders according to the
  notation of Wollan\cite{wollan55_pr100_545} as well as for charge
  and orbital diffraction of the CE-type low-temperature structure of
  {\pcmohalf}.  $(h,k,l)$ are the relative coordinates in the reciprocal
  lattice of the orthorhombic Pbnm setting. $h,k,l$ are integer unless
  mentioned otherwise. (color online)}
\begin{center}
\begin{tabular}{|l|c|c|c|}
\hline
spin     & \multicolumn{2}{|c|}{$h+k$} & $l$ \\
\hline
B-type & \multicolumn{2}{|c|}{even integer} & even integer \\
A-type & \multicolumn{2}{|c|}{even integer} & odd integer \\
C-type & \multicolumn{2}{|c|}{odd integer}  & even integer \\
G-type & \multicolumn{2}{|c|}{odd integer} & odd integer  \\
\hline
CE-type  & $h$&$k$&$l$\\
\hline
spin    & half-integer& half-integer  & odd integer\\
        & not integer& or integer & \\
charge & \multicolumn{2}{|c|}{h+k=odd integer}  & even integer \\
orbital & integer & half-integer & even integer \\
        &         & not integer  &             \\
\hline
\end{tabular}
\end{center}
\end{table}

\subsubsection{Diffraction patterns}
A transition at 250~K is attributed to the emergence of charge and/or
orbital order\cite{jirak85_jmmm53_153} from a disordered polaron
distribution at higher temperatures.  The charge and orbital order has
been explored experimentally by X-ray diffraction of the
Mn-K-edge\cite{zimmermann01_prb64_195133}.

The diffraction patterns for the low-temperature phase are listed
under the header CE-type in \tab{tab:diffracpatt}.  The diffraction
peaks are quoted as relative coordinates $(h,k,l)$ of the reciprocal
lattice vectors in the setting of the orthorhombic (Pbnm) crystal
structure obtained at room temperature.

\begin{figure}[htb]
\begin{center}
\includegraphics[width=\linewidth,clip=true]{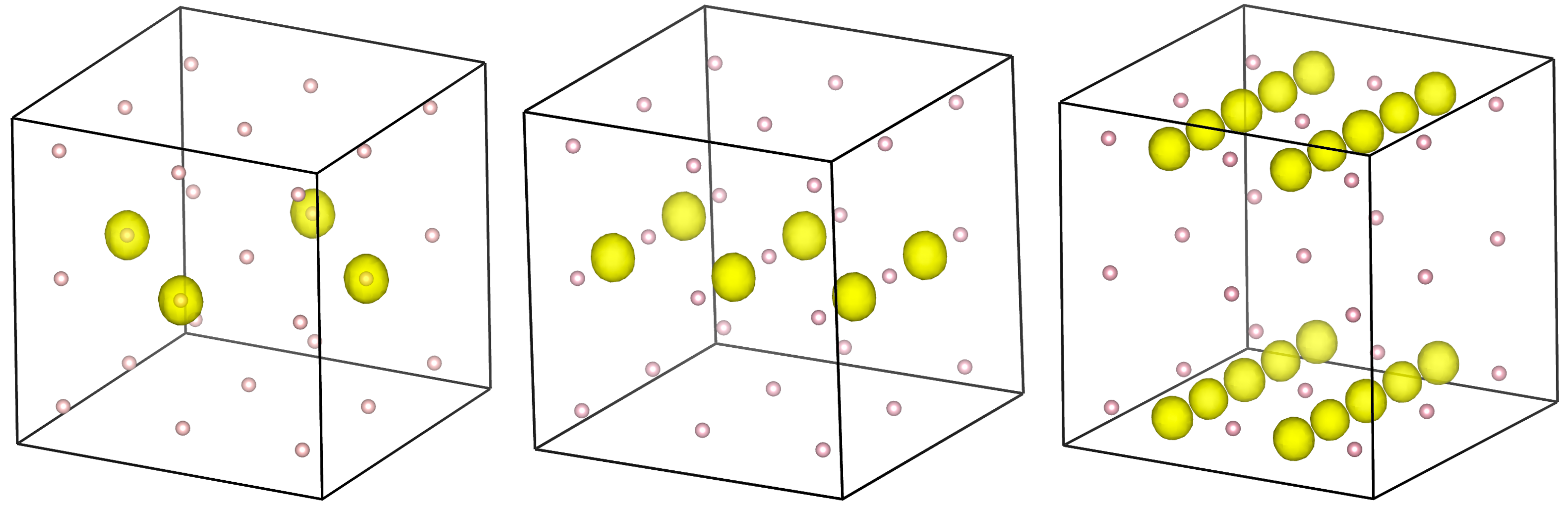}
\end{center}
\caption{\label{fig:cediffraction}Diffraction patterns for
  charge-order (left), orbital order (middle) and spin order (right)
  of the CE-type low-temperature phase of {\pcmohalf}.  The a-axis
  points right, the b-axis towards the back and the c-axis up. The
  small white spheres indicate points with integer $h,k,l$ in the Pbnm
  setting.  Reciprocal space is shown for $h,k,l\in[-1.25,1.25]$.
  (color online)}
\end{figure}

Zimmermann et al.\cite{zimmermann01_prb64_195133} exploited that the
diffraction spots of charge and orbital order can be distinguished
when scanning $(h,k,l)=(0,k,0)$ along the b-direction, the direction
of the zig-zag chain.  The dominant peaks of the Mn-K-edge 
with even integer $k$ are due to the pseudo-cubic atomic lattice of
Mn-sites. The charge order introduces additional diffraction peaks at
odd-integer $k$, and the orbital order produces diffraction spots at
half-integer (but not integer) values of $k$, as shown in
\fig{fig:cediffraction} and listed in \tab{tab:diffracpatt}.

At 175~K, {\pcmohalf} undergoes a N{\'e}el transition.
Neutron-diffraction studies\cite{jirak85_jmmm53_153} identify the
magnetic lines characteristic for the CE-type spin order for
{\pcmohalf}.

Our simulations reproduce the diffraction patterns due to charge,
orbital, and spin order for the CE-type ground state as shown in
\fig{fig:cediffraction}. 

The diffraction spots of other magnetic orders listed in
\tab{tab:diffracpatt} will be used to characterize the evolution of
the magnetic order following the light pulse:  B-type refers to a pure
ferromagnet, A-type refers to ferromagnetic planes that are stacked
antiferromagnetically in c-direction, C-type refers to ferromagnetic
Mn-lines running along the c-direction, which are
antiferromagnetically aligned with respect to neighboring strands. In
a G-type antiferromagnet, the Mn-sites are antiferromagnetic with
respect to all their neighbors. The magnetic orders can also be
described by their wave vector of the magnetization.  Expressed by the
pseudo-cubic lattice formed by the Mn-sites, the B-type order refers
to $\vec{k}=\frac{\pi}{d_{Mn-Mn}}(0,0,0)$, A-type refers to
$\vec{k}=\frac{\pi}{d_{Mn-Mn}}(0,0,1)$ C-type refers to
$\vec{k}=\frac{\pi}{d_{Mn-Mn}}(1,1,0)$ and G-type refers to
$\vec{k}=\frac{\pi}{d_{Mn-Mn}}(1,1,1)$.

\subsubsection{Charge order, orbital order and Jahn-Teller distortions}
The pattern of Jahn-Teller distortions in \fig{fig:wannierlikeorb} has
been attributed to checkerboard-like charge order in the ab-plane with
Mn$^{3+}$ ions at the central sites of each segment and
Mn$^{4+}$ ions at the corner sites of the zig-zag
chains.\cite{goodenough55_pr100_564}

More recently, this picture of charge order has been challenged:
Rather than deducing the charge state from the pattern of Jahn-Teller
distortions, experimental techniques such as core level spectroscopy,
(XANES, ELNES) and neutron diffraction measurements of the magnetic
moments provide a more direct access to the charge on the ions. These
experiments rule out a fully ionic picture and indicate the absence of
a considerable charge
disproportionation\cite{jirak85_jmmm53_153,garcia01_jpcm13_3229,grenier04_prb69_134419,jooss07_pnas104_13597,mierwaldt14_catalysts4_129}

This seeming contradiction between atomic structure and charge
distribution can be reconciled by considering orbital
polarization.\cite{sotoudeh17_prb95_235150} 
A Mn-ion has a complete orbital polarization,
when only one of the two spatial {\eg}~orbitals is occupied, while the
other is empty. Hereby, the shape and spin orientation of the occupied
orbital is not relevant.  When both spatial {\eg}~orbitals are
equally occupied, the atom lacks orbital polarization. 

To quantify the orbital polarization at site $R$, we determine the difference
$|f^{orb}_{1,R}-f^{orb}_{2,R}|$ of the orbital occupations
$f^{orb}_{\alpha,R}$, which are the eigenvalues of the spin-averaged
local density matrix $\bm{\rho}^{orb}_{R}$ with matrix elements
\begin{eqnarray}
\rho^{orb}_{\alpha,\beta,R}:=\sum_\sigma \rho_{\sigma,\alpha,R,\sigma,\beta,R}\;.
\end{eqnarray}

The orbital polarization $P_O$ is defined as
\begin{eqnarray}
P_O&:=&|f^{orb}_{1,R}-f^{orb}_{2,R}|
\nonumber\\
&=&\sqrt{
\Bigl(\rho^{orb}_{a,a,R}-\rho^{orb}_{b,b,R}\Bigr)^2+4\Bigl|\rho^{orb}_{a,b,R}\Bigr|^2
}
\end{eqnarray}
where $a$ and $b$ denote the two Mn {\eg}~orbitals.

Orbital polarization can be recognized indirectly via the resulting
Jahn-Teller distortions.  Mn-ions without orbital polarization do not
exhibit a Jahn-Teller distortion, irrespective of the number of
electrons in the {\eg}~shell.  Hence, there is a direct link between
Jahn-Teller distortions and orbital polarization. The connection to
the charge order is, however, indirect. It is present only in the case
of complete orbital polarization. This assumption is violated in
{\pcmohalf}\cite{sotoudeh17_prb95_235150}. The Jahn-Teller distortions
of the corner sites are small, not because of their charge, but because
of their lack of orbital polarization. The orbital polarization of the
corner sites is small because Wannier-like orbitals from two segments
of the zig-zag chain contribute equally to the two
{\eg}~orbitals.\cite{sotoudeh17_prb95_235150}

\begin{figure}[hbtp]
\begin{center}
\includegraphics[width=0.5\linewidth]{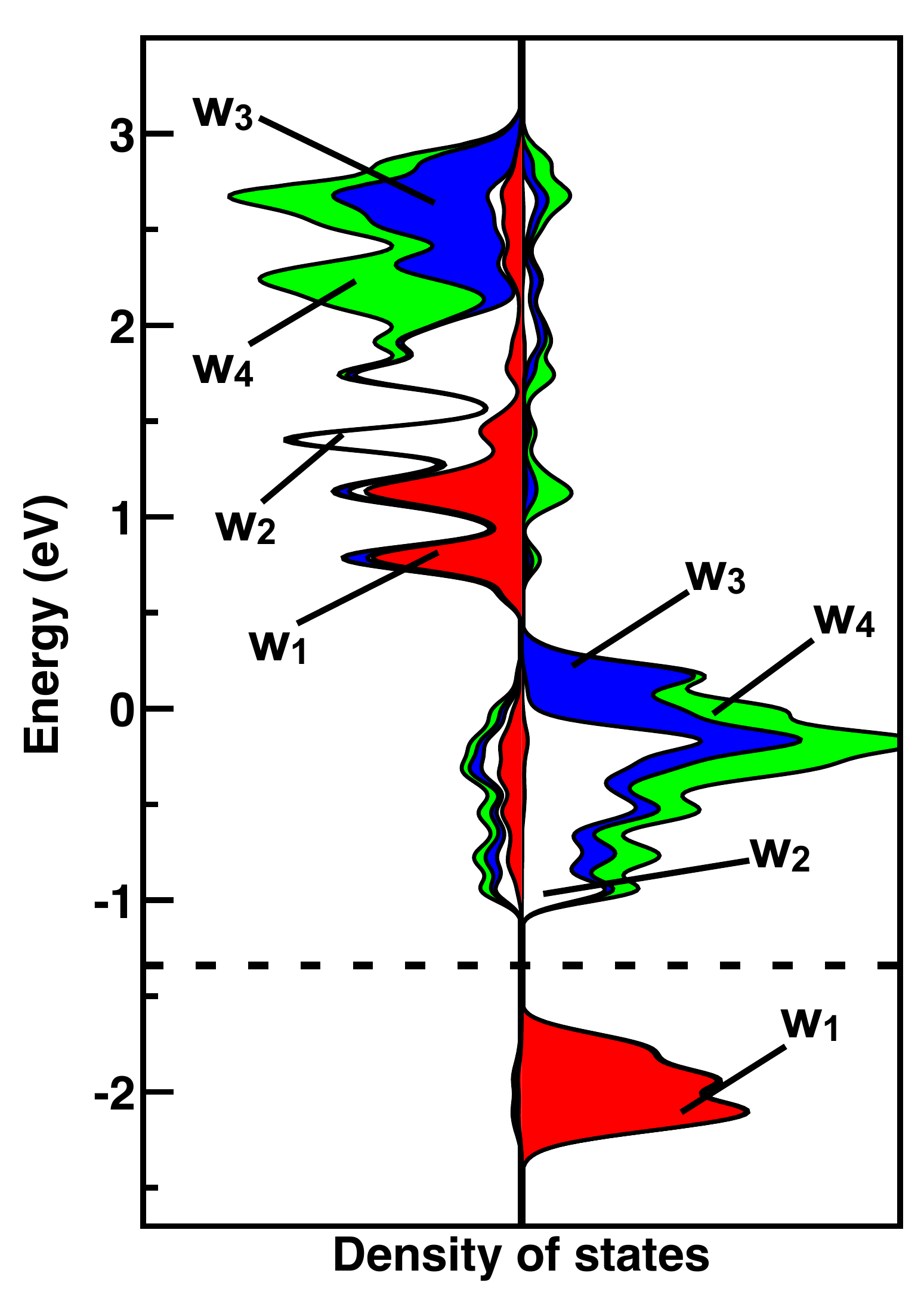}
\end{center}
\caption{\label{fig:dosce}Density of states of the ground state of
  {\pcmohalf} projected on the Wannier-like states $|w_1\rangle$ (red),
  $|w_2\rangle$ (white), $|w_3\rangle$ (blue) and $|w_4\rangle$
  (green). The axis of the majority-spin direction points right and
  that of the minority-spin direction points left. (color online)}
\end{figure}

As shown earlier\cite{sotoudeh17_prb95_235150}, the electronic structure of the
low-temperature phase of {\pcmohalf} can be rationalized using a
specific set of Wannier-like states formed from the Mn {\eg}~orbitals.
These states, shown in \fig{fig:wannierlikeorb}, are localized on
  specific segments of the zig-zag chains, which we denote, in the
  following, as trimers. The Wannier-like states are constructed as
  orthonormal eigenstates of a pseudo symmetry of a trimer, namely
  three orthogonal mirror planes through the central Mn-ion  of a
  trimer.  The functional form of the Wannier-like orbitals has been
  given in an earlier publication\cite{sotoudeh17_prb95_235150}.  The
  requirements given above determine the Wannier-like orbitals up to a
  single parameter, which governs the charge disproportionation
  between central and corner sites. With a suitable choice of this
  parameter, the first Wannier-like state $|w_1\rangle$ describes the
  occupied states almost perfectly. This can be seen in
  \fig{fig:dosce}, which shows that the occupied portion of the
  density of states can be attributed almost exclusively to
  $|w_1\rangle$.  To be specific, the one-particle-reduced density
  matrix of the {\eg}~states is well described by
\begin{eqnarray}
\rho_{\sigma,\alpha,R,\sigma',\beta,R'}&=&\sum_m
\langle\chi_{\sigma,\alpha,R}|w_{\sigma_m,1,m}\rangle
\delta_{\sigma,\sigma_m}\delta_{\sigma',\sigma_m}
\nonumber\\
&&\times\langle{w}_{\sigma_m,1,m}|\chi_{\sigma',\beta,R'}\rangle
\end{eqnarray}
where $|w_{\sigma,\alpha,m}\rangle$ is a Wannier-like orbital with
spin $\sigma$, spatial type $j$ with $j=\{1,2,3,4\}$ according to
\fig{fig:wannierlikeorb} and the index $m$ specifying a particular
trimer. $\sigma_m$ denotes the majority-spin direction of the trimer
with index $m$.

In our model calculations, the charge on the central site is 3.75~e
and that of the corner site is 3.25~e, which corresponds to a charge
disproportion of $q/e=3.5\pm\delta$ with $\delta=0.25$. This value
lies within the range of values obtained from various experimental
probes as discussed earlier\cite{sotoudeh17_prb95_235150}.

The shape of the orbital $|w_1\rangle$ is responsible for the orbital
order with strong orbital polarization on the central size  and
negligible orbital polarization on the corner sites.  Thus, the
electronic structure is consistent with both, the observed Jahn-Teller
pattern and the more direct measurements of the charge
state\cite{jirak85_jmmm53_153,garcia01_jpcm13_3229,grenier04_prb69_134419,jooss07_pnas104_13597,mierwaldt14_catalysts4_129}

When the charge order is described in terms of integral charge states,
they should be understood as oxidation states, which, per definition,
attribute electrons as a whole to the more electronegative
partner.\cite{goldbook}

This is, however, a definition rather than a detailed description of
an electron distribution.  The notion of integral charge states
Mn$^{3+}$ and Mn$^{4+}$ ions shall be understood in this context. The
real charge distributions in manganites are more subtle.

\section{Results and discussion}
\label{sec:results}

\subsection{Choice of the photon energy}
The photo-excitation in manganites occur through both d-to-d and
p-to-d transitions in the spectral energy range ${\sim}0.5$-$2.3$~eV
\cite{mildner15_prb92_35145,sotoudeh17_prb95_235150,loshkareva04_prb70_224406,hartinger06_prb73_024408,quijada98_prb58_16093}. While
the d-to-d transitions occur between Mn-3d states, the p-to-d
transitions occur between O-2p and Mn-d states
\cite{moskvin10_prb82_035106,sotoudeh17_prb95_235150,ifland17_njp19_063046}. The
transitions observed experimentally in the ${\sim}0.5$-$0.75$~eV
energy range are mainly dipole-allowed transitions from the occupied
to the unoccupied {\eg}~states.  While d-d transition on a single
Mn-site are dipole forbidden, there are dipole-allowed transitions,
which involve charge-transfer oscillations between different
Mn-sites.\cite{sotoudeh17_prb95_235150} In this work, we focus
entirely on transitions within the Mn {\eg}~shell.  The
charge-transfer transitions from O-p to Mn-d states dominate only at
comparatively higher energies
\cite{mildner15_prb92_35145,sotoudeh17_prb95_235150,loshkareva04_prb70_224406,hartinger06_prb73_024408,quijada98_prb58_16093}.

A quantity used to describe the excitation is the photon-absorption
density $D_{p}$, which is the total number of photons absorbed per
Mn-site and pulse. We calculate it as
\begin{eqnarray}
D_{p}{:=}\frac{\Delta E^{tot}_{f-i}}{N_{Mn}\hbar\omega}
\label{eq:defdph} 
\end{eqnarray}
 from the energy $\Delta{E}^{tot}_{f-i}$ added by the light-pulse to
 the system with $N_{Mn}$ Mn ions (see also
 \fig{fig:energyconservation}). The division by the photon energy
 $\hbar\omega$ and $N_{Mn}$ provides the number $D_{p}$ of absorbed
 photons per Mn-ion and pulse.

Another quantity, often used to characterize experiments, is the
pump-fluence $F_p$. It is the energy transmitted to the sample per
pulse and unit area.  The pump fluence determines together with the
pulse duration the intensity of the light-field.

The pump fluence $F_p$ is 
  \begin{eqnarray}
F_p=\frac{1}{2}|A_0|^2\omega^2c\varepsilon_0 \int dt\; g(t)^2
=\frac{1}{2}|A_0|^2\omega^2 c\varepsilon_0
\label{eq:pumpfluence}
 \end{eqnarray}
where $c$ is speed of light and $\varepsilon_0$ is vacuum
permeability. With $A_0$ we denote the amplitude of the vector
potential (see Eq.~\ref{eq:lightpulse}). The photon energy is
$\hbar\omega$. Due to the normalization of the pulse-shape function
$g(t)$, Eq.~\ref{eq:pulsesshapefunction}, the relation
Eq.~\ref{eq:pumpfluence} is independent of the pulse duration.

\begin{figure}[htp!]
\begin{center}
\includegraphics[width=\linewidth]{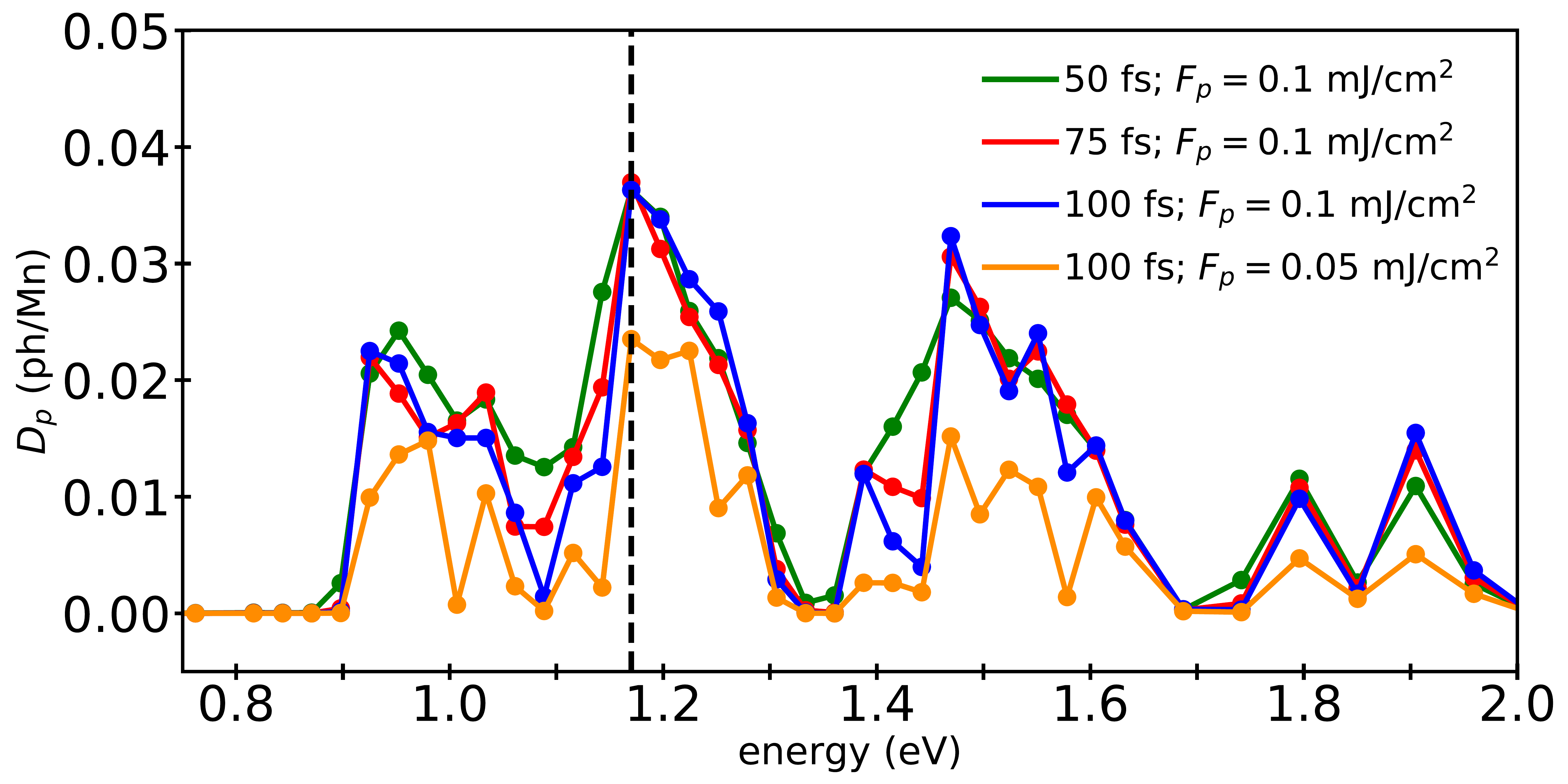}
\end{center}
\caption{\label{fig:absorptionspectra}Photon-absorption density
  $D_{p}$ defined in Eq.~\ref{eq:defdph} as function of the
  photon-energy $E=\hbar\omega$ for different intensities and pulse
  lengths. The dashed line indicates the photon-energy
  $\hbar\omega=1.17$~eV chosen for the simulations described below.}
\end{figure}

The spectral distribution of the photon-absorption density is shown in
\fig{fig:absorptionspectra}.

For the simulations discussed below, we have chosen the photon energy
equal to the absorption maximum of $\hbar\omega=$1.17~eV. The
position of the absorption maximum appears to be rather independent of
pulse length and light intensity.

\begin{figure}[htp!]
\begin{center}
\includegraphics[width=\linewidth,clip=true]{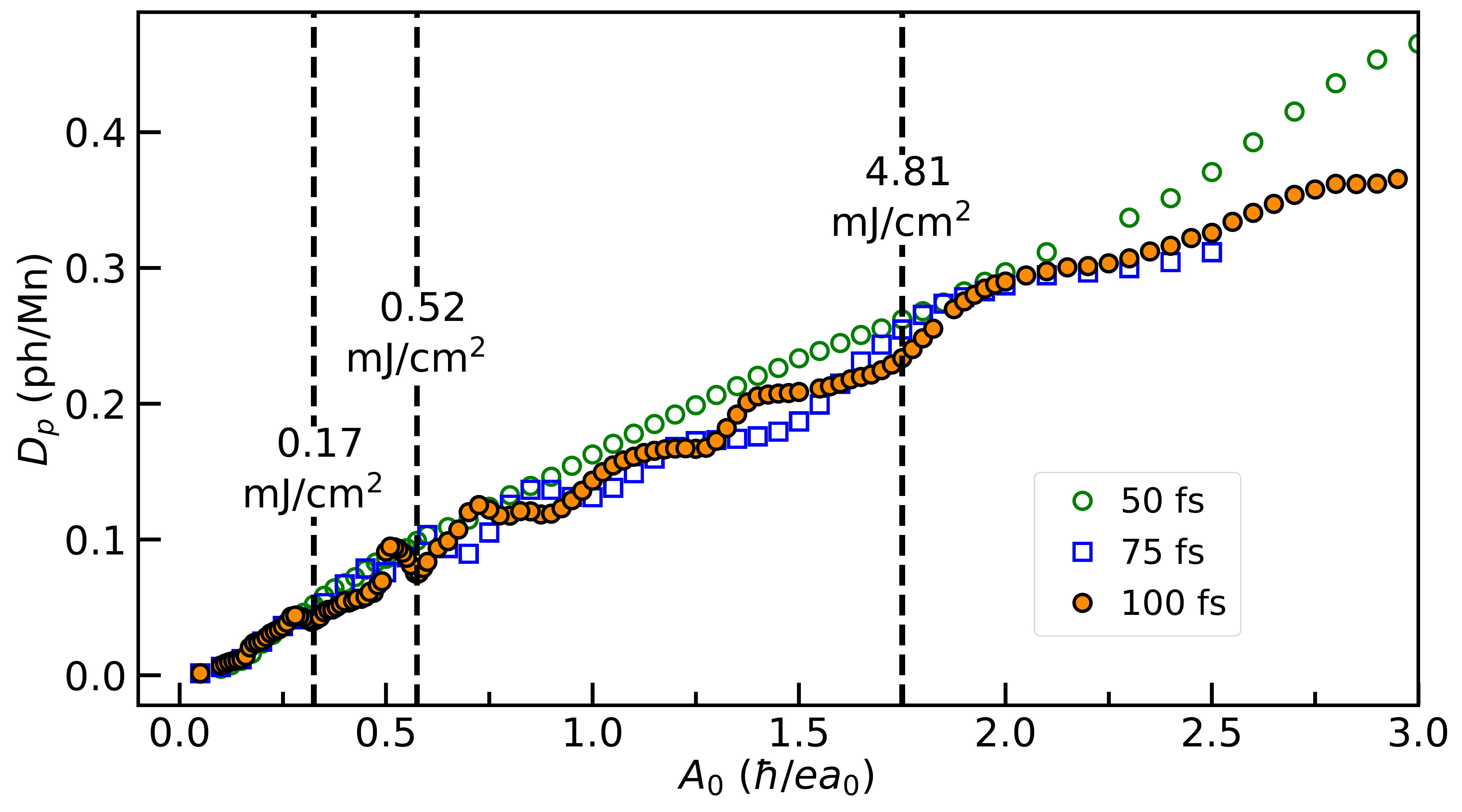}
\end{center}
\caption{\label{fig:absorptiondensity}Photon-absorption density
  $D_{p}$ defined in Eq.~\ref{eq:defdph} as function of the amplitude
  of the exciting field $A_0$ and a photon energy of
  $\hbar\omega=1.17$~eV. Also given are the pump fluences $F_p$ in
  units of $\rm{mJ/cm^2}$ according to Eq.~\ref{eq:pumpfluence} for
  the boundaries of the four regimes discussed below in
  section~\ref{sec:regimes}. (color online)}  
\end{figure}

\fig{fig:absorptiondensity} shows the photon-absorption density
$D_{p}$ as a function of the amplitude $A_0$ of the light field, as
defined in Eq.~\ref{eq:lightpulse}. 
At the largest fluences, every third electron is excited,
which explains the large changes of the magnetic and polaronic
microstructure observed in those simulations.

The photon-absorption density in \fig{fig:absorptiondensity}
  grows approximately linearly with the amplitude of the light
  field. This behavior differs from the low fluence regime, where the
  photon absorption grows quadratically with the amplitude. The
  approximate linear behavior can be attributed to damping and
  decoherence. For a two-state system, the optical Bloch equations
  with damping have a steady state solution with an excited-state
  population of $P_e(A_0)=A_0^2/(a+A_0^2)$, where $a$ is constant
  determined, among others, by detuning and friction
  parameters.\cite{steck19_book} $P_e$ has a point of inflection at a
  population of 25~\%, which explains the approximate linear behavior
  on the amplitude for the fluences studied here. Additional
  features seen in \fig{fig:absorptiondensity} can be attributed
  in parts to strongly damped Rabi oscillations.

\subsection{Regimes with distinct relaxation behavior}
\label{sec:regimes}
\begin{figure}[htp!]
\begin{center}
\includegraphics[width=\linewidth]{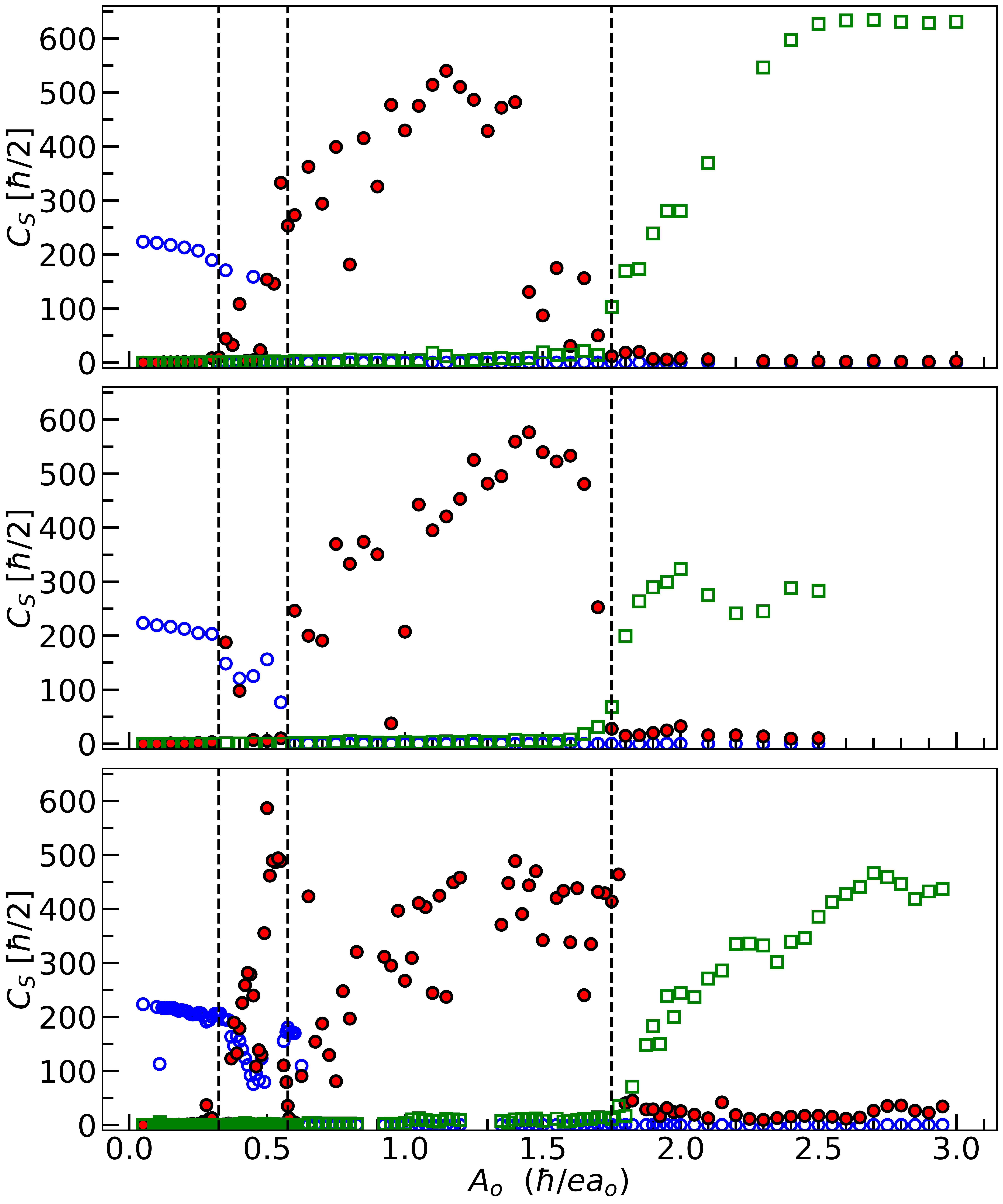}
\end{center}
\caption{\label{fig:distinctregions}Identification of the distinct
  regimes on the basis of the intensity of magnetic diffraction spots.
  The pulse length are 50~fs, 75~fs and 100~fs from top to bottom.
  Open blue circles show the minimum values of the spin-correlation
  function $C_S(\frac{1}{2},1,1)$ characteristic for the CE-type
  magnetic structure during the first 2~ps following the
  excitation. Red-filled circles are the maximum intensities
  $C_s(0,0,1)$ characteristic of the A-type magnetic structure and
  open green squares are maximum intensities $C_S(0,1,1)$ characteristic
  for the G-type structure. (color online)}
\end{figure}

 The relaxation following the optical absorption depends strongly on
 the pump-fluence $F_p$.  Based on the diffraction patterns, we
 identify four regimes with distinct relaxation behavior.  The
 diffraction intensities of the characteristic diffraction spots are
 shown in \fig{fig:distinctregions}.  The boundaries of the
 different regimes depend little on the pulse duration.

 The nature of these regimes are discussed in detail below. They can
 be characterized as follows:
\begin{enumerate}
\item In regime~I, the spin, charge, and orbital-order is
  preserved. Coherent phonons with long lifetime
  are observed.
\item In regime~II, the spin dynamics sets in, but the spin pattern
  relaxes back to the original state on a picosecond time scale. The
  charge order remains unaffected. Coherent phonons are present
    as in regime~I
\item In regime~III and beyond, the charge order is disrupted and the
  system is driven into a photo-induced ferromagnetic state.
\item In regime~IV, the system enters a photo-induced
  anti-ferromagnetic state.
\end{enumerate}

For the demonstration of the characteristic behavior in the four
fluence regimes, we selected the fluence values in \tab{tab:dpvsa0}.
\begin{table}[!hbt]
\caption{\label{tab:dpvsa0}Photon-absorption density $D_p$ and fluence
  $F_p$ for the $A_0$ values used in the graphs. (color online) }
\begin{center}
\begin{tabular}{|c|cccc|}
\hline
\hline
Regime                     & I     & II & III & IV\\
\hline
$A_0$($\rm{\hbar/(ea_0)}$) & 0.20  & 0.45   & 0.53 & 2.50 \\
$D_p$(ph/Mn)               & 0.025 & 0.058 & 0.094 & 0.326 \\
$F_p$(mJ/cm$^2$)             & 0.063 & 0.32  & 0.44 & 9.81 \\
\hline
\hline
\end{tabular}
\end{center}
\end{table}

\begin{figure}[htbp!]
\begin{center}
\includegraphics[width=\linewidth]{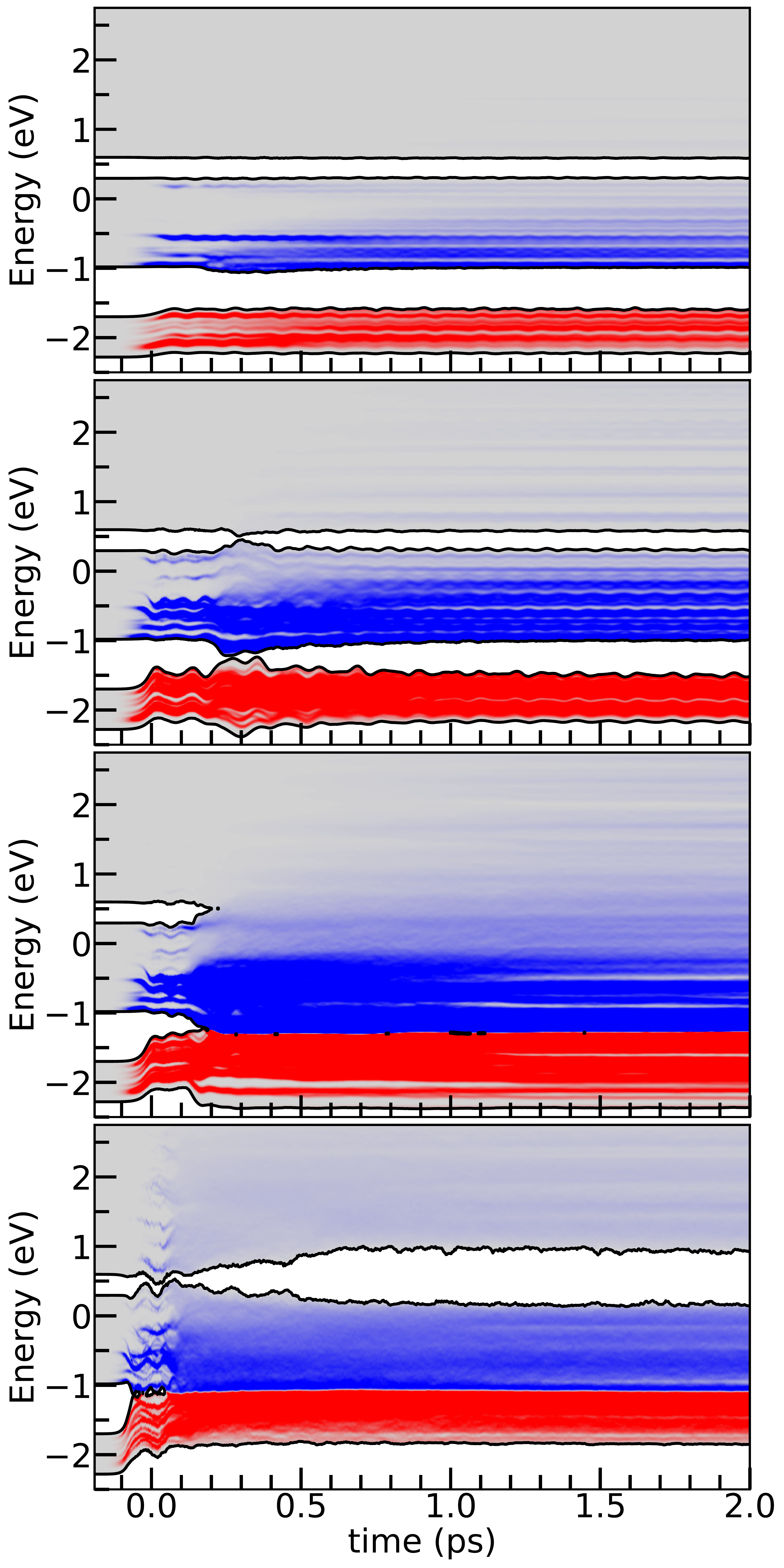}
\end{center}
\caption{\label{fig:dynamicsehdistrib}Time evolution of the electron
  and hole distributions induced by a femtosecond pulse of different
  intensities. Regime~I to IV from top to bottom, i.e.
  $A_0=0.20, 0.45, 0.53, 2.50$~$\rm{\hbar/(ea_0)}$. The instantaneous
  one-particle spectrum of Born-Oppenheimer energies is shown in
  gray.  The intensity of blue color indicates conduction electrons
  and that of red color indicates holes.}
\end{figure}

The time-dependent distributions of excited electrons and holes are
shown in \fig{fig:dynamicsehdistrib}. While the band structure for
regimes I and II are qualitatively similar, in regime~III the band gap
at the Fermi level collapses as a ferromagnetic metallic state is
formed. Also the band gap between minority and majority spins
collapses. The band gap between majority- and minority-spin orbitals
reoccurs in regime~IV, where the system evolves into a new
antiferromagnetic state. The antiferromagnetism is accompanied by a
smaller band width of majority- and minority-spin bands so that the gap
between them opens. Like in the ferromagnetic regime~III, the system
is metallic in regime~IV.

\subsection{Regime~I}
For the weak pump fluences of regime~I, the magnetic, charge and
orbital orders remain intact. The excitation can be described as
formation of electrons and holes in an essentially rigid band
structure. The electron-hole pairs are strongly coupled to breathing
modes and Jahn-Teller active phonons at the $\Gamma$-point.  As a
consequence, two coherent phonons with a long lifetime are excited.

The excitation can be rationalized using the Wannier-like states
introduced previously. They are shown in \fig{fig:wannierlikeorb} for
one segment of the zig-zag chain.  Unless mentioned otherwise, the
electric field of the light wave points along the b-direction of the
Pbnm unit cell, that is along to the zig-zag chains of the
ground-state magnetic structure.

\subsubsection{During the light pulse}
\begin{figure}[hbtp!]
\begin{center}
\includegraphics[width=\linewidth]{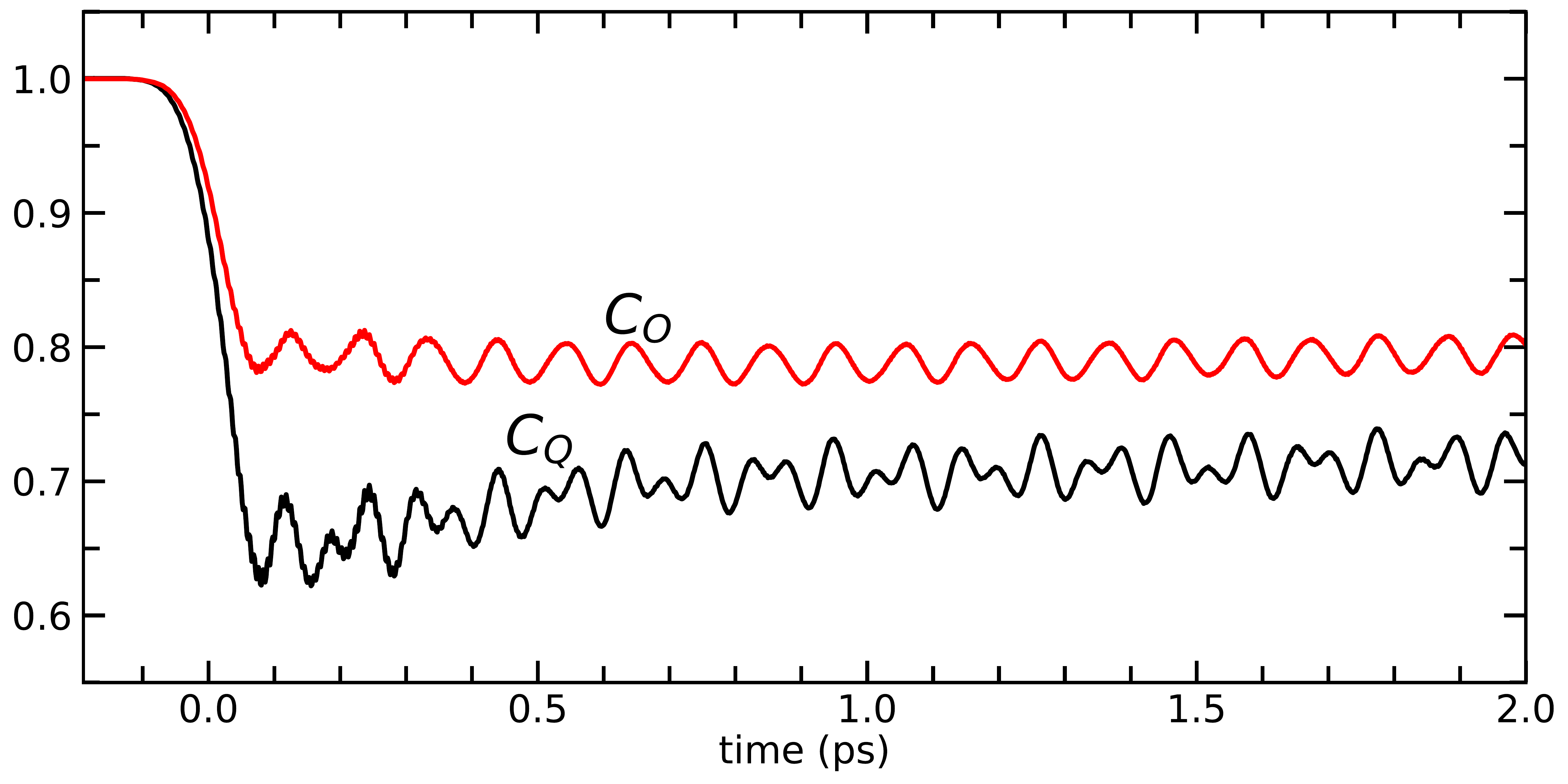}
\end{center}
\caption{\label{fig:region1I} Charge-order correlation $C_Q(1,0,0)$
  (black) and orbital-order correlation $C_O(0,\frac{1}{2},0)$ (red)
  as function of time for regime~I with
  $A_0=0.2$~$\rm{\hbar/(ea_0)}$. The correlations are scaled each so
  that their initial value is unity.  The correlations can be
  represented well by a superposition of two harmonic oscillations
  with frequency $\nu_1=10$~THz and $\nu_2=16$~THz, which are the
  frequencies of the two coherent phonons. (color online) }
\end{figure}

\begin{figure}[htp!] 
\begin{center}
  \includegraphics[width=\linewidth]{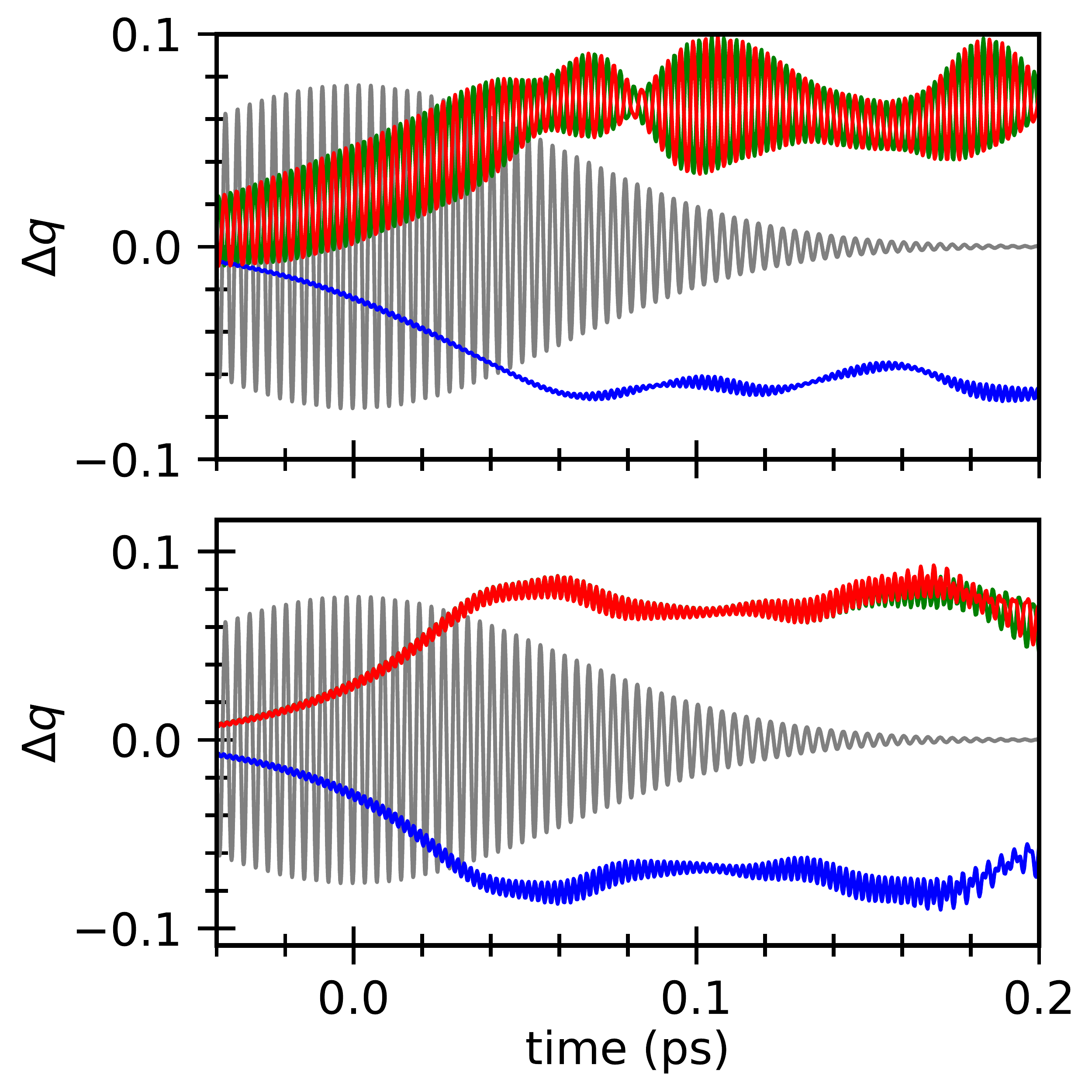}
\end{center}
\caption{\label{fig:initialico_regime1}Short-term dynamics for
  polarization along the a-axis (top) and along the b-axis (bottom) in
  regime~I with $A_0=0.20$~$\hbar/(ea_0)$.  Shown are the deviations
  $\Delta q$ of the charges from a trimer of the CE-type ground state
  as function of time: central site (blue) and corner sites
  (red/green). The instantaneous amplitude (arbitrary scale) of the
  light pulse is shown in grey. (color online)} 
\end{figure}

In the initial phase of the excitation, i.e. during the 100~fs light
pulse, charge and orbital order drop to lower values. Furthermore,
long-lived oscillations, discussed below, are initiated.  On top of
these effects, high-frequency oscillations of the electronic system
are induced that, however, decay after few tenths of
picoseconds. These oscillations are apparent in the charge-order and
orbital-order diffraction peaks in \fig{fig:region1I} and
\fig{fig:initialico_regime1}.

The only dipole-allowed transitions are between the bonding orbital
$|w_1\rangle$ and the non-bonding orbital $|w_2\rangle$ as well as
between latter, $|w_2\rangle$, and the antibonding orbital
$|w_3\rangle$ with the same spin direction shown in
\fig{fig:wannierlikeorb}.  There are no dipole-allowed transitions to
$|w_4\rangle$. Furthermore, there are no dipole-allowed transitions
between Wannier-like orbitals from different segments of the zig-zag
chains.

There is only one type of dipole-allowed transitions from the filled
states. It lifts an electron from the majority-spin bonding orbital
$|w_1\rangle$ to the non-bonding orbital $|w_2\rangle$ of the same
segment and with the same spin.

\begin{figure}[hbtp]
\begin{center}
\includegraphics[width=0.9\linewidth,clip=true]{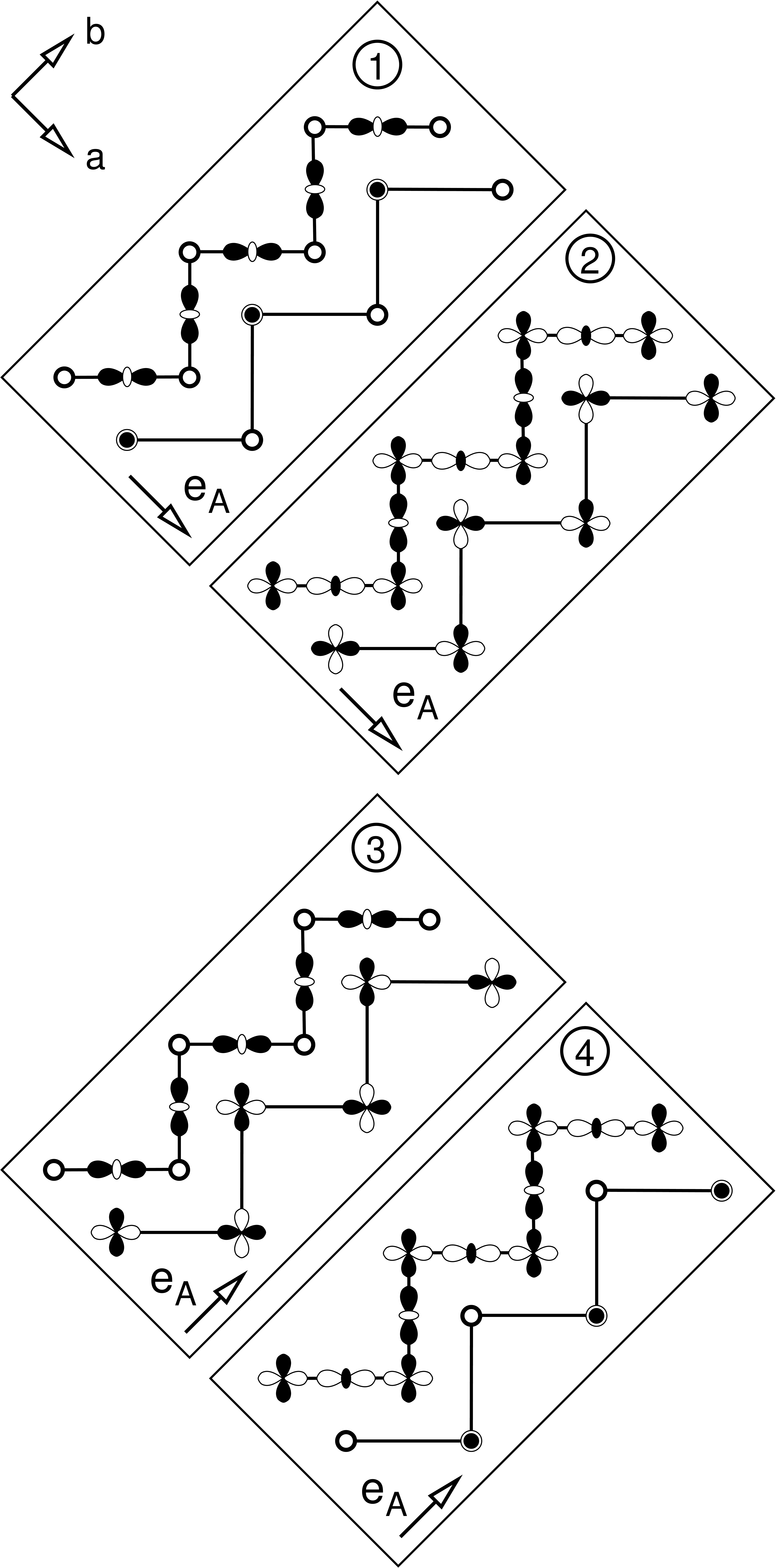}
\end{center}
\caption{\label{fig:ceblochwaves}Schematic drawing of Bloch waves
  involved in the photoexcitation with the electric field
  $(\vec{E}||\vec{e}_A)$ polarized perpendicular $(\vec{E}||\vec{a})$
  (1) and (2) or along $(\vec{E}||\vec{b})$ (3) and (4) to the zig-zag
  chains of the CE-type spin order. Each box shows the initial state
  with $|w_1\rangle$-character in the upper left and the final state
  with $|w_2\rangle$-character in the lower right. For each
  polarization two pairs of initial and final states, one without and
  one with sign change from one segment to the other, are shown. The
  induced charge density is related in first order to the product of
  initial and final state.}
\end{figure}

The nature of the high-frequency charge oscillation on a segment of
the zig-zag chain is rationalized via the Bloch waves of $|w_1\rangle$
and $|w_2\rangle$ character shown in \fig{fig:ceblochwaves}, which are
connected by the optical excitation. The excitation depends on the
polarization of the light.  Two representative Bloch waves of the
initial state with $|w_1\rangle$ character and the corresponding final
state with $|w_2\rangle$ character are shown schematically in
\fig{fig:ceblochwaves} for each polarization in the ab-plane.  The
product of initial and final state wave functions is proportional to
the first-order change of the charge density, which is in turn
responsible for the dipole oscillation that couples to the light
field.

For an electric field along the a-axis, i.e. perpendicular to the
zig-zag chains, large dipole oscillations between the two corner sites
are visible in \fig{fig:initialico_regime1}. The charges of the two
corner sites (red/green in \fig{fig:initialico_regime1}) oscillate
out-of-phase and with the frequency of the light field. They describe
the oscillating charge transfer between the corner sites. This is
consistent with the Bloch waves (1) and (2) in \fig{fig:ceblochwaves}
for $\vec{e}_A||\vec{a}$. The product of initial, $|w_1\rangle$
derived, states and final, $|w_2\rangle$ derived, Bloch waves lead to
charge contributions with alternating sign on the corner sites of a
zig-zag chain. The central site (blue in \fig{fig:initialico_regime1})
has a smaller oscillation with twice the frequency of the
light-field. This oscillation is due to the rescaling of the
$|w_1\rangle$ contribution required to maintain a normalized overall
wave function, while $|w_2\rangle$ is mixed in.

When the electric field is polarized along the b-axis, i.e. parallel
to the zig-zag chain, we observe in \fig{fig:initialico_regime1} only
oscillations with small amplitude and with the doubled frequency of
the light wave.  The two corner sites oscillate in-phase with the
doubled frequency. The charges on the central sites oscillate out of
phase with the corner sites.  This is consistent with the Bloch waves
(3) and (4) in \fig{fig:ceblochwaves}: The product of $|w_1\rangle$-
and $|w_2\rangle$-derived waves has contributions only on the corner
sites. However, there, the product of
two orthogonal orbitals is formed, which does not contribute to the
net charge. The net charge dipole coupling to the light field is not
apparent in the bulk. It would show up at the surface of the
material. Thus, only the second-order charge oscillations describing
the charge transfer from the central site to the corner sites is
visible in the bulk.  The charge oscillations
between the corner sites are initially absent and only kick in
at later times.

\subsubsection{Orbital order}
Let us now turn from the high-frequency electronic excitations during
the light pulse, to the changes that persist beyond the light pulse.

\begin{figure}[htp!] 
\begin{center}
  \includegraphics[width=\linewidth]{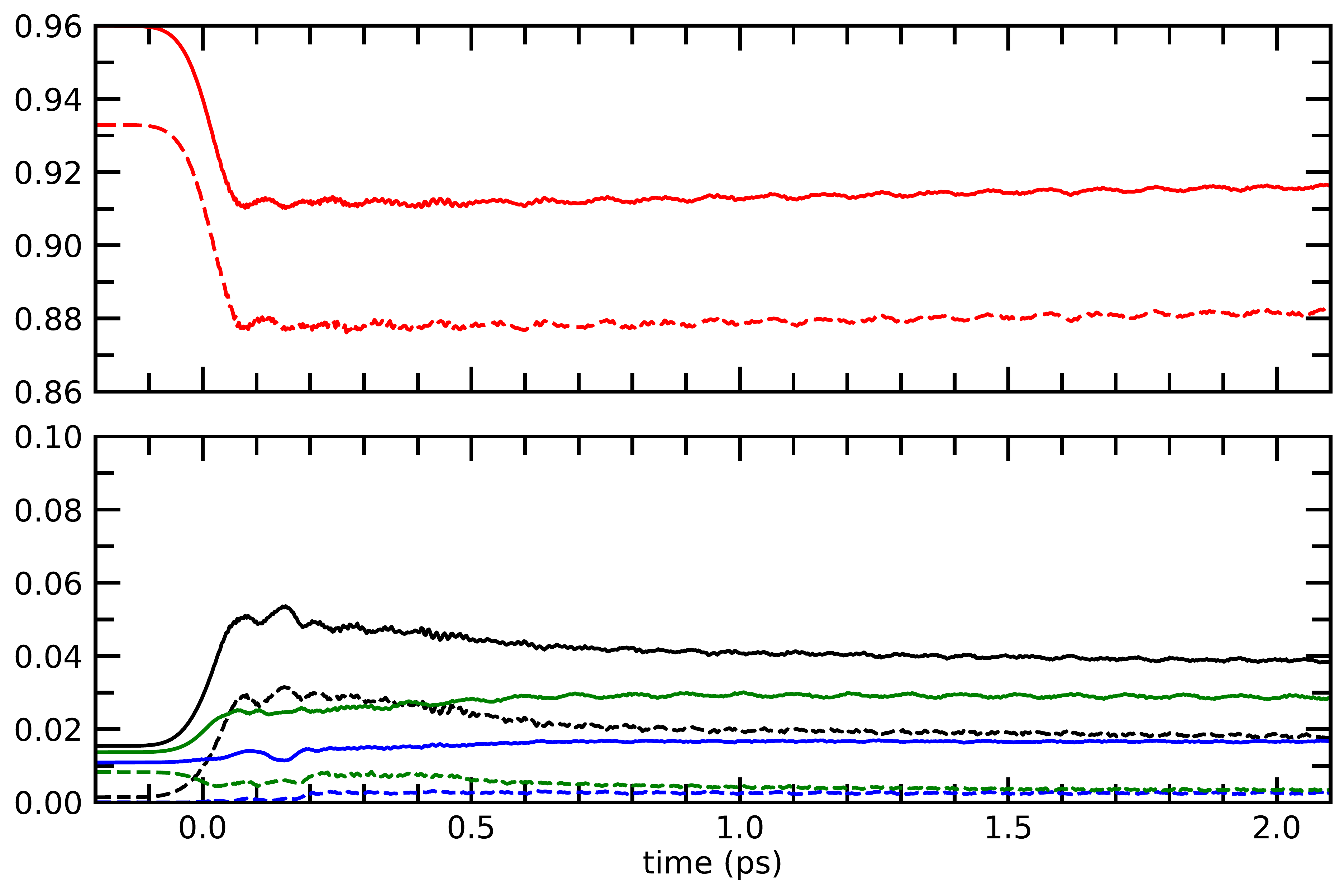}
\end{center}
\caption{\label{fig:region1F} 
Total occupancies $F^{tot}_{j}(t)$ (full lines) of Wannier-like
orbitals $|w_j\rangle$ and their spin polarization $F^{spin}_j(t)$
(dashed lines) as function of time for fluence regime~I with
$A_0=0.20$~$\rm{\hbar/(ea_0)}$.  Occupancies of $|w_{1}\rangle$ (red),
$|w_{2}\rangle$ (black), $|w_{3}\rangle$ (blue), and $|w_{4}\rangle$
(green). (color online) }
\end{figure}

For the analysis, we expand the time-dependent one-particle wave
functions $|\psi_n(t)\rangle$ in Wannier-like orbitals
\begin{eqnarray}
 |\psi_{n}(t)\rangle&=&\sum_{j,\sigma, m} |w_{j,\sigma, m}\rangle
 \langle w_{j,\sigma, m}|\psi_{n}(t)\rangle
 \label{eq:15}
\end{eqnarray}
The Wannier-like orbitals $|w_{j,\sigma, m}\rangle$ have spin $\sigma$
($\sigma\in\{\uparrow,\downarrow\}$) and belong to the trimer $m$ of
the unit cell. Index $j$ ($j\in 1,2,3,4$) selects one of the four
Wannier-states from a given trimer according to
\fig{fig:wannierlikeorb}.

The instantaneous occupancy $F_j^{tot}(t)$ of the $j$-th
  Wannier-like state $|w_{j}\rangle$ is
\begin{eqnarray}
 F^{tot}_j&=&\frac{2}{N_{Mn}}\sum_m\sum_\sigma Q_{j,m,\sigma,\sigma}
  \label{eq:17a}
\end{eqnarray}
and its spin polarization is
\begin{eqnarray}
 F^{spin}_j&=&\frac{2}{N_{Mn}}
\nonumber\\
&&\hspace{-1.5cm}
\times
\sum_m
\sqrt{(Q_{j,m,\uparrow,\uparrow}-Q_{j,m,\downarrow,\downarrow})^2+4
Q_{j,m,\uparrow,\downarrow}Q_{j,m,\downarrow,\uparrow}}
  \label{eq:17b}
\end{eqnarray}
with the number of Mn-sites $N_{Mn}$  and
\begin{eqnarray}
Q_{j,m,\sigma,\sigma'}&=&
\sum_n 
\langle w_{j,\sigma,m}|\psi_{n}\rangle f_n
\langle\psi_{n}|w_{j,\sigma',m}\rangle\;.
\end{eqnarray}

As shown in \fig{fig:region1F}, the dipole-allowed optical transitions
from $|w_{1}\rangle$ to $|w_2\rangle$ dominate at the low pump
fluences of regime~I. Thus, the occupancy of $|w_2\rangle$ grows at
the expense of $|w_{1}\rangle$ during the light pulse, while the
occupancies of $|w_{3}\rangle$ and $|w_{4}\rangle$ remain small.
After the light-pulse, the occupation of $|w_2\rangle$ remain almost
constant, which is one sign of the preservation of the ordered state
of the material.

Often, e.g. \cite{beaud14_naturematerials13_923}, the excitation is
attributed to an onsite d-to-d transition at the central Mn-site of a
trimer segment, which formally is the Mn$^{3+}$-ion.  The picture,
which emerges from our calculations, is more subtle. Rather than a
dipole-forbidden excitation on central Mn-ion from $|w_1\rangle$ to
$|w_4\rangle$, the excitation is a $|w_1\rangle$-to-$|w_2\rangle$
charge-transfer excitation, which displaces electrons from the central
Mn-ions to the corner Mn-ions. This transition does not exist in the
limit of complete charge order. The description in terms of
Wannier-like orbitals in \fig{fig:wannierlikeorb} explains the strong
optical absorption and it has consequences on the coherent modes
described below.

\subsubsection{Charge order}
The $|w_1\rangle$-to-$|w_2\rangle$ re-population rearranges electrons
from the central to the corner sites, which reduces the charge
disproportionation between central and corner sites.  The reduced
charge disproportionation reflects on the charge-order correlation
$C_{Q}(0,1,0)$ shown in \fig{fig:region1I}. The light pulse induces a
sharp drop of the charge-order diffraction intensity $C_Q(1,0,0)$ from
the initial value. Then, the intensity oscillates around this reduced
intensity, with little sign of recovery during our simulation.

In our simulation, the orbital-order peak $C_{O}(0,\frac{1}{2},1)$ has
a characteristic frequency of 10~THz, while the charge-order
diffraction peak $C_Q(1,0,0)$ exhibits one frequency at 10~THz and a
second one at 16~THz.

\subsubsection{Coherent vibrations}
\begin{figure}[htp!]
\begin{center}
\includegraphics[width=\linewidth]{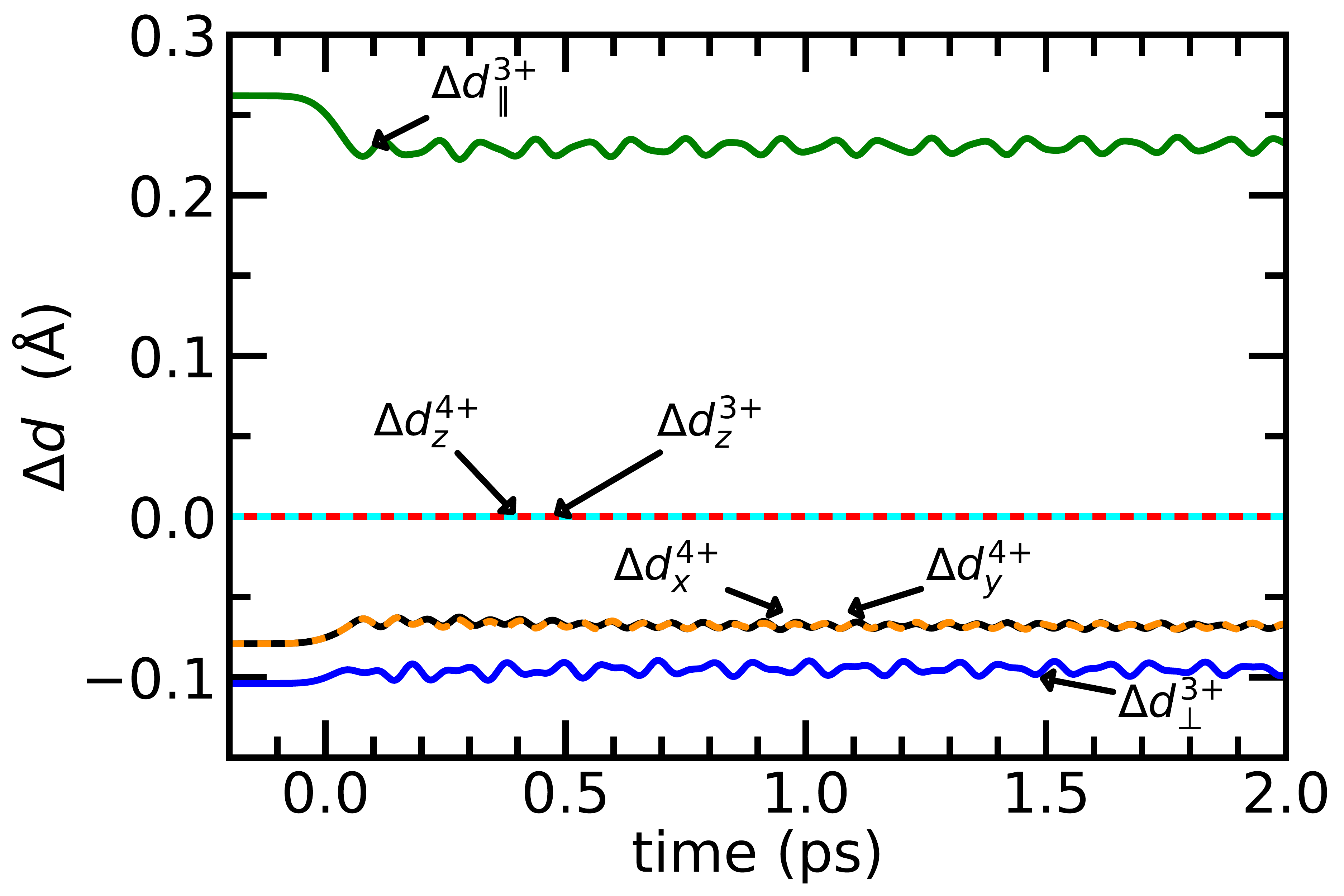}
\end{center}
\caption{\label{fig:region1ph}Phonon modes of region I with
  $A_0=0.20$~$\rm{\hbar/(ea_0)}$.  Shown are the expansions $\Delta d$
  of the oxygen distances along the octahedral axes. The superscript
  $3+$ refers to the formal charge state of the central Mn ion in a
  trimer of the zig-zag chain of the CE-type magnetic structure, while
  $4+$ refers to the corner site. For the central octahedron, the
  expansion along the trimer axis is $\Delta d^{3+}_{||}$ and the
  expansion in the ab-plane perpendicular to the trimer axis is
  $\Delta d^{3+}_{\perp}$. The planar displacements at the corner
  atoms along and in the ab-plane perpendicular to the trimer axis
  $\Delta d^{4+}_x$ and $\Delta d^{4+}_y$ are identical. The
  expansions in c-direction, indicated by a subscript $z$
  vanish. Visible is the initial reduction of the Jahn-Teller
  distortion, followed by the coherent breathing mode with 16~THz on
  the corner sites and a Jahn-Teller mode at the central site with
  10~THz. The displacements are averaged over equivalent atoms. (color
  online)}
\end{figure}

The removal of electrons from the central site during the
$|w_1\rangle$-to-$|w_2\rangle$ transition reduces its Jahn-Teller
distortion. The sudden change excites a Jahn-Teller
mode with $\nu=10$~THz, which affects
predominantly the central site. The displacements as function of time
are shown in \fig{fig:region1ph}.

The charge transfer from central to the corner sites reduces the
charge order and thus induces a planar breathing mode on the corner
sites with a frequency of $\nu=16$~THz.  Note, that an electron
addition to a Mn site populates the Mn-O antibonds, which, in turn,
expands the nearest neighbor distances.  The expansion of the corner
sites also affects the Jahn-Teller vibration on the central site.

On the time scale of a few picoseconds, the vibrations do not
dissipate significantly. Furthermore, the phonon modes are fully
coherent on the picosecond time scale of our calculation.

We attribute the lack of dissipation to the absence of heat
conduction.  In an experiment, a limited spot is illuminated and the
energy can escape from the illuminated region by heat conduction. In
our simulation, this process is prohibited, because the infinite
material is illuminated homogeneously.

Another reason for an unexpectedly slow dissipation is the specific
non-equilibrium state at hand. The coherent phonons are in contact
with a phonon bath that is extremely cold. Thus, the collision
probability of the coherent phonon with another one is extremely
small.

Pump-probe reflectivity measurements\cite{matsuzaki09_prb79_235131} of
$\rm{Nd_{1/2}Ca_{1/2}MnO_3}$, another manganite with the CE-type
ground state, provided frequencies with 2.5~THz ($82\rm{cm^{-1}}$,
6.7~THz ($224\rm{cm^{-1}}$), 10.2~THz ($339\rm{cm^{-1}})$ and 14.1~THz
($469~\rm{cm^{-1}}$).
A coherent vibration with 14~THz has been experimentally observed also
for $\rm{La_{1/2}Ca_{1/2}MnO_3}$\cite{lim05_prb71_134403} and
{\pcmohalf}
\cite{beaud14_naturematerials13_923,esposito18_prb97_14312} for a weak
photo-excitation, i.e. below a photo-absorption density of
$D_{p}<0.01$, and attributed to Jahn-Teller modes.

The two highest frequencies measured in $\rm{Nd_{1/2}Ca_{1/2}MnO_3}$
\cite{matsuzaki09_prb79_235131} at 10.2~THz
and 14.1~THz agree very well with those in our simulations. This
suggests a different assignment of the coherent vibrational modes: the
mode previously assigned to a octahedral rotation mode at 10~THz, is
in our simulation a Jahn-Teller oscillation on the central site of a
segment. The mode assigned as Jahn-Teller mode at 16~THz, is in our
simulation a symmetric breathing mode at the corner sites. Overlapping
with the 10~THz vibration, we also find the antisymmetric expansion of the
corner sites along the trimer axis. This latter vibration, however,
is not coupled to the optical excitation. It should be noted that
displacements of the oxygen ions in our model, denoted as the
Jahn-Teller and breathing modes, also have a small implicit component
from octahedral tilting.

Vibrations observed in the low-frequency range
2.4-7~THz\cite{lim05_prb71_134403,matsuzaki09_prb79_235131,beaud14_naturematerials13_923,raiser17_aenm7_1602174,esposito18_prb97_14312},
have been attributed to A-type ion motion and rotational modes of
MnO$_6$ octahedra. Our model does not describe these low-frequency
modes because it does not contain explicit A-type ions. Pure
octahedral rotations are not included, because our model does not
describe oxygen vibrations perpendicular to the oxygen bridge.

Our description of the coherent phonons differs from that given
earlier\cite{matsuzaki09_prb79_235131} as being due to an
instantaneous melting of the charge and orbital order.  The picture
emerging from our simulations is that of a mechanistic rather than a
thermal process.

\subsubsection{Magnetic order}
In the fluence regime~I, the magnetic order is preserved: The spin
angles deviate by about 10 degree from the ideal ground-state
arrangement.

\subsection{Regime~II: Transient changes of the magnetic order}
\label{secs:region_2} 
In the fluence regime~II, the Jahn-Teller-active phonons respond
similar to regime~I. After an initial period of about 200~fs, however,
also the spin order is perturbed.  The spin system relaxes back into
the ground state within few picoseconds, while the reduction of charge
and orbital order persists much longer.

\begin{figure}[htp!]
\begin{center}
\includegraphics[width=\linewidth]{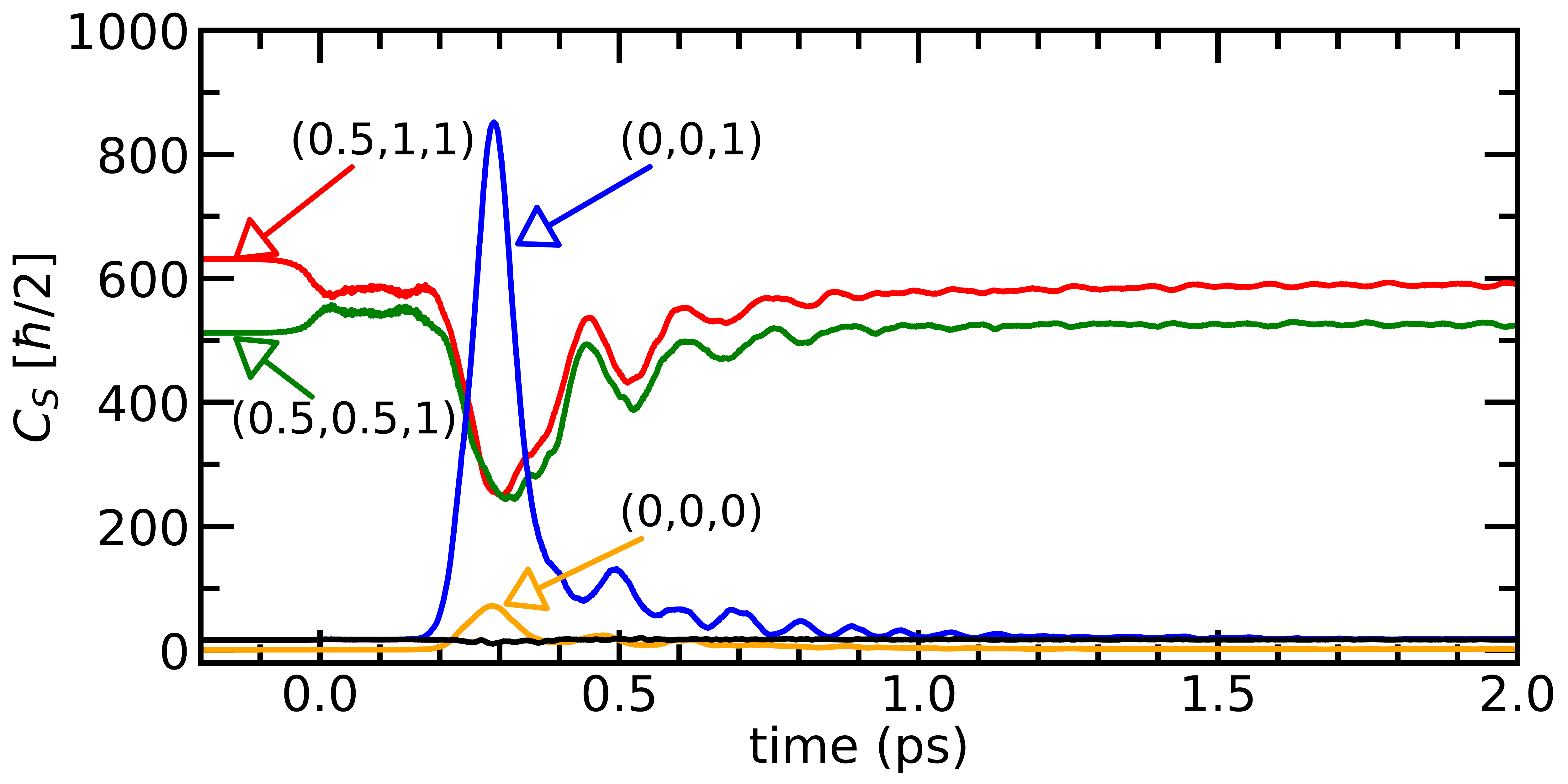}
\end{center}
\caption{\label{fig:region2mag}Spin-correlation function of regime~II
  with $A_0=0.45$~$\rm{\hbar/(ea_0)}$.  The correlation function
  $C_S(\frac{1}{2},1,1)$ and $C_S(\frac{1}{2},\frac{1}{2},1)$ are
  characteristic for the CE-type ground state.  The correlation
  function $C_S(0,0,1)$ is characteristic for A-type magnetic
  structure. The ferromagnetic (B-type) peak $C_S(0,0,0)$ is shown in
  orange. The CE-type structure recovers after a transient
  period. (color online)}
\end{figure}

\begin{figure}[htb]
\begin{center}
\includegraphics[width=\linewidth,clip=true]{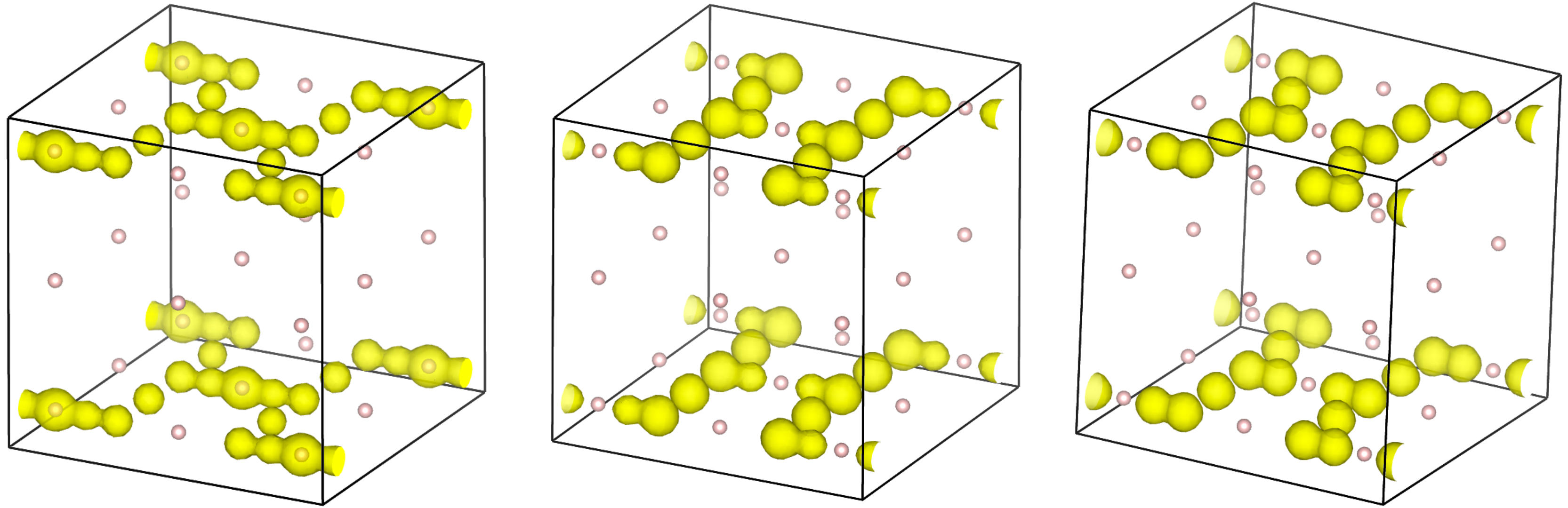}
\end{center}
\caption{\label{fig:spindiffraction_regimeII}Spin-diffraction patterns
  in regime~II with $A_0=0.45$~$\rm{\hbar/(ea_0)}$. At about 0.29~ps
  (left), the magnetic structure is a superposition of patterns of the
  CE-type and the A-type antiferromagnetic structure.  At 0.5~ps
  (middle), a spin-wave emerges, which is similar to obtained in
  equilibrium with $\delta=15$~\% excited electrons (right). (color
  online)}
\end{figure}

\begin{figure}[htp!]
\begin{center}
\includegraphics[width=\linewidth]{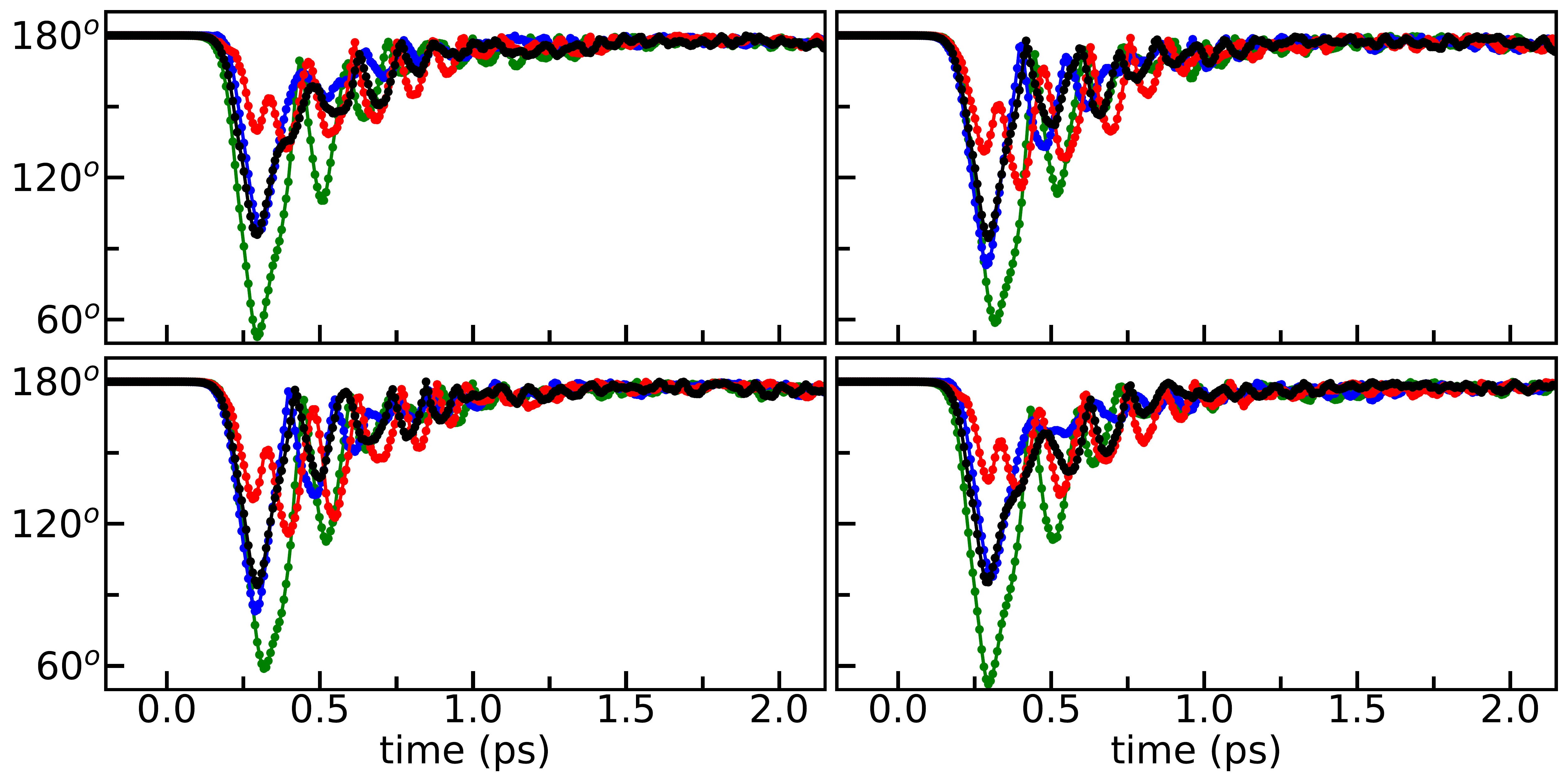}
\includegraphics[width=\linewidth]{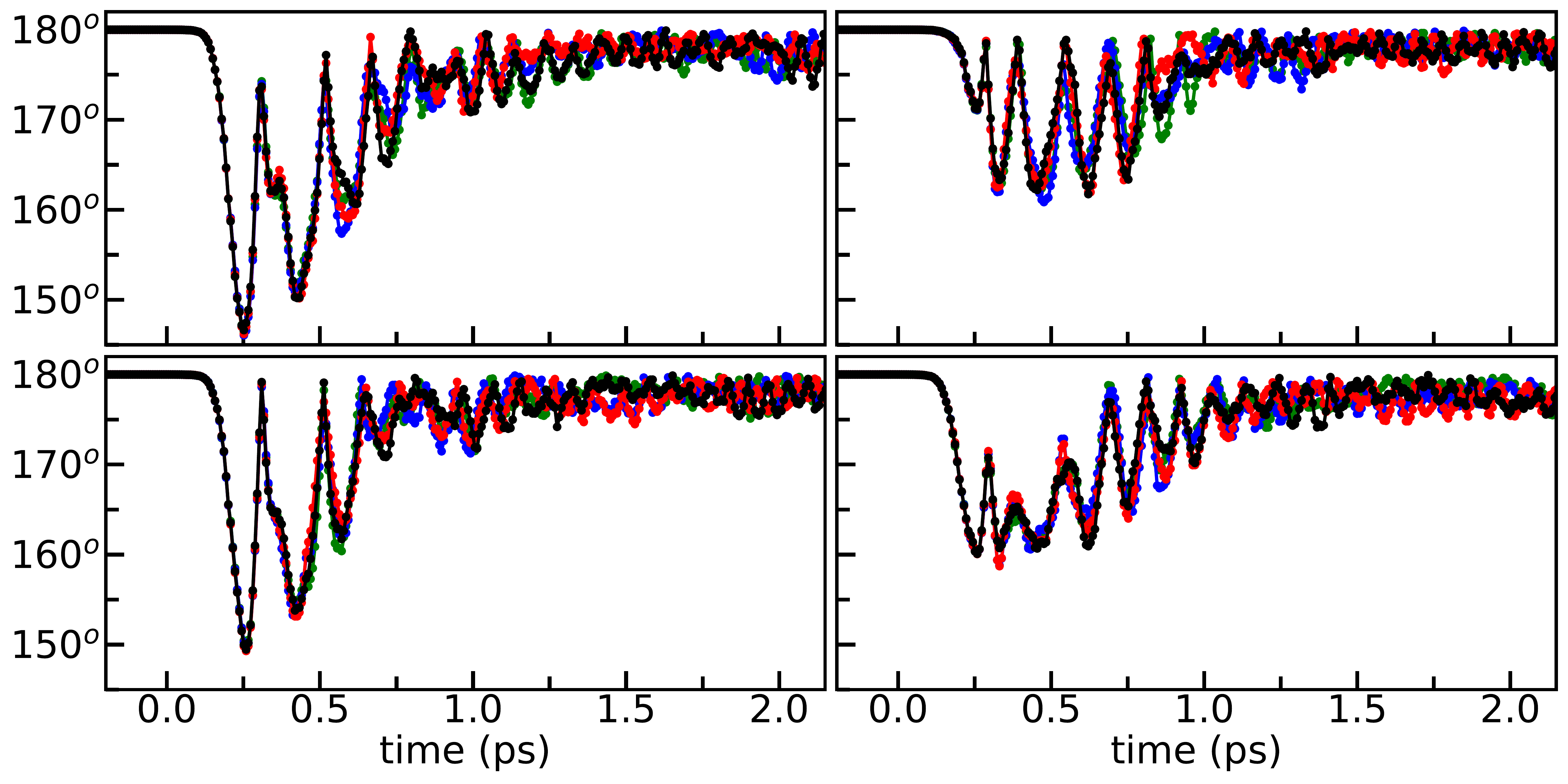}
\includegraphics[width=\linewidth]{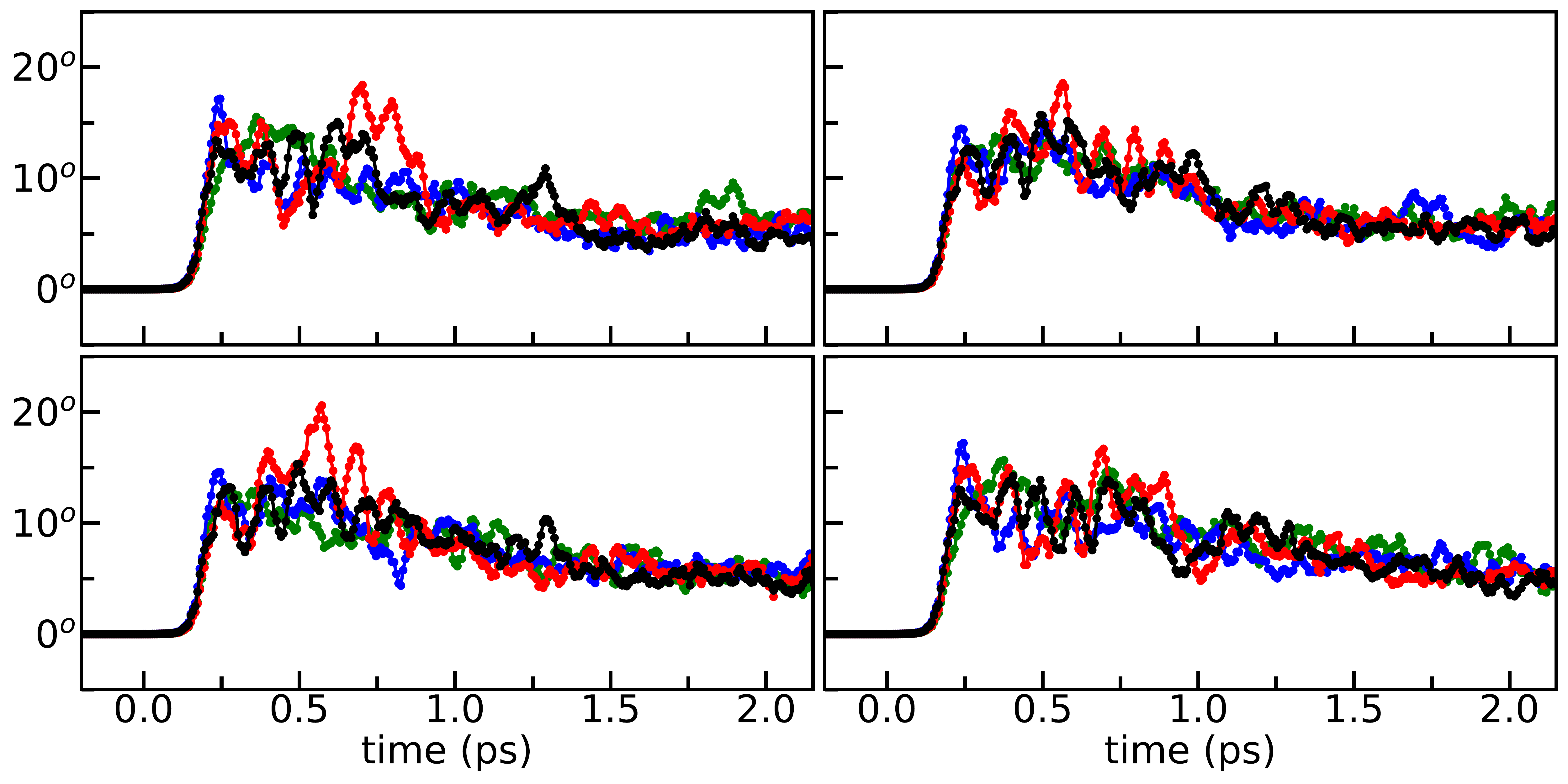}
\end{center}
\caption{\label{fig:region2S}Spin correlations in regime~II with
  $A_0=0.45$~$\rm{\hbar/(ea_0)}$ as function of time.  $\Phi_{n,n'}$
  is the angle between the spins $\langle\vec{S}\rangle_n$ of
  neighboring zig-zag chains $C_n$ and $C_{n'}$. The mean angle of a
  chain is $\langle\vec{S}\rangle_n=\frac{1}{N}\sum_{j\in C_n}
  \vec{S}_j$, where $N$ is the number of Mn-sites in the chain.  The
  top four figures show the angles $\Phi_{n,n'}$ for neighboring
  chains in the same ab plane. The middle four figures show the angles
  for neighboring chains stacked along the $c$-direction.  The bottom
  four figures show the ferromagnetic spin correlations within a chain
  $\Delta_n=\sqrt{ \frac{1}{N}\sum_{j\in C_n} \Bigl[
      \angle\Bigl(\vec{S}_j,\langle\vec{S}\rangle_n\Bigr) \Bigr]^2
  }$. (color online) }
\end{figure}

\subsubsection{Magnetic order}
As shown in \fig{fig:region2mag}, the spin correlations of the initial
CE-type order drop to lower values at about 0.2~ps, while a prominent
but short-lived signal of an A-type magnetic order shows up. This
signal lives for about 0.2~ps before it dies out again. The
spin-diffraction pattern during this period, which reflects the
superposition of CE-type and A-type spin patterns, is shown in the
left graph of \fig{fig:spindiffraction_regimeII}.

As shown in \fig{fig:region2S}, the angle of antiferromagnetic
neighbors changes by up to 90$^\circ$.  Besides some fluctuations, the
ferromagnetic order within the zig-zag chain is preserved in the
fluence regime~II. What is affected most, is the spin correlation
between neighboring chains in the ab plane.  The onset time decreases
with increasing fluence.

\begin{figure}[htp!]
\begin{center}
\includegraphics[width=\linewidth]{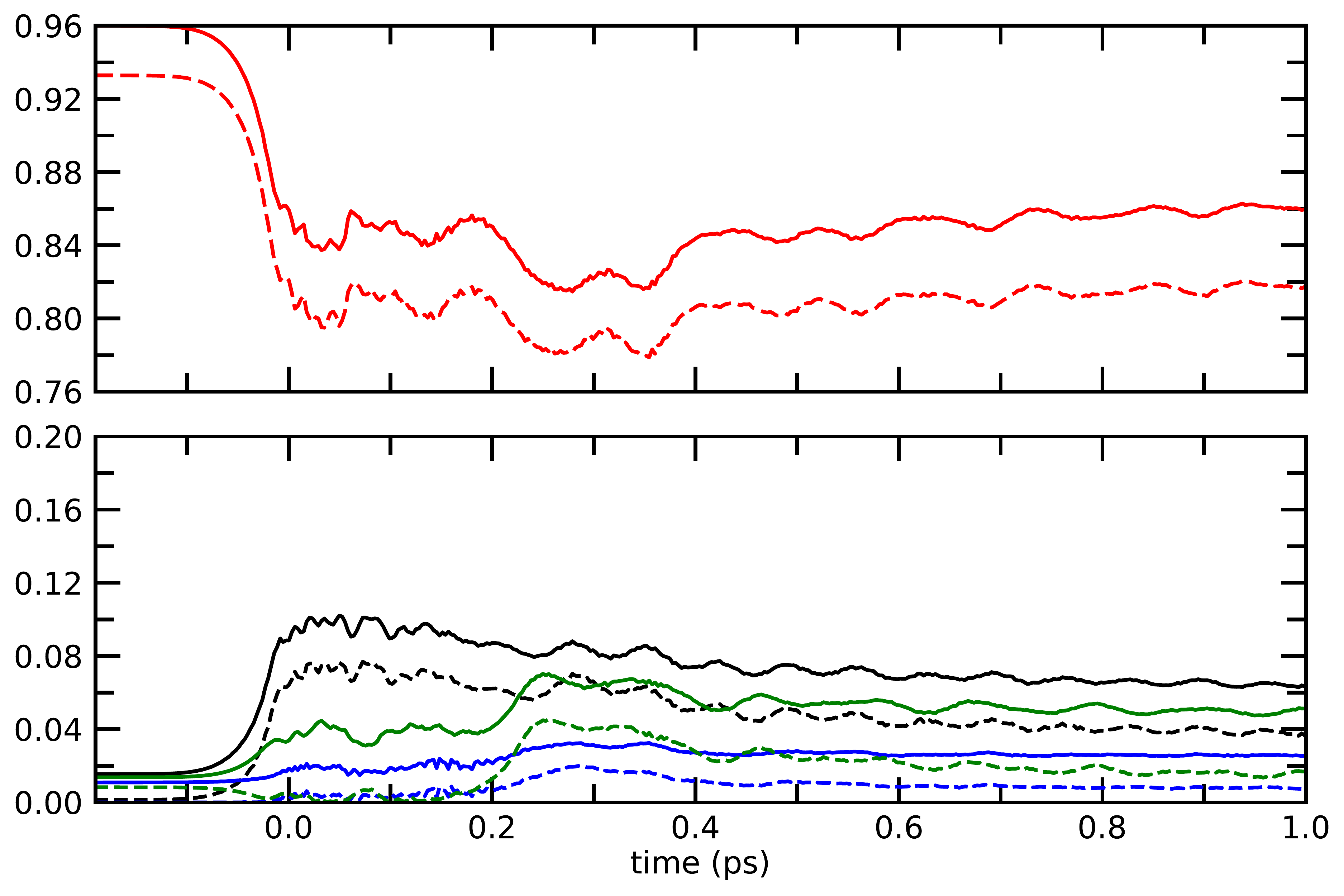}
\end{center}
\caption{\label{fig:region2F} Total occupancies $F^{tot}_{j}(t)$ (full
  lines) of Wannier-like orbitals $|w_j\rangle$ and their spin
  polarization $F^{spin}_j(t)$ (dashed lines) as function of time for
  fluence regime~II with $A_0=0.45$~$\rm{\hbar/(ea_0)}$.  Occupancies
  of $|w_{1}\rangle$ (red), $|w_{2}\rangle$ (black), $|w_{3}\rangle$
  (blue), and $|w_{4}\rangle$ (green). (color online) }
\end{figure}

We attribute the response of the spin system to the optically-induced
intersite spin transfer (OISTR)\cite{dewhurst18_nanolett18_1842}
caused by the coupling of the majority-spin $|w_2\rangle$ states with
minority-spin states $|w_1\rangle$ and $|w_4\rangle$ of a neighboring
zig-zag chain: the time-dependent populations of the relevant
Wannier-like orbitals are shown in \fig{fig:region2F}. The
$|w_2\rangle$ state is populated by the photo excitation. It is
located at the corner sites and has lobes pointing towards the central
atom of a neighboring zig-zag-chain. Thus, there is a spatial overlap
of the $|w_2\rangle$ orbitals with the minority-spin $|w_1\rangle$ and
$|w_4\rangle$ orbitals of a neighboring chain.  The excitation into
the $|w_2\rangle$ orbital is therefore accompanied by a spin transfer
between neighboring zig-zag chains in the ab-plane.  The
delocalization of electrons among the antiferromagnetically coupled
zig-zag chains changes the magnetization of the {\eg}~electrons, which
in turn acts onto the classical spins describing the {\ttg}~electrons
and which causes transient or permanent changes of the magnetization
pattern.

In the period from 0.2 to 0.4~ps in \fig{fig:region2F}, we observe a
transfer of weight from $|w_1\rangle$ orbitals of one chain to the
majority spin $|w_4\rangle$ of a neighboring chain. This is a
consequence of the transient magnetic transition, which takes place
during this period. The hybridization between these orbitals becomes
possible because the spin orientation of neighboring chains deviate
from $180^\circ$.

In order to obtain a better understanding of these spin fluctuations,
we investigated the spin structure as function of the ratio of excited
electrons.  For this purpose, we reduce the occupation for the
$\frac{1}{2}N_{Mn}$ occupied states from $1$ to $1-\delta$ and we
increase the occupation of the first $\frac{1}{2}N_{Mn}$ unoccupied
states from $0$ to $\delta$. Then, we investigate the ground state as
function of $\delta$, where all degrees of freedom except strain are
relaxed.

The original CE-type spin order is preserved up to $\delta=11$~\%.
For larger $\delta$, a non-collinear but co-planar spin-wave structure
emerges, where every two antiferromagnetic zig-zag chains in the
ab-plane pair up and form a finite spin-angle with the spin axis of
the next pair. The ferromagnetic spin order within the zig-zag chains
and the strict antiferromagnetic coupling in c-direction remain
intact.  The angle of the spin axes grows with increasing $\delta$
until the angle approaches 90$^\circ$. At this point,
$\delta\approx16$~\%, the system collapses into a ferromagnetic
metallic state.

Hence, we interprete the spin fluctuation observed in the simulations
as the onset of a N{\'e}el transition, which, in this case, is driven
by photo-excited rather than by thermal
$|w_1\rangle\rightarrow|w_2\rangle$ electron-hole pairs.  In this
context, the OISTR concept may be generalized to a temperature-induced
intersite spin transfer (TISTR).

\subsubsection{Charge order}
\begin{figure}[htp!]
\begin{center}
\includegraphics[width=\linewidth]{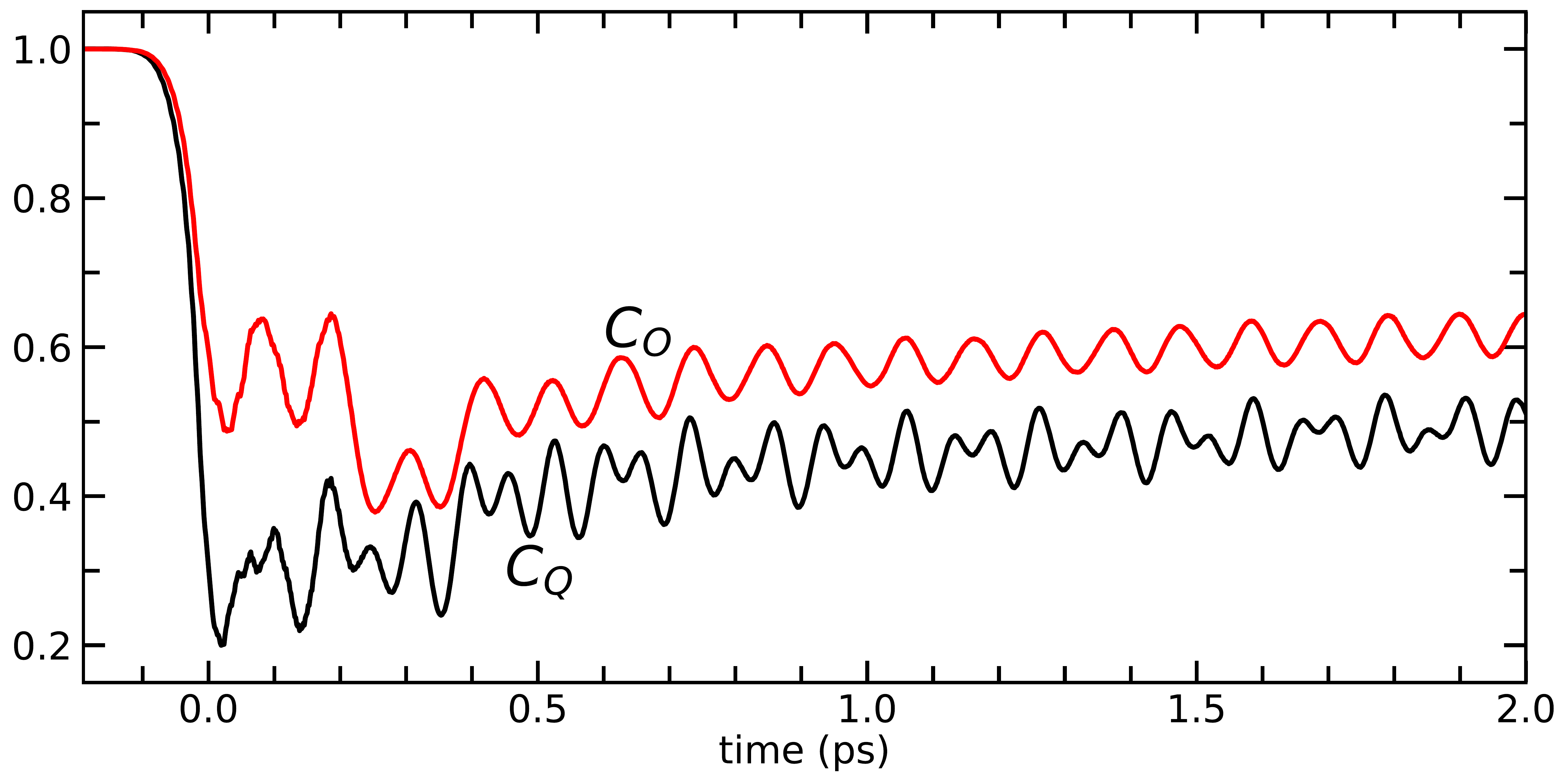}
\end{center}
\caption{\label{fig:region1I_X}Charge-order correlation $C_Q(1,0,0)$
  (black) and orbital-order correlation $C_O(0,\frac{1}{2},0)$ (red)
  as function of time for regime~II with
  $A_0=0.45$~$\rm{\hbar/(ea_0)}$. The correlations are scaled each so
  that their initial value is unity. (color online)}
\end{figure}

\begin{figure}[htp!]
\begin{center}
\includegraphics[width=\linewidth]{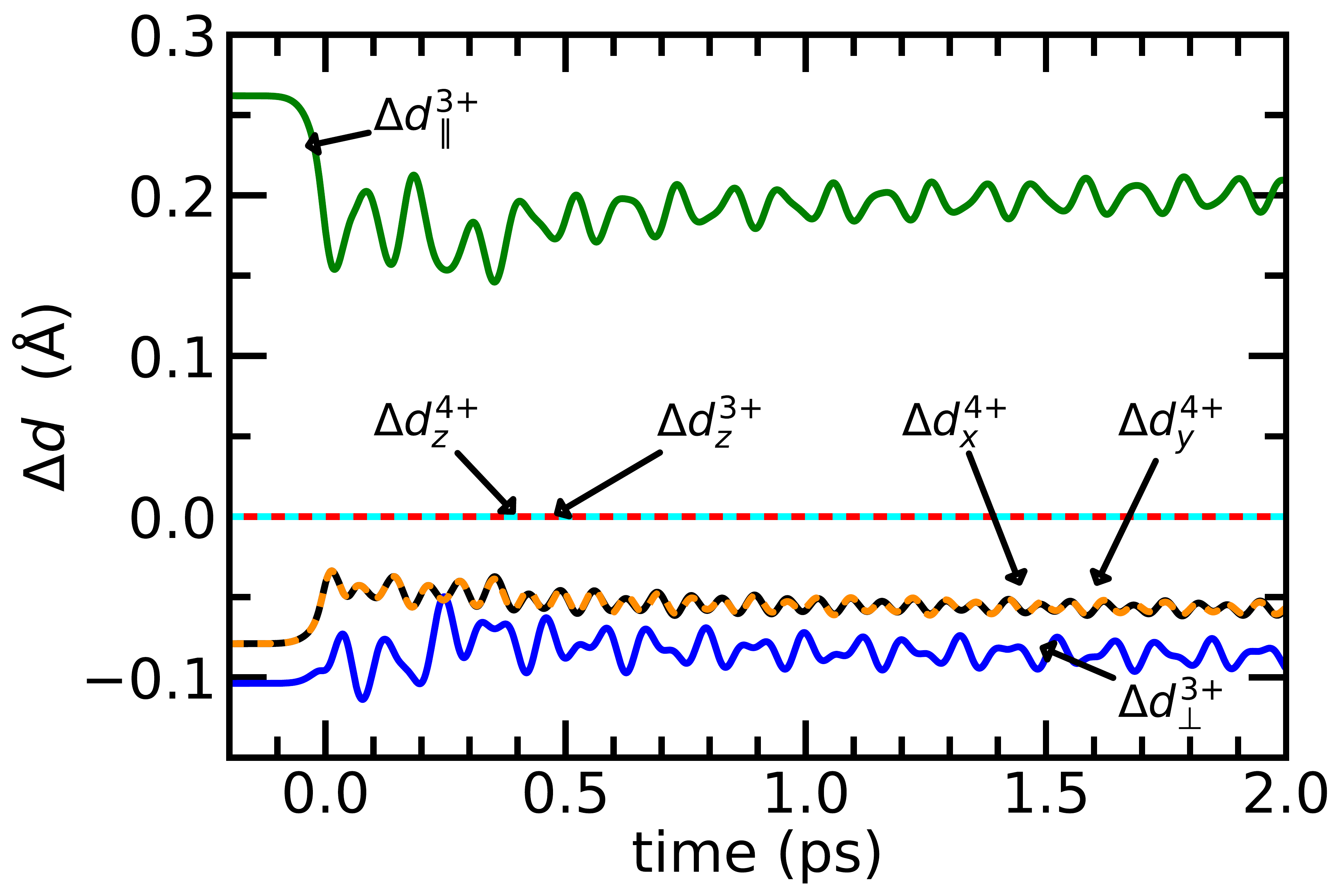}
\end{center}
\caption{\label{fig:region2ph}Phonon modes of regime~II with
  $A_0=0.45$~$\rm{\hbar/(ea_0)}$.  For a description of the
  symbols see \fig{fig:region1ph}. (color online)}
\end{figure}

Charge and orbital order, \fig{fig:region1I_X}, as well as coherent
phonons, \fig{fig:region2ph}, show the same behavior as in regime~I,
albeit with larger amplitude. The oscillations of the phonon
displacements and the correlation functions for charge and orbital
order are long-lived. Unlike regime~I, a slow decay of the
oscillations is noticeable in regime~II.

The coherent phonons, charge order, and orbital order are only little
affected by the transient change of the magnetic structure. The reason
is that the charge order and orbital order remain intact during the
transient change of the magnetic correlations.

Coherent phonons, charge order, and orbital order, are strongly coupled
via electron-phonon coupling. They are due to the same physical
mechanism, namely the $|w_1\rangle$ to $|w_2\rangle$ excitation. Thus,
they are expected to exhibit similar decay properties.

\subsection{Regime~III: Photo-induced ferromagnetism}
\label{secs:region_3}
In regime~III, the system undergoes a photo-induced phase transition,
which converts the CE-antiferromagnetic order into a ferromagnetic metallic
state without charge and orbital order.

\subsubsection{Magnetic order}
\begin{figure}[htp!]
\begin{center}
\includegraphics[width=\linewidth]{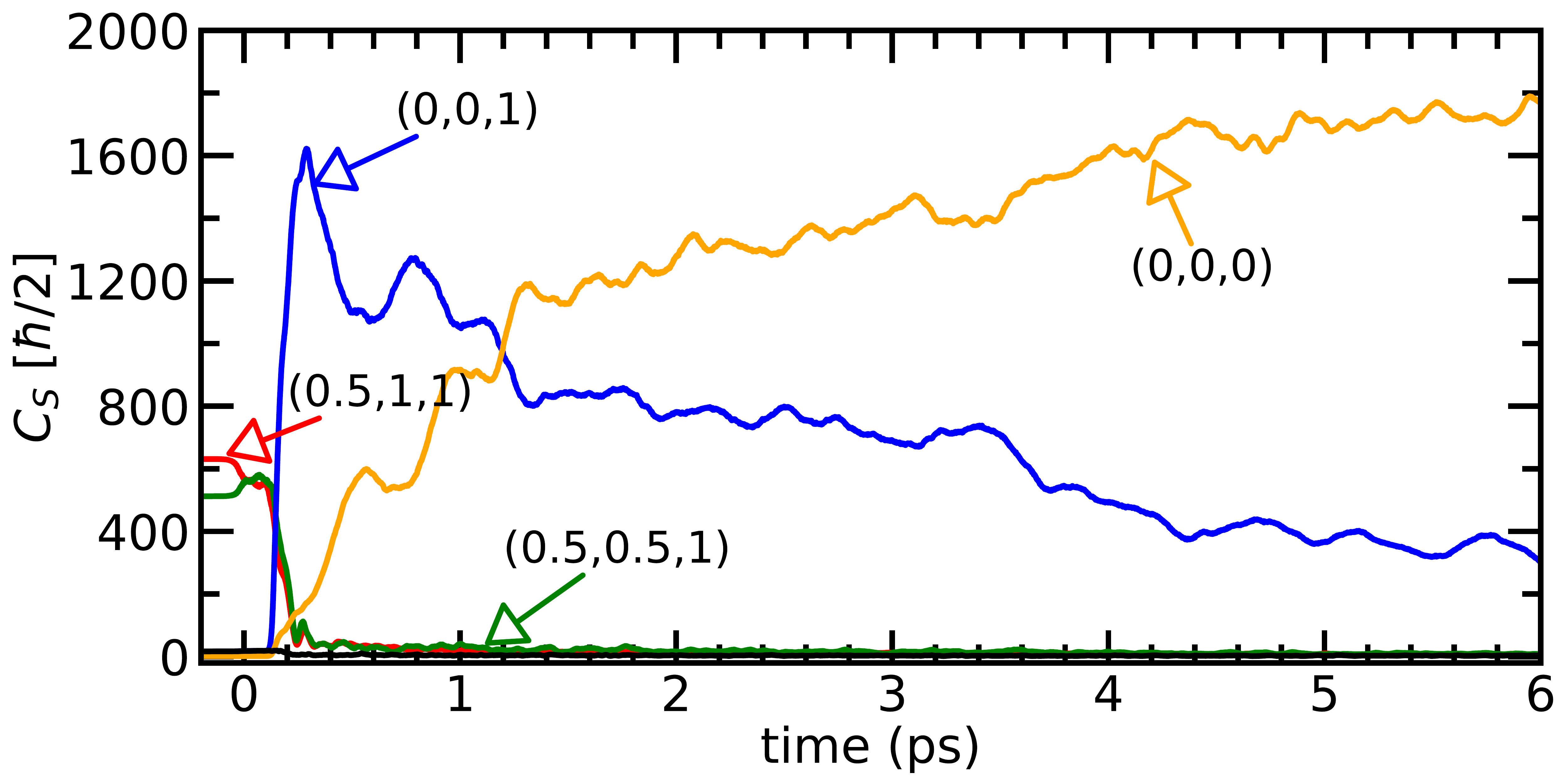}
\end{center}
\caption{\label{fig:region3mag}Spin-diffraction intensity of regime
  III with $A_0=0.53$~$\rm{\hbar/(ea_0)}$.  The peaks $C_S(0.5,1,1)$
  and $C_S(0.5,0.5,1)$ shown in red, respectively green, are
  characteristic for the CE-type ground state.  The peak $C_S(0,0,1)$
  shown in blue is characteristic for A-type magnetic structure. The
  ferromagnetic (B-type) peak $C_S(0,0,0)$ is shown in orange.  The
  original CE-type magnetic pattern is quickly destroyed, while an
  A-type magnetic pattern emerges. The latter evolves over time into a
  the ferromagnetic phase. (color  online)}
\end{figure}

\begin{figure}[htp!]
\begin{center}
\includegraphics[width=\linewidth]{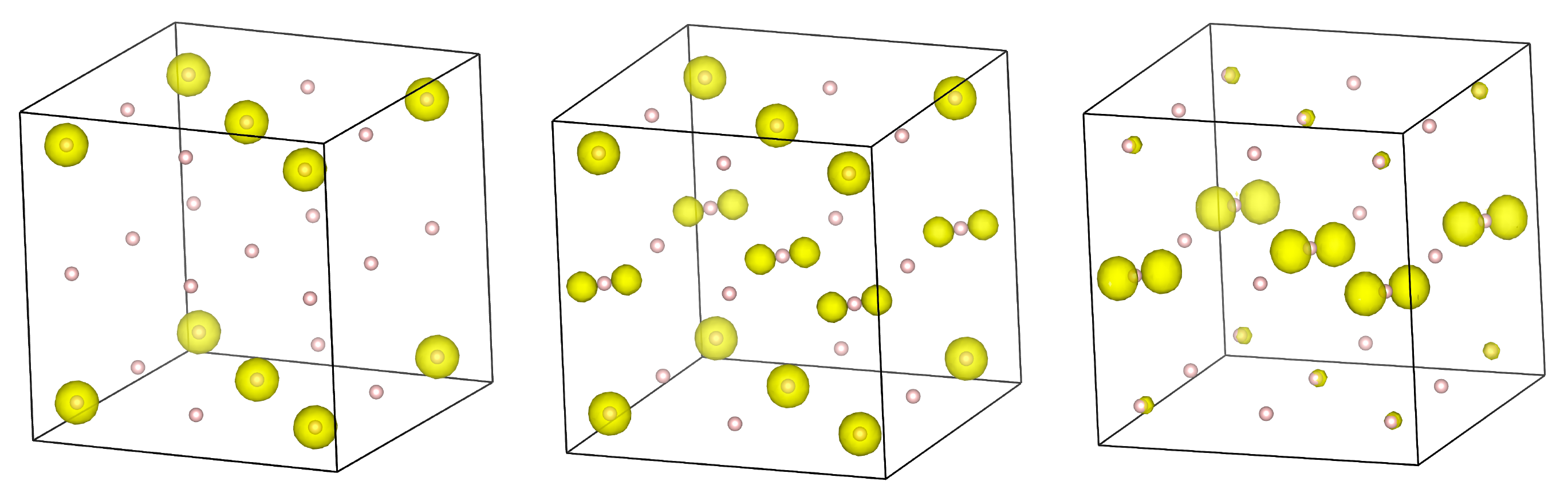}
\end{center}
\caption{\label{fig:region3diffr}Magnetic diffraction patterns of
  region III with $A_0=0.53$~$\rm{\hbar/(ea_0)}$ at 0.3~ps, 1.2~ps and
  6.1~ps .  The a-axis points right, the b-axis towards the back and
  the c-axis up. The small white spheres indicate points with integer
  $h,k,l$ in the Pbnm setting.  Reciprocal space is shown for
  $h,k,l\in[-1.25,1.25]$.  At 0.3~ps the diffraction pattern is
  dominated by a A-type pattern. At 1.2~ps the diffraction pattern
  exhibits spots from both A and B-type, while at 6.1~ps the
  diffraction pattern is ferromagnetic, i.e. B-type. The double spots
  are a sign of ferromagnetic magnetic domains, respectively a
  long-wave length spin wave, rather than a pure ferromagnet. (color
  online)}
\end{figure}

Initially, the antiferromagnetic correlation between the zig-zag
chains is perturbed rather similar to regime~II, leading to an A-type
magnetization as seen in \fig{fig:region3mag}.  Unlike regime~II,
however, the A-type diffraction pattern persists for several
picoseconds.  During this time, the diffraction pattern of the
ferromagnet builds up until it replaces the A-type diffraction
pattern altogether.

The ferromagnetic state obtained is not fully established in the
simulation: The non-relativistic Schr\"odinger equation employed in
the simulations conserves the total spin. As a result, the system evolves
into a state that is better characterized as a spin wave or a lattice
of ferromagnetic domains.

Nevertheless, the diffraction pattern obtained is very similar to the
ferromagnetic structure.  For each diffraction spot of the
ferromagnetic structure, we do not obtain a single spot, but a set of
two ``twin-peaks''. The two peaks are located at the supercell
reciprocal-space vectors adjacent to those of the ideal ferromagnet as
seen in \fig{fig:region3diffr}.  The displacement of the twin peaks
from the diffraction spot of a true ferromagnet is governed by the
size of our supercell, which limits the wave length of the spin wave,
respectively the domain size. 

The ferromagnetic spin correlation function in \fig{fig:region3mag}
exhibits a finite signal by considering the contribution from the
immediate neighborhood of the specified reciprocal lattice vector. The
signal at the center of the spot is zero.

We envisage that a larger supercell leads to larger domains and
thus to twins-peaks that are even closer together, making them
indistinguishable by experiment.

\subsubsection{Charge order}
\begin{figure}[htp!]
\begin{center}
\includegraphics[width=\linewidth]{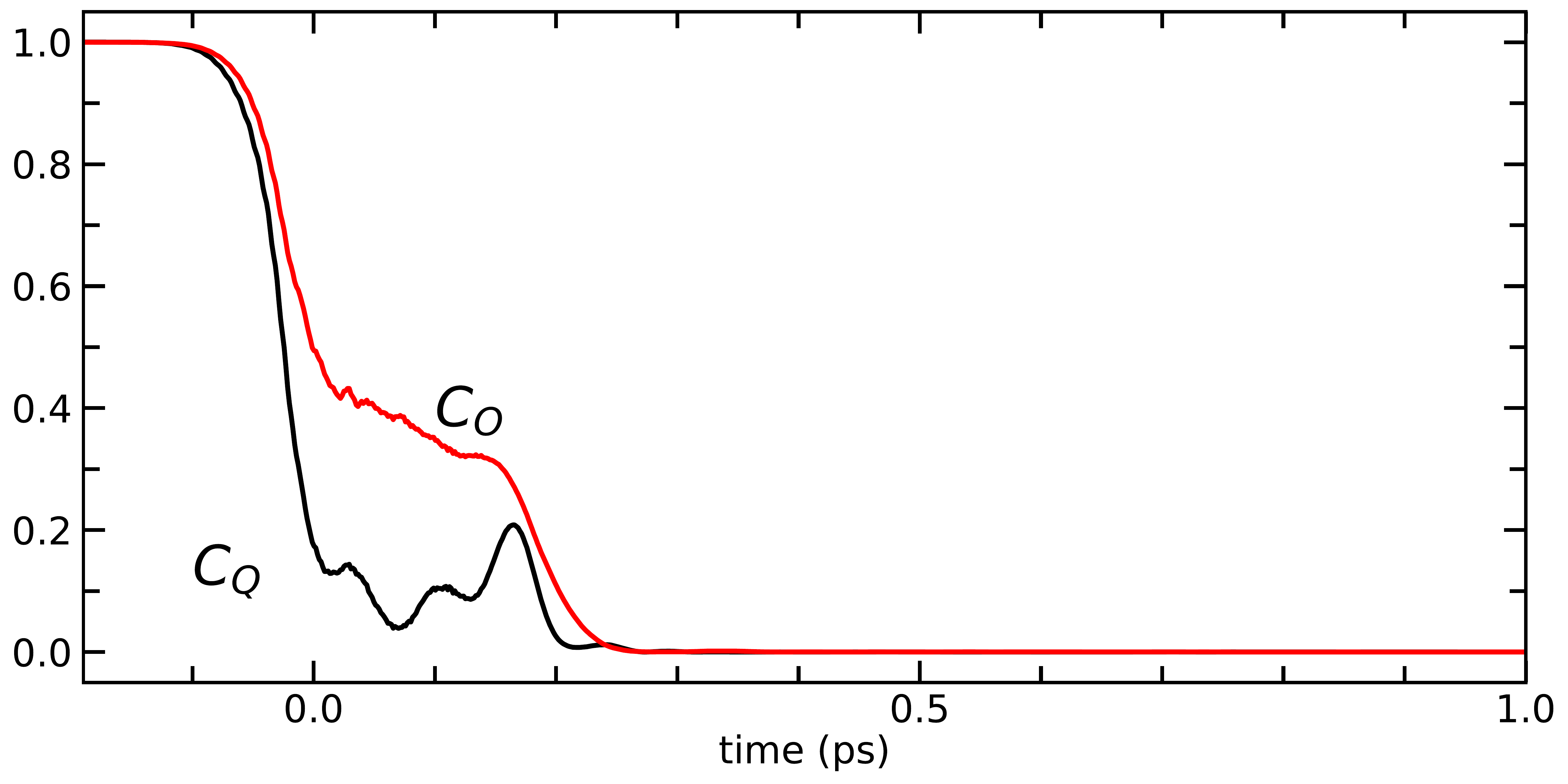}
\end{center}
\caption{\label{fig:region3I} Charge-order correlation $C_Q(1,0,0)$
  (black) and orbital-order correlation $C_O(0,\frac{1}{2},0)$ (red)
  as function of time for regime~III with
  $A_0=0.53$~$\rm{\hbar/(ea_0)}$. The correlations are scaled each so
  that their initial value is unity.  (color online)}
\end{figure}

\begin{figure}[htp!]
\begin{center}
\includegraphics[width=\linewidth]{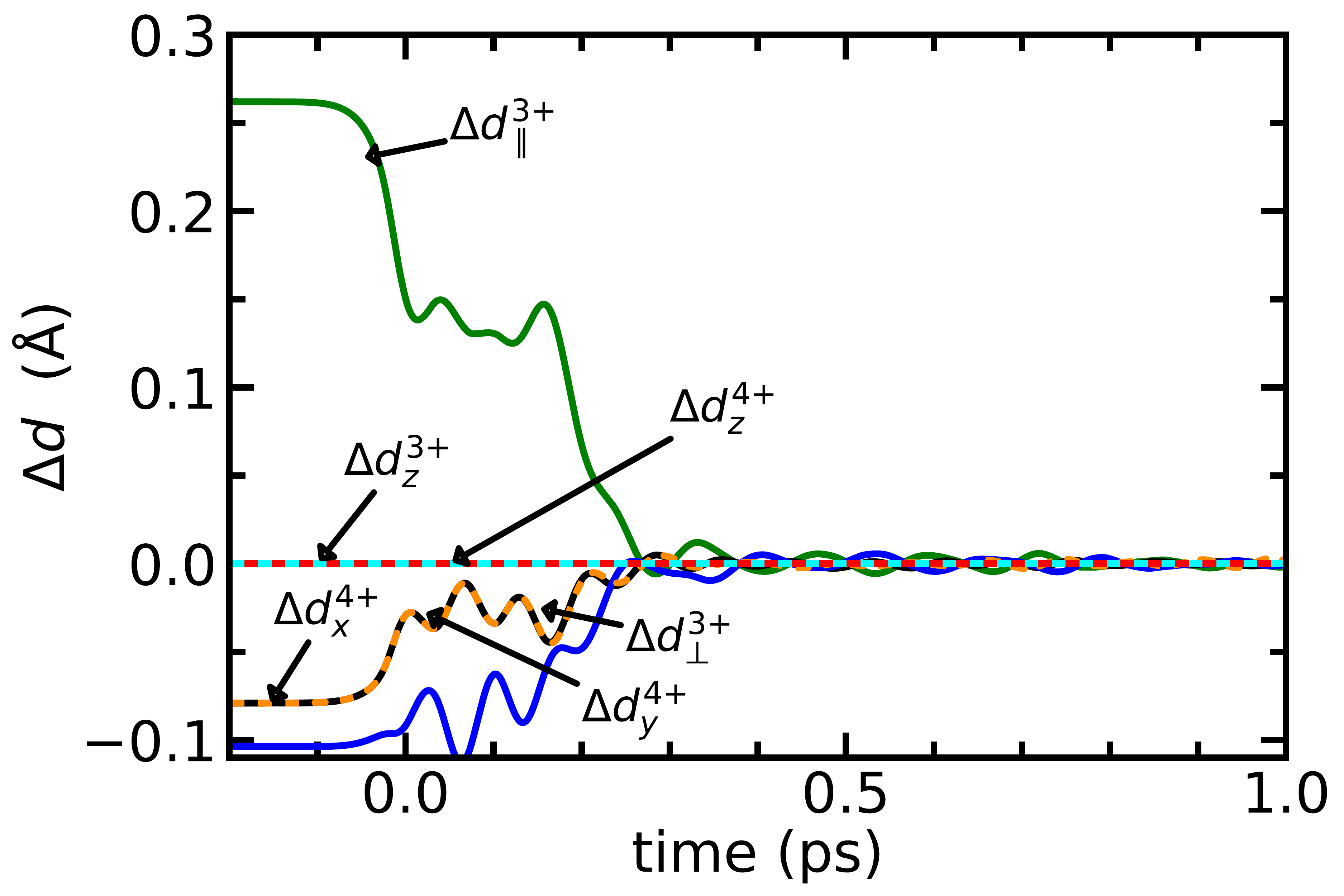}
\end{center}
\caption{\label{fig:region3ph}Phonon modes of region III with
  $A_0=0.53$~$\rm{\hbar/(ea_0)}$.  For a description of the symbols
  see \fig{fig:region1ph}. (color online)}
\end{figure}

As shown in \fig{fig:region3I}, the charge-order correlation $C_Q$ and
the orbital-order correlation $C_O$ are completely wiped out after
about 0.2~ps. The loss of orbital order makes the system metallic as
seen in \fig{fig:dynamicsehdistrib}.  The loss in charge and orbital
order is also reflected in attenuation of the phonon displacements
shown in \fig{fig:region3ph}.

We attribute the ferromagnetic order to a mechanism in the spirit of
the double-exchange
picture.\cite{zener51_pr82_403,anderson55_pr100_675,millis96_prb54_5405}
The origin of the band gap in the ground state of {\pcmohalf} is the
formation of Zener polarons. These Zener polarons are also the origin
of charge and orbital order.  Due to the re-population of electrons
across the band gap, the stabilization due to Zener polarons is lost
and another, competing mechanism can take over. A ferromagnetic
alignment of the spins lowers the kinetic energy of the electrons,
which can now spread over a large area: An electron with a given spin
is effectively excluded from orbitals of Mn-ions with opposite spin. A
Mn-ion with opposite spin thus leads to an energy cost. Thus, it is
favorable, when all spins align ferromagnetically. In other words, a
configuration of ferromagnetic spins produces a larger effective band
width of the majority-spin configuration.  The larger band width
stabilizes electrons which populate the lower half of the
majority-spin band formed.

\subsubsection{Experiment}
A recent ultrafast pump-probe experiment \cite{esposito18_prb97_14312}
carried out on {\pcmox} at 100~K with different pump fluences showed
that the characteristic charge- and orbital-order reflection peaks of
the CE-type ground state disappear for larger fluences
$F_p{>}2.5$~$\rm{mJ/cm^2}$. 

The experimental study with the same material class by Li et
al.\cite{li13_nature496_69} revealed a photo-induced ferromagnetic
state within about 120~fs above the threshold fluence
$F_p=2.4$~$\rm{mJ/cm^2}$.  A rise in magnetization has also been
measured by Zhou et al.\cite{zhou14_scientificreports4_4050}.

The measured threshold fluence of $F_p=2.5$~$\rm{mJ/cm^2}$ translates
via Eq.~\ref{eq:pumpfluence} into an amplitude
$A_0=1.26~\hbar/(ea_0)$.  In our simulations, the loss of charge and
orbital order sets in with $A_0=0.58~\hbar/(ea_0)$
($F_p=0.52$~$\rm{mJ/cm^2}$) as seen in \fig{fig:distinctregions} from the
appearance of the A-type magnetic order, which finally converts into
the B-type (ferromagnetic) order. Our simulations and experiments
produce the transition in the same fluence range. The remaining
difference of a factor five in the threshold fluence may be
attributed, for example, to the different photon energies.

 It is worth mentioning that the ferromagnetic states observed in
 our study for region III is expected to persist on longer timescale
 hinting towards its possible long lifetime. Similar long-lived states
 are recently observed in {\pcmox} series within the
 charge-ordered region  of the phase diagram
 \cite{raiser17_aenm7_1602174}.

\subsection{Regime~IV: Non-collinear antiferromagnet}
\label{secs:region_4}
\begin{figure}[htp!]
\begin{center}
\includegraphics[width=\linewidth]{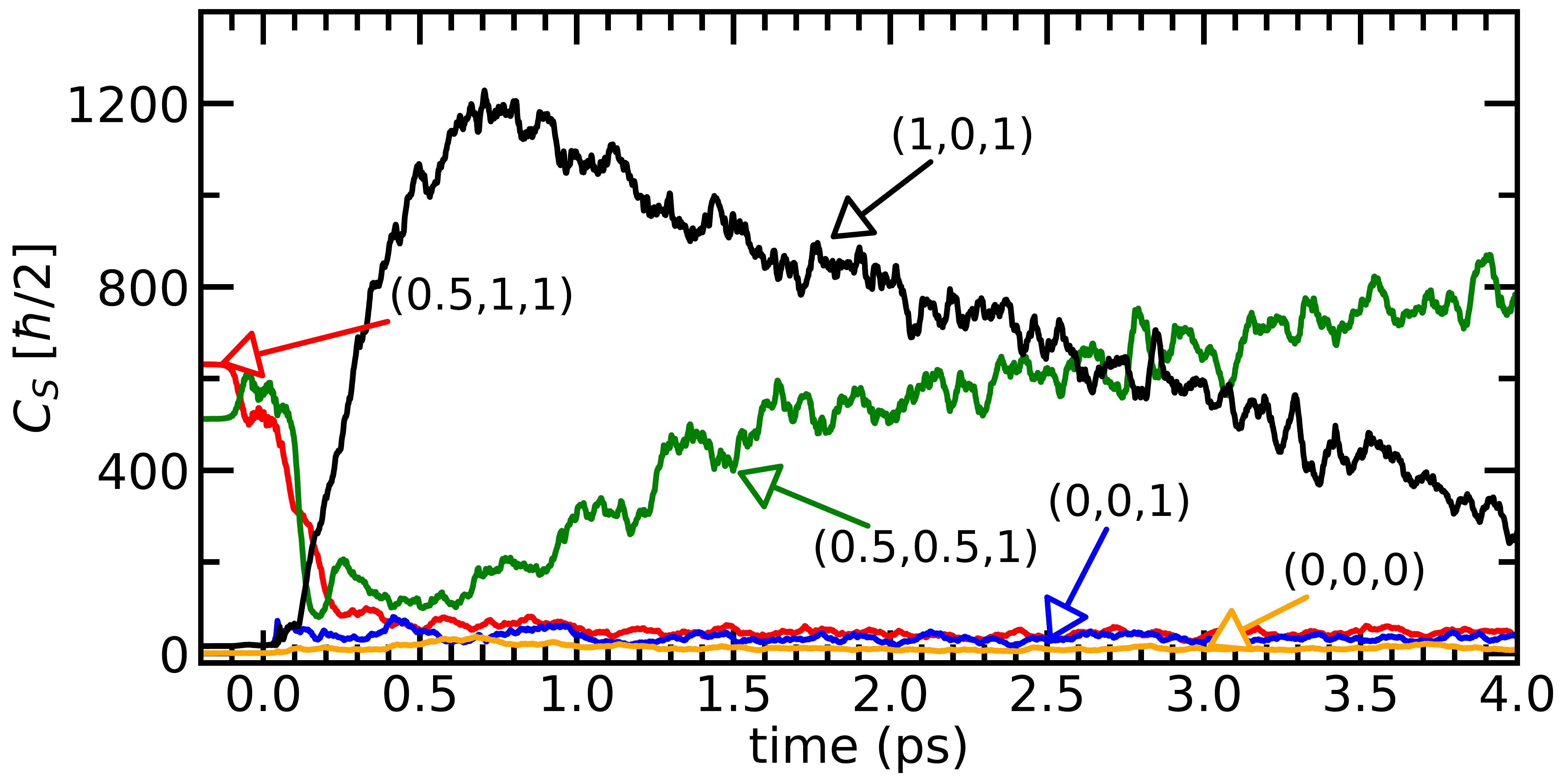}
\end{center}
\caption{\label{fig:region4mag}Spin-diffraction intensity of regime~IV
  with $A_0=2.50$~$\rm{\hbar/(ea_0)}$.  The peaks $hkl=(0.5,1,1)$ and
  $(0.5,0.5,1)$ shown in red, respectively green, are characteristic
  for the CE-type ground state.  The ferromagnetic (B-type) peak
  $C_S(0,0,0)$ is shown in orange. The peak with $hkl=(1,0,1)$ is
  characteristic for G-type magnetic structure. Over time, a new,
  non-collinear magnetic structure (see Fig. 25) emerges evidenced by
  the occurrence of the peak $hkl=(0.5,0.5,1)$. (color online)}
\end{figure}

In regime~IV with the highest fluence, the system evolves first into a
G-type antiferromagnet as shown in \fig{fig:region4mag}, rather than
forming an A-type antiferromagnet as in regime~III. After about
1.5~ps, the diffraction spots of the G-type structure fall off again
in favor of a more complicated structure with non-collinear magnetic
order. We note that this regime may be difficult to access
experimentally due to the limited stability of the material.

\begin{figure}[htp!]
\begin{center}
\includegraphics[width=\linewidth]{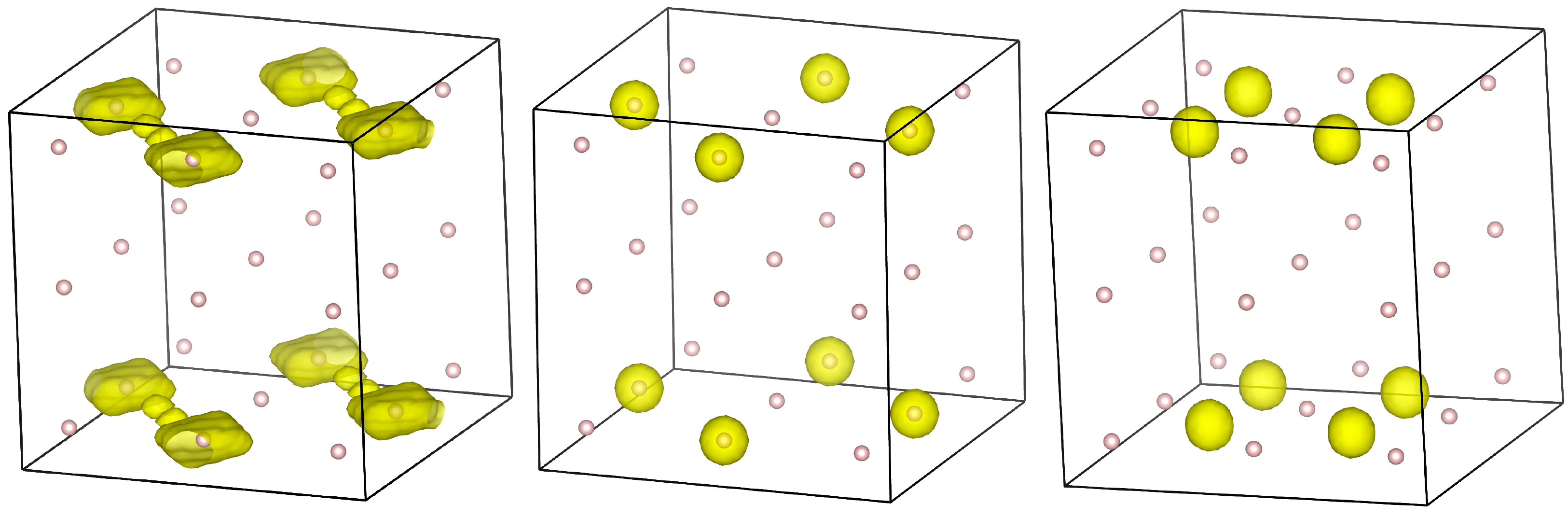}
\end{center}
\caption{\label{fig:region4diffr}Magnetic diffraction patterns of
  region IV with $A_0=2.50$~$\rm{\hbar/(ea_0)}$ at 0.3~ps, 0.8~ps and 6.2~ps.
  At 0.3~ps and 0.8~ps the diffraction pattern is dominated by a
  G-type pattern.  At 6.2~ps a new diffraction pattern occurs, which
  can be attributed to a non-collinear spin structure described in the
  text and in \fig{fig:region4diffrmodel}. Reciprocal space is shown
  for $h,k,l\in[-1.25,1.25]$. The a-axis points right, the b-axis
  towards the back and the c-axis up. The small spheres indicate
  points with integer $hkl$ in the Pbnm setting. (color online)}
\end{figure}

\begin{figure}[htp!]
\begin{center}
\includegraphics[width=0.49\linewidth]{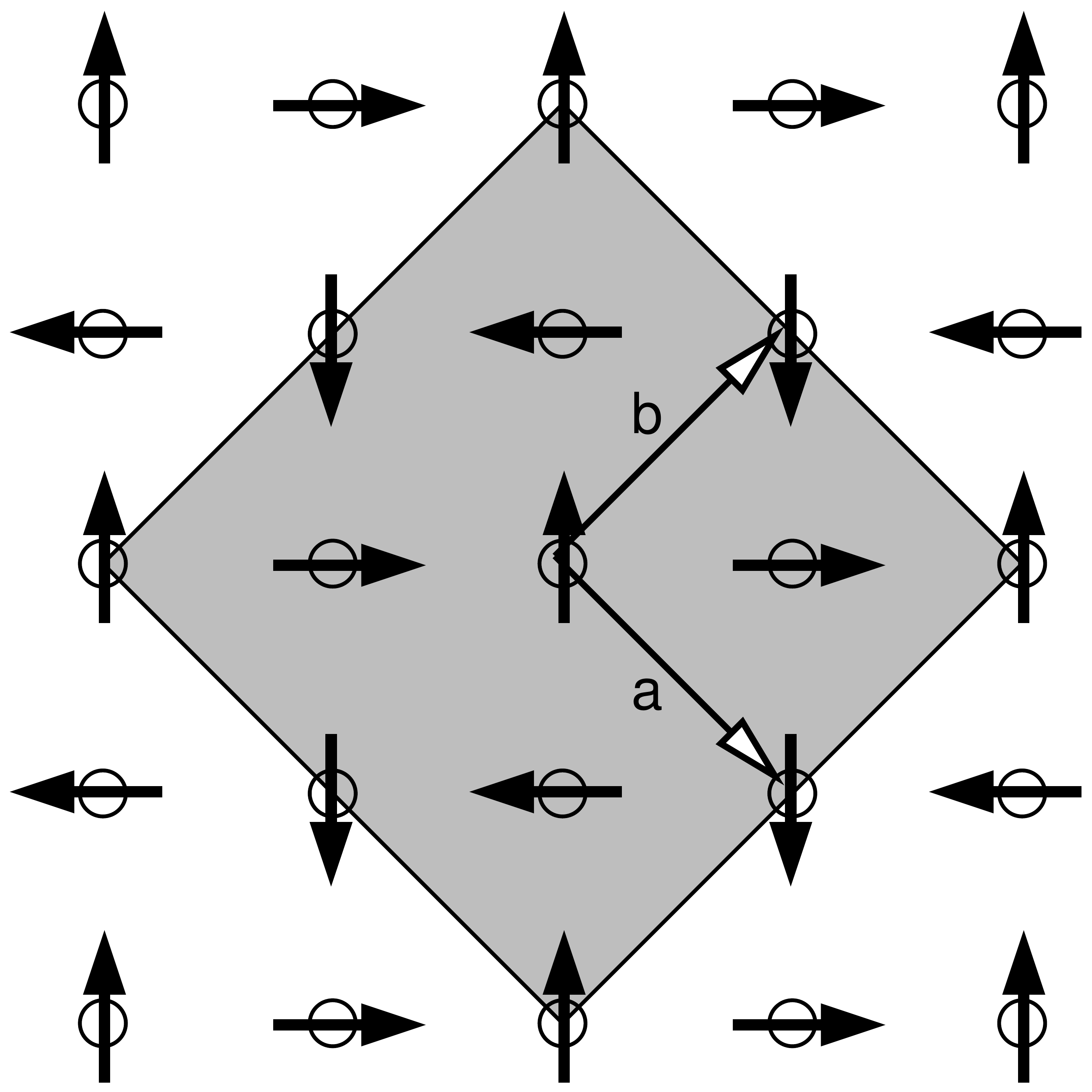}
\includegraphics[width=0.49\linewidth]{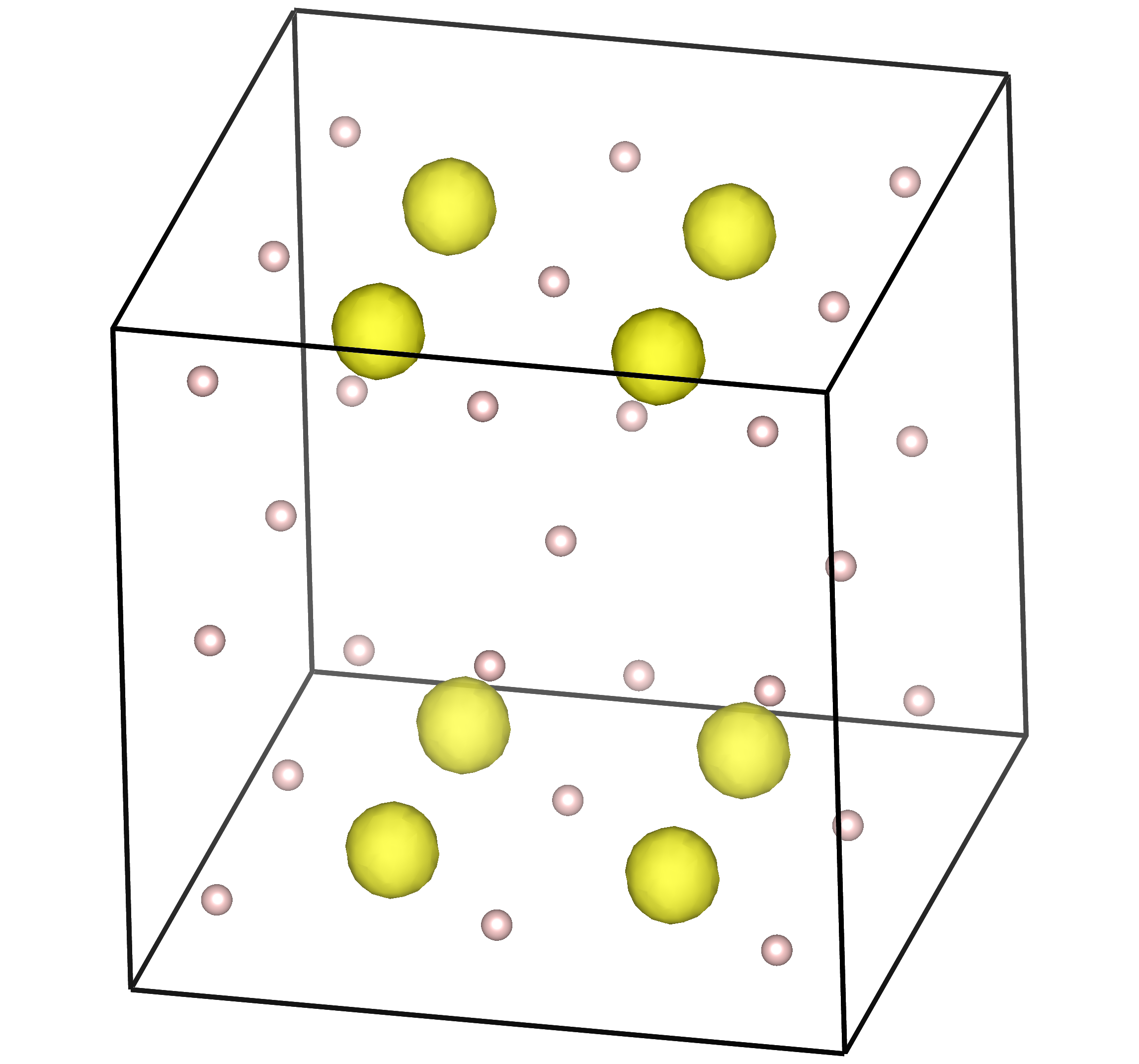}
\end{center}
\caption{\label{fig:region4diffrmodel}Idealized model for the spin
  distribution reached in regime~IV beyond 2~ps. The non-collinear
  spin structure in the ab-plane of the Pbnm setting is shown on the
  left. Shown is the grid of Mn sites with the orientation of the
  {\ttg}~spins. Planes are stacked antiferromagnetically in the
  c-direction. On the right, the corresponding magnetic diffraction
  pattern is shown. The diffraction pattern is nearly
  indistinguishable from the one shown in \fig{fig:region4diffr},
  which is obtained in regime~IV at 6.2~ps after the light pulse. The
  a-axis points right, the b-axis towards the back and the c-axis
  up. The small spheres indicate points with integer $hkl$ in the Pbnm
  setting. Reciprocal space is shown for $h,k,l\in[-1.25,1.25]$.  (color online)}
\end{figure}

The diffraction patterns calculated for characteristic times along the
trajectory are shown in \fig{fig:region4diffr}. The spin structure of
the final state, which emerges at approximately 1.5~ps after the
light-pulse has been extracted on the basis of the real space
spin-correlation function. It is shown in its idealized form in
\fig{fig:region4diffrmodel}. All spins are perpendicular to its
neighbors in the ab-plane and antiferromagnetic in the
c-direction. This model produces the spin diffraction pattern of the
final spin configuration of regime~IV. To our knowledge this
configuration has not been investigated before in the context of
manganites.

\begin{figure}[htp!]
\begin{center}
\includegraphics[width=\linewidth]{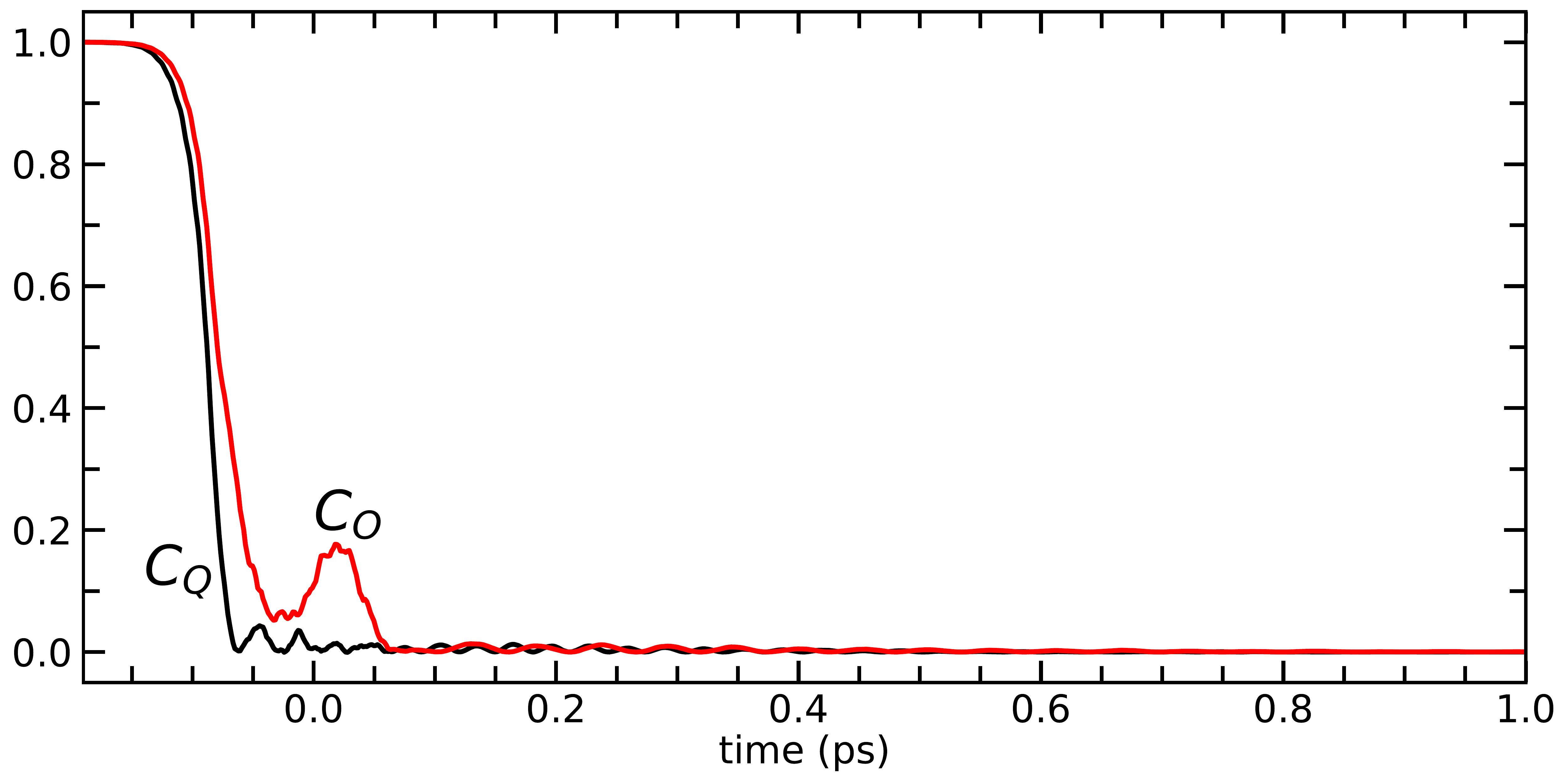}
\end{center}
\caption{\label{fig:region4I_X}Charge-order correlation $C_Q(1,0,0)$
  (black) and orbital-order correlation $C_O(0,\frac{1}{2},0)$ (red)
  as function of time for regime~IV with
  $A_0=2.5$~$\rm{\hbar/(ea_0)}$. The correlations are scaled each so
  that their initial value is unity.  (color online)}
\end{figure}

\begin{figure}[htp!]
\begin{center}
\includegraphics[width=\linewidth]{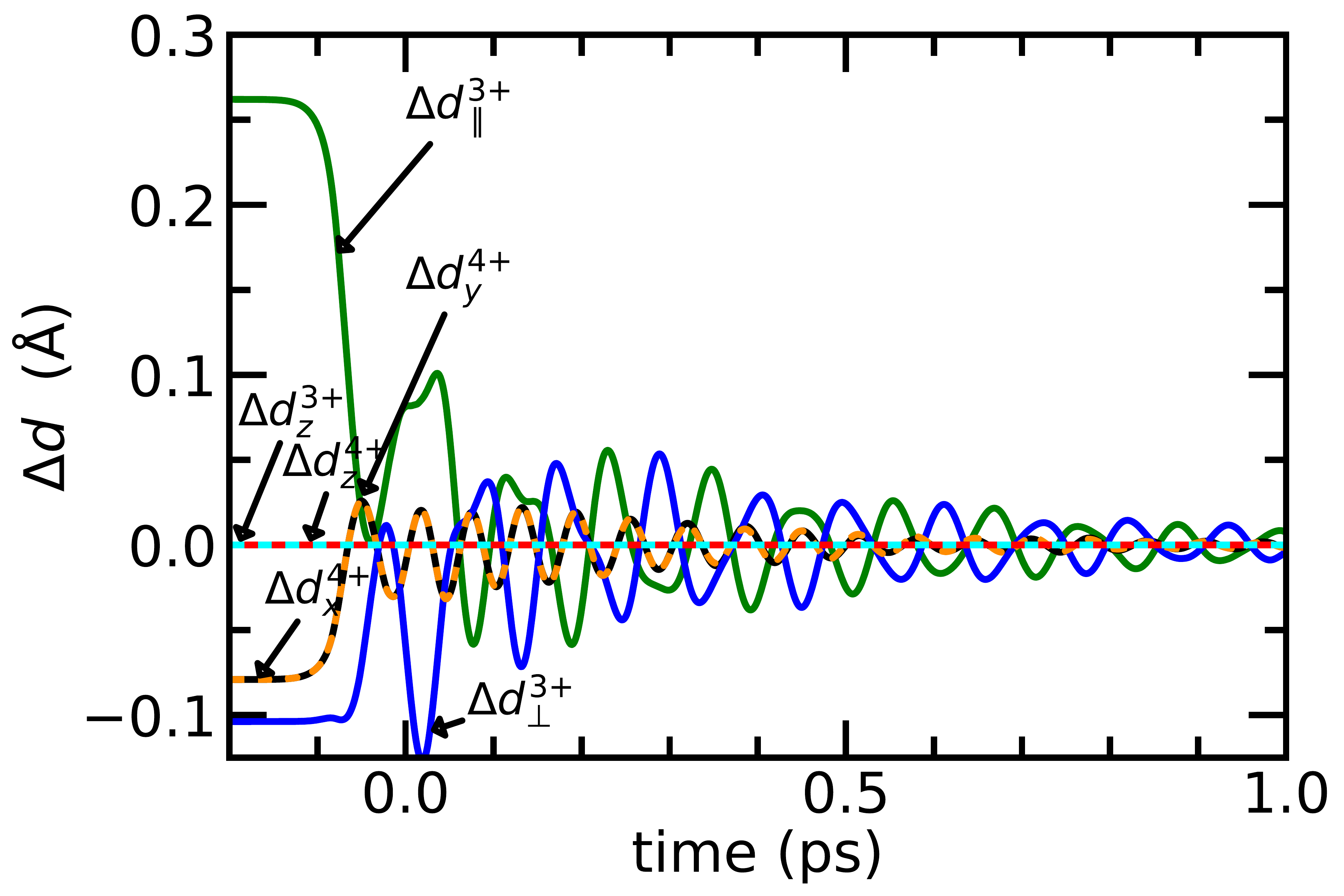}
\end{center}
\caption{\label{fig:region4ph}Phonon modes of region IV with
  $A_0=2.50$~$\rm{\hbar/(ea_0)}$.  For a description of the symbols
  see \fig{fig:region1ph}. (color online)} 
\end{figure}

The charge and orbital order is destroyed almost immediately, that is
during the light pulse as shown in \fig{fig:region4I_X}.  This
destruction of the orbital order excites phonons, that, however,
dissipate on a picosecond time scale as seen in \fig{fig:region4ph}.
The amplitude of the phonon vibrations is considerably larger than
that in regime~III.

We attribute the transition with increasing fluence from a ferromagnet
in regime~III to the non-collinear antiferromagnetic structure in
regime~IV to a mechanism analogous to that described
earlier.\cite{ono17_prl119_207202} The double-exchange mechanism favors
ferromagnetism through the increase in band width only, when a majority
of electrons populate the lower half of the majority-spin {\eg}~states.
Thus, increasing the fluence beyond a certain point switches off the
double-exchange mechanism again, so that a antiferromagnetic structure can
develop. The reduction of the band width of majority-spin and
minority-spin electrons opens a band gap between them, which is
seen in \fig{fig:dynamicsehdistrib}.  Compared to the minimal
model\cite{ono17_prl119_207202}, our system evolves
into a more complicated non-collinear antiferromagnetic structure.

\subsection{Thermalization}
\label{secs:thermalization}
One of the questions of interest is how thermal equilibrium is
established from the excited state. 

Therefore, we investigated the evolution of the temperatures of the
subsystems.

Below, we consider the quasi temperatures described in 
appendix~\ref{sec:temperatures}.  As shown in \fig{fig:temps},
the light pulse immediately raises the temperature of the electronic
system to high temperatures, i.e. several thousands of Kelvin, while
the temperatures of the phonon and spin systems remain low in
comparison. Thus, a state far from equilibrium is formed. The state is
analogous to that of a non-thermal (cold) plasma, where the electrons
reach $10^4$~K, while the ions remain near room temperature.

While the temperature of the phonon system remains cold, the coherent
phonons of regime~I and II are strongly coupled to the electronic
subsystem and reach comparable temperatures.  When we attribute the
complete thermal energy of the ions considered to the two phonon
modes, the resulting temperature of the two modes is comparable to the
electronic temperature.

\begin{figure}[htp!]
\begin{center}
\includegraphics[width=\linewidth]{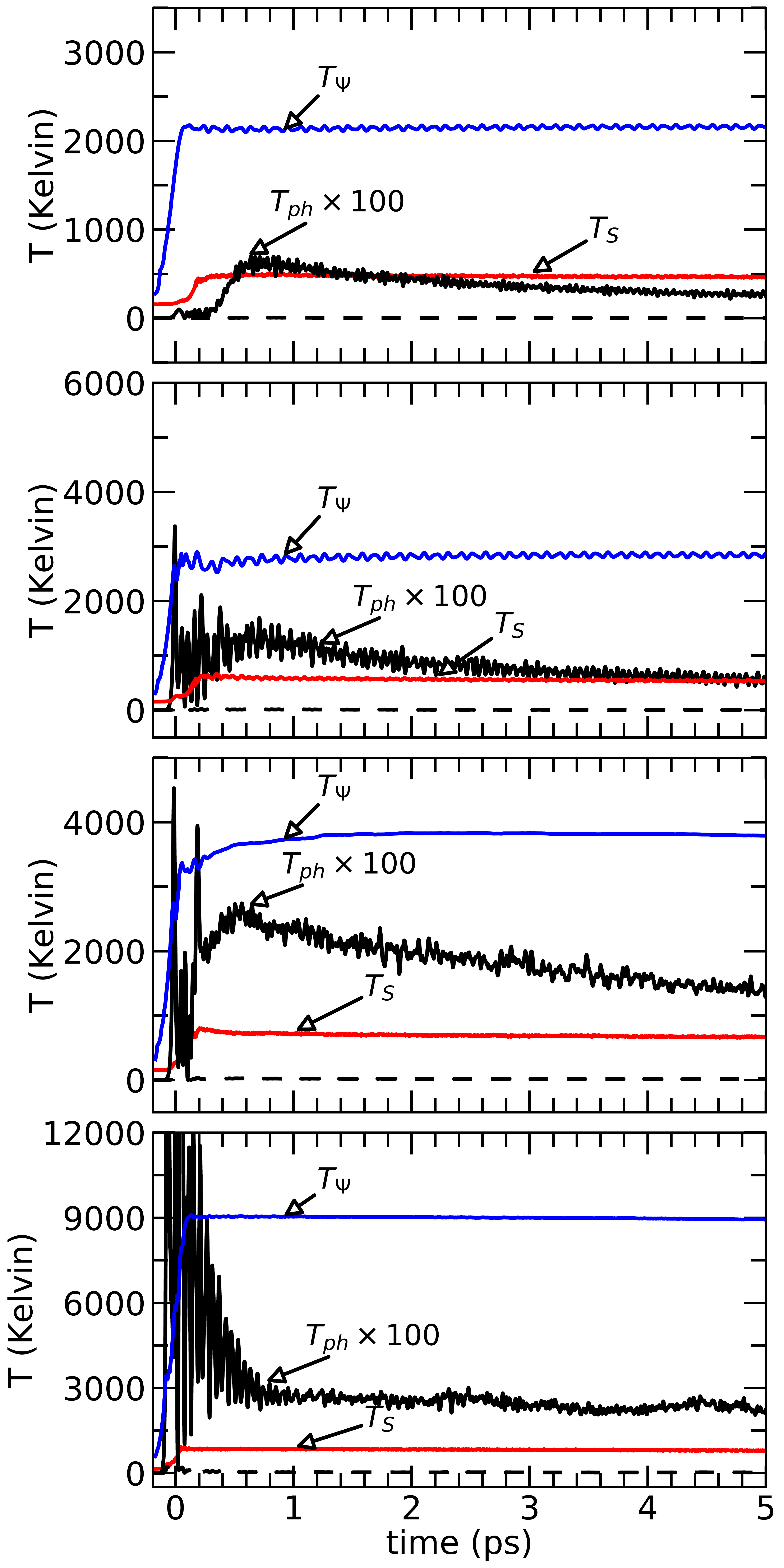}
\end{center}
\caption{\label{fig:temps}Temperatures of the electron (blue), spin
  (red)and phonon (black) subsystems as function of time.  The phonon
  temperatures are scaled by 100. From top to bottom are the
  representative examples from regime~I to IV,
  i.e. $A_0=0.20,0.45,0.53$ and $2.50$. (color online)}
\end{figure}

\subsubsection{Non-equilibrium distribution of the electrons}
In particular, the channel of the electrons is of interest because
the energy from this channel is most easily put into practical
use, such as in a solar cell.

Our simulations may shed light onto the workings of the Boltzmann
equation. For this purpose, we inspect the emergence of a
distribution, i.e. the occupations, as function of energy and we
compare the distribution obtained in our calculation with the Fermi
distribution. The approach to a Fermi distribution is one of the
common assumptions made for the Boltzmann equation.

In order to explore the approach to the Fermi distribution, we choose
a representation of ${\rm atanh}\big(1-2\bar{f}_j\big)$ versus energy
$\epsilon^{BO}_j$. In this representation, a Fermi distribution maps
onto a straight line with slope $1/k_BT_\psi$ and zero $\mu_\psi$.

The Born-Oppenheimer energies $\epsilon^{BO}_j$, their occupations
$\bar{f}_j$ and the Born-Oppenheimer wave functions
$|\phi_j^{BO}\rangle$ are defined in
appendix~\ref{sec:electrontemperature}.

\begin{figure*}[htp!]
\begin{center}
\includegraphics[width=0.49\linewidth]{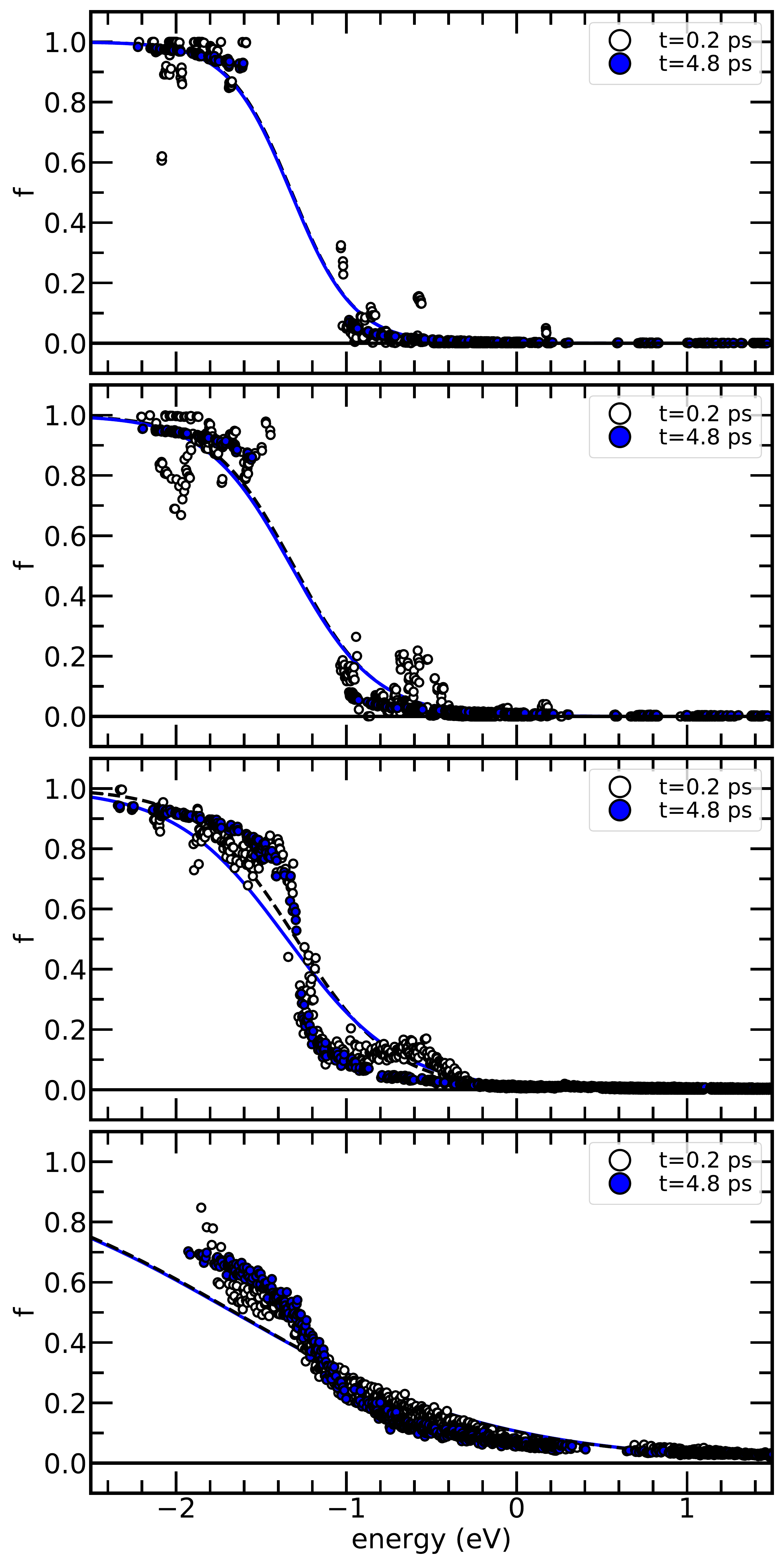}
\includegraphics[width=0.49\linewidth]{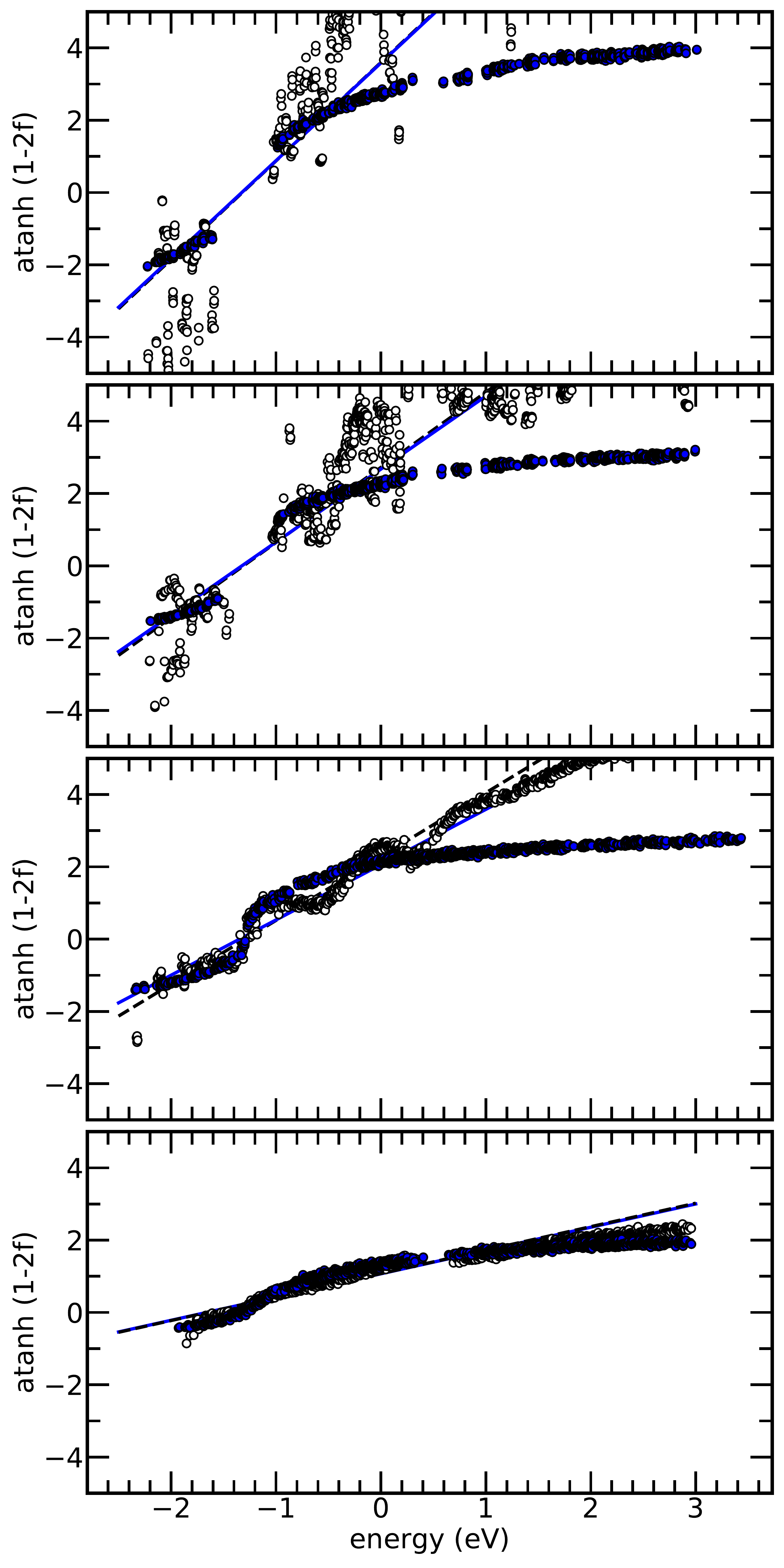}
\end{center}
\caption{\label{fig:distributions}Distributions of electron
  occupations $\bar{f}$ at different times, namely t=0.2~ps (open
  circles) and 4.8~ps (blue-filled circles). From top to bottom are
  the representative examples from regime~I to IV, i.e. $A_0=0.20,
  0.45, 0.53$ and $2.50$~$\rm{\hbar/(ea_0)}$.  The left figure shows
  the occupations (circles) versus Born-Oppenheimer energy
  $\epsilon^{BO}_j$. The dashed black line is the Fermi distributions
  at 0.2~ps obtained from the energy and particle-number sum
  rules. The full blue line is the Fermi distribution at 4.8~ps. On
  the right, the occupations $\bar{f}$ are transformed by
  $\rm{arctanh}(1-2\bar{f})$ which maps a Fermi distribution to a
  straight line with slope $1/(k_BT_\psi)$ and zero $\mu_\psi$.
  (color online)}
\end{figure*}

The occupations for different time slices and for the four regimes are
shown in \fig{fig:distributions}.

Initially, that is 0.2~ps after the maximum of the light pulse, the
occupations do not lie on a continuous function of the one-particle
energies but scatter wildly. This is expected because the occupations
of the electron states are dominated by the ground-state occupations,
the photon energy and the dipole matrix elements.

Already after an initial period of about 0.5~ps after the pulse
maximum, the occupations form a continuous function of the
energy. This indicates that the thermalization between electrons of
the same energy is very efficient.

This result is specific to the choice of one-particle orbitals and
energies, namely the Born-Oppenheimer states.

However, on the time scale of our simulation, the system does not
approach a Fermi distribution. Rather, occupations of electrons
further away from the chemical potential deviate further from integral
occupations than a Fermi distribution: The electrons further away from
the Fermi level seem to be ``hotter'' than those close to the Fermi
level. Surprisingly, the tails of the distribution become even flatter
with time, that is they seem to deviate even further from a Fermi
distribution.

We attribute this behavior to the strong coupling between different
subsystems:  Due to the dynamics of the spin and phonon systems, the
electrons experience a time-dependent Hamiltonian, that constantly
drives the electrons out of their equilibrium distribution.
For the one-particle basis $|\phi^{BO}_j(t)\rangle$ used here, the
approach to a Fermi distribution is not a requirement.

\subsubsection{Cold-plasma model}
\begin{figure}[htp!]
\begin{center}
\includegraphics[width=\linewidth]{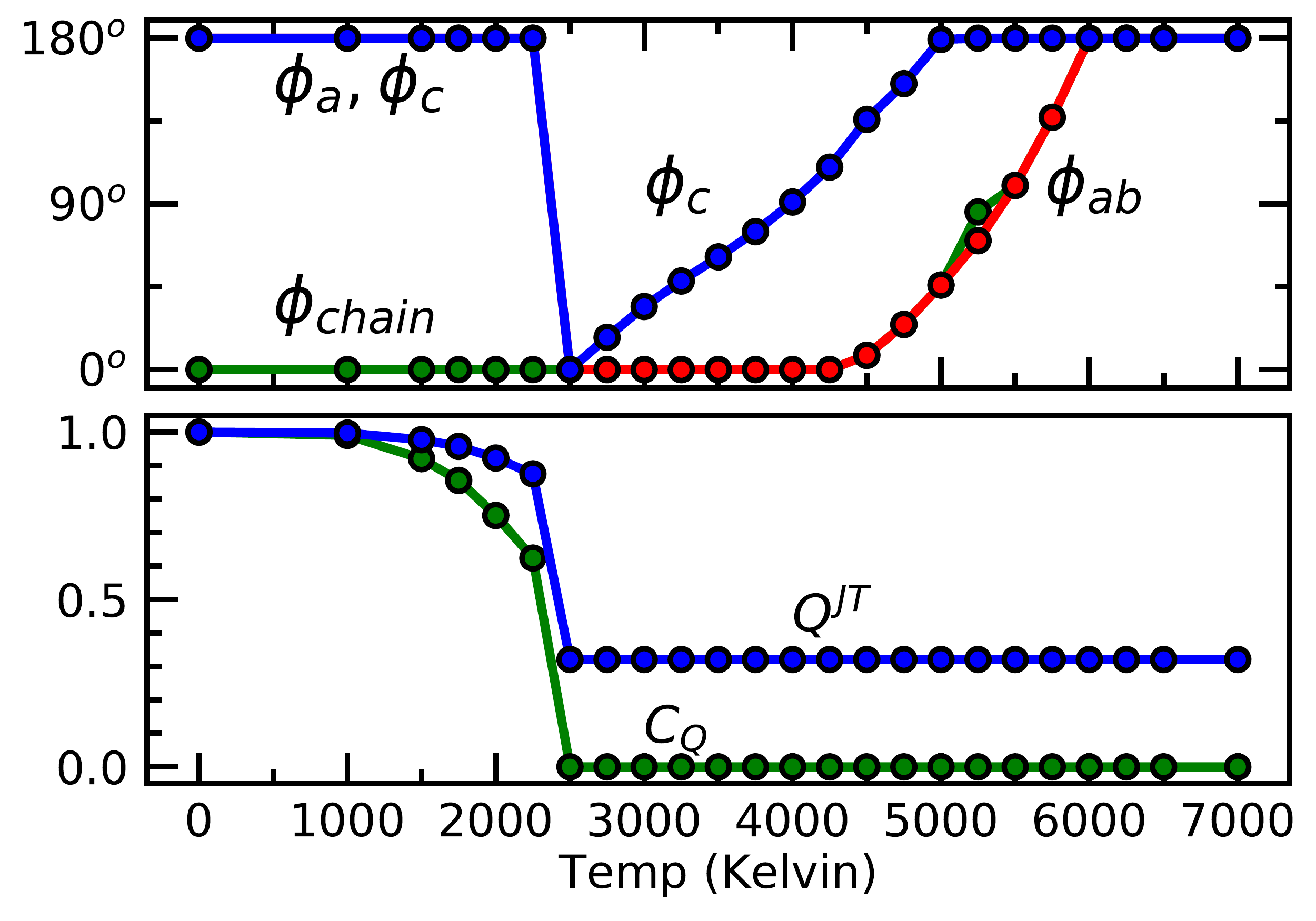}
\end{center}
\caption{\label{fig:hotelectronphasses}Non-equilibrium phase diagram
  with hot electrons and cold spins and phonons. Top: spin
  angles. $\phi_c$ is the spin angle between spins in the $c$
  direction of the Pbnm setting. $\phi_{ab}$ is the spin angle of
  neighboring Mn-sites in the ab-plane. Below $T=T_1$, the angle
  $\phi_{ab}$ is replaced by one angle $\phi_a$ perpendicular to the
  zig-zag chains of the CE-magnetic structure and $\phi_{chain}$ is
  the angle within the zig-zag chains. Bottom figure: Charge-order
  correlation function $C_Q$ and Jahn-Teller distortion $Q^{JT}$ of
  the central site as function of temperature. Both are divided by
  their zero-temperature value. (color online)}
\end{figure}

On the picosecond time scale after an excitation, we find a large
disparity between the high temperature of the electrons on the one
hand and the low temperatures of spins and phonons on the other
hand. This suggests that the optically accessed states are the result
of thermodynamic equilibrium of the electron system alone. A
quasi-equilibrium state such as this has been assumed
earlier\cite{zeiger92_prb45_768}.

In order to test this conjecture, we investigated the phase diagram by
increasing the temperature of the electrons, while spins and phonons
are kept at zero temperature. That is, spins and phonons are optimized
for each electron temperature. The lattice constants are kept equal to
the values before excitation, because they are usually too slow on the
short time scale under consideration.

As shown in \fig{fig:hotelectronphasses}, we find four different
temperature ranges, which, however, do not directly correspond to the
ranges of different excitation behavior.
\begin{itemize}
\item For $T<T_1$ with $T_1\approx 2500$~K, we obtain the
  charge-ordered phase with CE-type magnetic order. Charge-order
  correlations and the corresponding Jahn-Teller distortions on the
  central site vanish at $T_1$ with an approximate square-root
  behavior $\sim(T_1-T)^\alpha$ with $0<\alpha<1$. 
\item $T_1<T<T_2$ with $T_2=4200$~K: At $T=T_1$ the charge and orbital
  order is completely lost, and the system transforms abruptly from
  the CE-type antiferromagnetic structure into a ferromagnet. The
  system is a pure ferromagnet only at $T=T_1$. For $T>T_1$ the spin
  angle $\phi_c$ between adjacent ab-planes increases with an
  approximate square-root-like behavior towards increasing
  temperatures, i.e. $\phi_c\sim(T-T_1)^\beta$ with $0<\beta<1$. The
  spin orientation alternates between two values from plane to plane.
\item $T_2<T<T_3$ with $T_3=5000$~K: At $T=T_2$, the spins in the
  ab-planes become non-collinear. The angle $\phi_{ab}$ between
  adjacent spins in the ab-plane increases approximately linearly from
  $0^\circ$ to $180^\circ$ as the temperature is raised from $T=T_2$
  to $T=T_3$.
\item $T>T_3$: At $T=T_3$, both spin angles, $\phi_{ab}$ and $\phi_c$,
  are $180^\circ$, which corresponds to the G-type magnetic
  order. This is the favorable high-temperature phase for the
  temperature range explored.
\end{itemize}

It is important to note that the phase diagram described here has
little to do with the equilibrium phase diagram of the material. The
phases described above are extreme non-equilibrium states, because
spins and phonons are at $T=0$.

We can identify the excitation regimes~I and II with the temperature
range $T<T_1$. The ferromagnetic state obtained in regime~III can be
attributed to the range $T_1<T<T_2$. A non-zero spin angle $\phi_c$
between the ferromagnetic planes has not been apparent in our
time-dependent simulation. We expect this to be a fluctuating quantity
that is averaged out.

In regime~IV, we find configurations which are non-collinear in the
ab-plane. Interestingly, the non-collinear state with
$\phi_{ab}=90^\circ$ shown in \fig{fig:region4diffrmodel} is a typical
state obtained for a range of fluences, while in
\fig{fig:hotelectronphasses} it is just one point in a region with
continuously changing angles $\phi_{ab}$. At even higher fluences,
also the G-type structure is encountered.

\section{Summary}
 \label{sec:summary}
The optical excitation of half-doped $\rm{Pr_{0.5}Ca_{0.5}MnO_3}$ has
been simulated to study the physical interplay between electronic,
spin and lattice degrees of freedom in response to a femtosecond light
pulse. The simulations use Ehrenfest dynamics, in which electrons and
spins follow the time-dependent Schr\"odinger equation while the
nuclei proceed on a classical trajectory.

Femtosecond excitations with various intensities and pulse lengths are
studied. The pulse acts on the charge-ordered, low-temperature phase
with CE-type antiferromagnetism. 

Four different intensity regimes with qualitative different behavior
could be identified.
\begin{itemize}
\item In regime~I, the electron-band structure remains essentially rigid.  The
  electron-hole distribution excitation transfers weight from the
  central Mn ions of the zig-zag chain to the corner sites. The dipole
  oscillations shuffle charge between two adjacent corner sites.  Two
  coherent phonons with long life time are excited as result of the
  electron-phonon coupling. 
\item In regime~II, the spins react and rearrange into short-lived
  A-type  antiferromagnetic structure. The
  ground-state CE-type antiferromagnetic structure is recovered within
  a picosecond. The coherent phonons, present also in regime~I,
  survive this transition.
\item In regime~III, charge and orbital orders are destroyed
  within few hundred femtoseconds and a ferromagnet is formed. In
  contrast to regime~I and II, the coherent phonons are damped out
  rapidly. Due to spin conservation, the ferromagnet is not directly
  accessible. Rather, an A-type antiferromagnet is formed, which
  evolves over several picoseconds into a ferromagnet having domains
  compatible with the size of our simulation cell.
\item In regime~IV, charge and orbital orders are immediately
  destroyed as in regime~III, but now a G-type antiferromagnet is
  formed rather than an A-type antiferromagnet. Over time, the system
  evolves into a new non-collinear spin structure with neighboring
  spins having 90$^\circ$ angles in the ab-plane.
\end{itemize}

The transient magnetic state observed in regime~II may shed light onto
the thermal N{\'e}el transition of {\pcmohalf} at 175~K. In regime~II,
the system maintains the orbital and charge order, but it modifies the
spin correlations of neighboring zig-zag chains of the CE-type spin
structure in the ab-plane. Analogously, the N{\'e}el transition may be
due to a melting of the antiferromagnetic correlations between the
zig-zag chains, while maintaining the ferromagnetic order within the
chains.  When the ferromagnetic order within the chains melts at
higher temperature, the integrity of the chains with their orbital and
charge order is destroyed as in regime~III and IV.

The long lifetime of the magnetic orders in regime~III and IV may
qualify for the concept of ``hidden phases''. Hidden
phases\cite{ichikawa11_naturematerials10_101} are states with unique
order which can not be accessed thermodynamically. It must be noted
however, that the time scales covered in our simulations are short
compared to those studied experimentally.

In order to make contact with thermodynamics, we estimated the
temperatures of the individual subsystems, namely electrons, spins and
phonons. The temperature of the electronic subsystem raises quickly to
several thousand Kelvin, while phonon and spin degrees of freedom
remain relatively ``cold''. An exception are the coherent phonon
modes, which initially reach the temperature of the electrons before
dissipating their energy into other degrees of freedom.

Following this concept of hot electrons and cold phonons and spins, we
have been able to identify the phases accessed by optical excitation
with those obtained by raising only the electron temperature.

\begin{acknowledgments}
Financial support from the Deutsche Forschungsgemeinschaft (SFB 1073)
through Projects B02, B03 and C03 is gratefully acknowledged. We are
grateful to Michael Ten Brink, Salvatore Manmana and Stefan Kehrein
for fruitful discussions.
\end{acknowledgments}
  
\appendix
\label{secs:apen}
\section{Numerical integration of time-dependent Schr\"odinger equation}
\label{secs:apen1}
To solve the time-dependent Schr\"odinger equation for wave functions
and spinors, we use the second-order differencing scheme proposed by
A. Askar and Cakmak.\cite{askar78_jcp68_2794,kosloff88_jpc92_2087}.

Given the wave function $|\psi(0)\rangle$ at time $t=0$ and the
time-dependent Hamiltonian $\hat{H}(t)$, the wave function
$|\psi(t)\rangle$ can be obtained as
$|\psi(t)\rangle=\hat{U}(t,0)|\psi(0)\rangle$ using the propagator
\begin{eqnarray}
\hat{U}(t',t)=\mathcal{T}_D\exp\left(-\frac{i}{\hbar}\int_t^{t'} d\tau\;
\hat{H}(\tau)\right)\;.
\end{eqnarray}
$\mathcal{T}_D$ is Dyson's time-ordering
operator\cite{dyson49_pr75_486}, which rearranges all operators in a
product into ascending time order from right to left.

With the time step $\Delta$, subsequent wave functions of a time
sequence are related by
\begin{eqnarray}
|\psi_n(\Delta)\rangle-|\psi_n(-\Delta)\rangle&=&
\Bigl(\hat{U}(\Delta,0)-\hat{U}(-\Delta,0)\Bigr)|\psi_n(0)\rangle
\nonumber\\
&&\hspace{-1cm}
=
-\frac{2i\Delta}{\hbar}\hat{H}(0)|\psi_n(0)\rangle+O(\Delta^3)
\;.
\end{eqnarray}

The error is reduced  by splitting off the dynamical phase
using the corresponding energy expectation value
$E_n(t):=\langle\psi_n(t)|\hat{H}(t)|\psi_n(t)\rangle$
\begin{eqnarray}
&&\e{\frac{i\Delta}{\hbar}E_n(0)}|\psi_n(\Delta)\rangle
-\e{-\frac{i\Delta}{\hbar}E_n(0)}|\psi_n(-\Delta)\rangle
\nonumber\\
&=&
-\frac{2i\Delta}{\hbar}\Big(\hat{H}(0)-E_n(0)\Big)|\psi_n(0)\rangle+O(\Delta^3)
\;.
\end{eqnarray}

This leads to the following iterative scheme
\begin{eqnarray}
&&|\psi_n(t+\Delta)\rangle=|\psi(t-\Delta)\rangle
\e{-\frac{2i\Delta}{\hbar}E_n(t)}
\nonumber\\
&&-
\frac{2i\Delta}{\hbar}\Big(\hat{H}(t)-E_n(t)\Big)|\psi(t)\rangle
\e{-\frac{i\Delta}{\hbar}E_n(t)}
+O(\Delta^3)\;.
\label{eq:a4}
\end{eqnarray}
These equations of motion are time-inversion symmetric per
construction. 

However, the equations of motion produce besides the correct solution
also a spurious partial solution which changes sign in
each iteration. This implies that, over time, the wave function will
pick up a contribution from the spurious solution.  In order to purify
the solution, we interrupt the simulation at regular time intervals
and perform a correction step.  In the correction step, we filter out
the spurious partial solution.
\begin{eqnarray}
|\psi'(t)\rangle&=&
|\psi(t)\rangle
+\frac{1}{4}\Bigl(
|\psi(t+\Delta)\rangle\e{\frac{i}{\hbar}E_n(t)\Delta}
\nonumber\\&&
-2|\psi(t)\rangle
+|\psi(t-\Delta)\rangle\e{-\frac{i}{\hbar}E_n(t)\Delta}
\Bigr)
\end{eqnarray}
and perform a Gram-Schmidt orthonormalization on the one-particle
wave functions for the two subsequent time steps used in the next iteration.

We use a time step of $\Delta\approx10^{-3}$~fs. Correction steps
are performed every 20 time steps.

\begin{figure}[hbtp]
\begin{center}
\includegraphics[width=\linewidth]{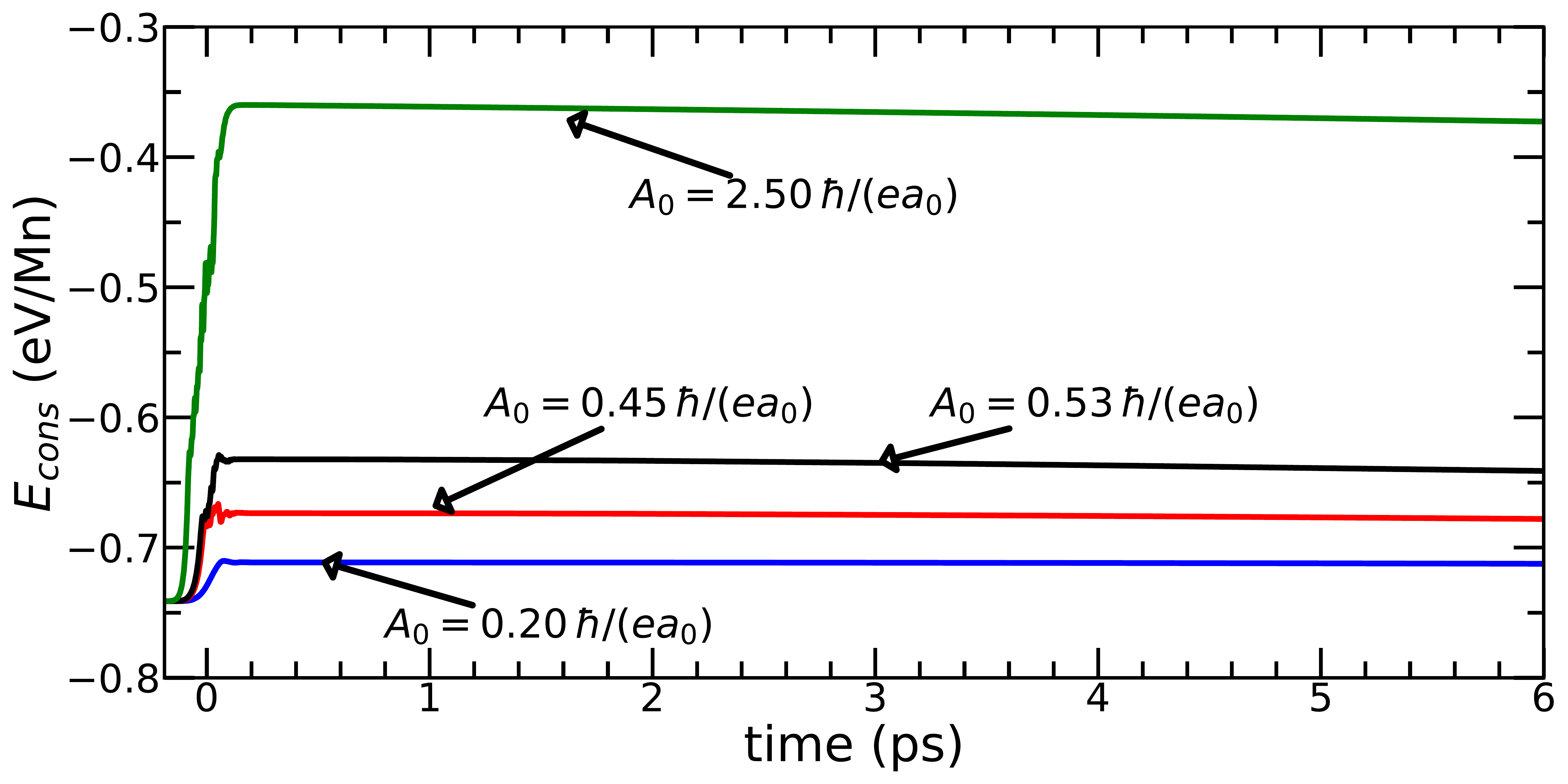}
\end{center}
\caption{\label{fig:energyconservation}Energy conservation for three
  intensities of the excitation. The initial rise is due to the
  excitation. The largest violation of energy conservation is a slow
  dissipation of the quantum systems. (color online)} 
\end{figure}

The energy conservation is shown in \fig{fig:energyconservation}
for different light amplitudes $A_0$.

\section{Spin dynamics}
\label{sec:spindynamics}
The dynamics of the spins $\vec{S}_R$ describing the {\ttg}~electrons
requires special attention. While the spin dynamics is intrinsically
of quantum nature, we want to keep all three {\ttg}~electrons of a
given Mn ions strictly collinear.  For this purpose, we map the spin
vector $\vec{S}_i$ onto a normalized, complex-valued, two-component
spinor $\left( \begin{array}{c} a_{\uparrow,R} \\ a_{\downarrow,R}
  \\ \end{array} \right)$ such that
 \begin{eqnarray}
\vec{S}_R&=&\frac{3\hbar}{2}
\left(\begin{array}{c}
a_{\uparrow,R}^*a_{\downarrow,R}+a_{\downarrow,R}^*a_{\uparrow,R}\\
-ia_{\uparrow,R}^*a_{\downarrow,R}+ia_{\downarrow,R}^*a_{\uparrow,R}\\
a_{\uparrow,R}^*a_{\uparrow,R}-a^*_{\downarrow,R}a_{\downarrow,R}
\end{array}\right)\;.
\label{eq:7}
 \end{eqnarray} 

The magnetic moment $\vec{m}_S$ of the {\ttg}~shell is anti-parallel
to its spin direction, namely
$\vec{m}_S=-m_S\left(\frac{3\hbar}{2}\right)^{-1}\vec{S}$. The scalar
$m_S:=|\vec{m}_S|$ is defined as the absolute value of the magnetic
moment.

The equation of motion is derived from the Lagrangian
\begin{eqnarray}
\mathcal{L}&=&
i\hbar\sum_{\sigma,\alpha,R,n}f_n
\psi^*_{\sigma,\alpha,R,n}\dot{\psi}_{\sigma,\alpha,R,n}
+i\hbar\sum_{\sigma,R}a^*_{\sigma,R}\dot{a}_{\sigma,R}
\nonumber\\
&&+\frac{1}{2}\sum_{j=1}^{N_{O}} M_O\dot{R}_j^2-E_{pot}[\psi,S[a],R]\;.
\end{eqnarray}

The spinors $(a_{\uparrow,R},a_{\downarrow,R})$ evolve under the
time-dependent Schr\"odinger equation
 \begin{eqnarray}
&& i\hbar\partial_t
\left( \begin{array}{c}
 a_{\uparrow,R} \\ a_{\downarrow,R} 
\end{array} \right) 
\nonumber\\
&=&m_{S}
\left( \begin{array}{cc} B_{z,R} &B_{x,R}-iB_{y,R}
\\B_{x,R}+iB_{y,R}& -B_{z,R}\end{array}\right)
 \left( \begin{array}{c} a_{\uparrow,R} \\ a_{\downarrow,R} \end{array} \right)
\nonumber\\
\label{eq:6}
\end{eqnarray}
with
\begin{eqnarray}
m_S\vec{B}_R&=&\left(\frac{3\hbar}{2}\right)^{-1}J_{AF}\sum_{R'\in{NN_R}}\vec{S}_{R'}
\nonumber\\
&&\hspace{-0.5cm}
-J_{Hund}\sum\limits_{\alpha}
\left(\begin{array}{c}
\rho_{\downarrow,\alpha,R,\uparrow,\alpha,R}  
+\rho_{\uparrow,\alpha,R,\downarrow,\alpha,R}
\\
-i
\rho_{\downarrow,\alpha,R,\uparrow,\alpha,R}  
+i\rho_{\uparrow,\alpha,R,\downarrow,\alpha,R}
\\
\rho_{\uparrow,\alpha,R,\uparrow,\alpha,R} 
-\rho_{\downarrow,\alpha,R,\downarrow,\alpha,R} 
\end{array}\right)
\;.
\label{eq:effmagfield}
\end{eqnarray}
The summation index $R'\in NN_R$ runs over nearest-neighbor sites
of site $R$. The first term in Eq.~\ref{eq:effmagfield} describes the
antiferromagnetic coupling with neighboring spins.  The second
term in Eq.~\ref{eq:effmagfield} describes Hund's coupling between {\ttg} and
{\eg}~electron on the same site.  $J_{AF}$ is the
antiferromagnetic spin-coupling parameter and $J_{Hund}$ is the Hund's
coupling parameter.

The dynamical equation Eq.~\ref{eq:6} is equivalent to
\begin{eqnarray}
\partial_t\vec{S}_R=\frac{2m_S}{\hbar}\vec{B}_R\times\vec{S}_R
\end{eqnarray}
where $\times$ denotes the vector product.

\section{Strain dynamics}
\label{sec:straindynamics}
In the present study, the scale factors $g_x$, $g_y$ and $g_z$ are
dynamical variables, which describe long-wave length acoustic modes
that are responsible for the strain effects in manganites
\cite{esposito18_prb97_14312,lim05_prb71_134403}.  We enforce $g_x=g_y$.

In order to describe the sound wave observed in experiment, we introduce a
classical kinetic energy $\frac{1}{2}M_g\sum_{m=1}^3 \dot{g}_m^2$,
which determines the Newton's equations of motion for the scale
factors $g_m$.

In thin-film experiments, the wave vector of a sound wave
perpendicular to the film is quantized, which results in a standing
wave with a characteristic frequency.  The sound wave modulates the
optical density of the material, which can be detected by the optical
absorption measurements.\cite{jang10_prb81_214416,zeiger92_prb45_768,thomsen84_prl53_989}

We adjusted the fictitious mass $M_g$ to simulate this effect. In our
model, a sound wave is excited in a material without {\eg}~electrons
with $\vec{q}=0$ and with polarization along the c-direction.  $M_g$
is chosen so that our model material oscillates with the same
frequency as the film in experiment\cite{raiser17_aenm7_1602174}, namely
$\approx25$~GHz.

\section{Peierls substitution}
\label{secs:apen2}
In this section, we give a brief derivation of the
Peierls-substitution
method\cite{peierls33_zp80_763,hofstadter76_prb14_2239}.

The electric field $\vec{E}=-\partial_t\vec{A}$ of the light pulse is
expressed by a vector potential
 \begin{eqnarray}
  \vec{A}(\vec{r},t)=
\frac{\vec{e}_A}{2}
\Big(A_0\e{i(\vec{k}\vec{r}-\omega t)}+A_0^*\e{-i(\vec{k}\vec{r}-\omega t)}\Big)g(t)
  \label{eq:8}
 \end{eqnarray}  
$\vec{e}_A$ is the polarization direction of the vector
potential and $g(t)$ is an envelope function, which is normalized so that
\begin{eqnarray}
\int dt\; |g(t)|^2=1\;.
\label{eq:normgoft}
\end{eqnarray} 

The electrons experience a Hamiltonian of the form
\begin{eqnarray}
\hat{H}[\vec{A}]
=\frac{1}{2m_e}\Big(\hat{\vec{p}}-q\vec{A}(\hat{\vec{r}},t)\Big)^2+V(\hat{\vec{r}})
\label{eq:a5}
 \end{eqnarray}
where $q$ is the electron charge, $m_e$ its mass, and $V$ is the
lattice potential.

The Hamilton matrix elements are evaluated in a basisset of local
orbitals centered at positions $\vec{R}_\alpha$, that have the
vector potential explicitly built in.  From a regular basisset of
local orbitals $|\chi_\alpha\rangle$, field-dependent basis functions
\begin{eqnarray}
\langle \vec{r} |\tilde{\chi}_\alpha \rangle
=\text{exp} \Big[ \frac{i}{\hbar}q 
\int_{\vec{R}_{\alpha}}^{\vec{r}} d\vec{r'}\vec{A}(\vec{r'},t) \Big] 
\langle\vec{r}|\chi_\alpha\rangle
\label{eq:a6}
 \end{eqnarray}
are constructed\cite{peierls33_zp80_763,hofstadter76_prb14_2239}.  The
integral of the vector potential is path dependent: we choose a
straight line from the central atom $\vec{R_\alpha}$ to the position
$\vec{r}$.

Substituting the above ansatz Eq.~\ref{eq:a6}, we obtain 
\begin{eqnarray}
&&\langle \vec{r}|\Big(\hat{\vec{p}}-q\vec{A}(\hat{\vec{r}},t)\Big)|\tilde{\chi}_\alpha \rangle
\nonumber\\
&&=\text{exp} 
\Big[ \frac{i}{\hbar}q \int_{\vec{R}_{\alpha}}^{\vec{r}} d\vec{r'}
\vec{A}(\vec{r'},t) \Big] \langle\vec{r}|\hat{\vec{p}}|\chi_{\alpha}\rangle
\label{eq:a7}
 \end{eqnarray}

From Eq.~\ref{eq:a5} and Eq.~\ref{eq:a7}, we obtain
\begin{eqnarray}
\langle\tilde{\chi}_\alpha|\hat{H}|\tilde{\chi}_\beta \rangle
&=&
\e{-i\Phi_{\alpha,\beta}(t)}
\langle \chi_{\alpha}|
\e{\frac{i}{\hbar}q\mathcal{F}_{\alpha,\beta}(\hat{\vec{r}},t)} \hat{H}
|\chi_\beta\rangle 
\label{eq:a9}
 \end{eqnarray}
where 
\begin{eqnarray}
\Phi_{\alpha,\beta}(t):=\frac{q}{\hbar}
\int_{\vec{R}_\alpha}^{\vec{R}_\beta}  d\vec{r'}\vec{A}(\vec{r'},t) 
\end{eqnarray}
is the Peierls phase. Furthermore, we define the small quantity
$\mathcal{F}_{\alpha,\beta}(\vec{r},t)$, which appears in the above
Eq.~\ref{eq:a9}, as
 \begin{eqnarray}
 \mathcal{F}_{\alpha,\beta}(\vec{r},t)
&:=&\int_{\vec{R}_{\alpha}}^{\vec{R}_{\beta}} d\vec{r'}
\vec{A}(\vec{r'},t) 
+\int_{\vec{R}_\beta}^{\vec{r}} d\vec{r'}\vec{A}(\vec{r'},t) \nonumber\\ 
 & & +  \int_{\vec{r}}^{\vec{R}_\alpha} d\vec{r'}\vec{A}(\vec{r'},t),
 \label{eq:a10}
 \end{eqnarray}
$\mathcal{F}_{\alpha,\beta}(\vec{r},t)$ is a magnetic flux through
 triangle with corners at $\vec{R}_{\alpha},\vec{R}_{\beta}$ and
 $\vec{r}$.

The time-dependent Schr\"odinger equation for a wave function
$|\psi_n(t)\rangle=\sum_\beta|\tilde{\chi}_\beta(t)\rangle c_{\beta,n}(t)$
obtains the form
\begin{eqnarray}
 \sum_\beta \e{-i\Phi_{\alpha,\beta}(t)}
[
\tilde{O}_{\alpha,\beta}(t)i\hbar\partial_t-\tilde{H}_{\alpha,\beta}(t)]
c_{\beta,n}(t)=0
 \label{eq:a12}
\end{eqnarray}  
with
  \begin{eqnarray}
 \tilde{H}_{\alpha,\beta}(t)&:=& \langle\chi_{\alpha}| 
\e{\frac{i}{\hbar}\mathcal{F}_{\alpha,\beta}(\hat{\vec{r}},t)}
\Bigl(\hat{H}[\vec{0}]
+q\int_{\vec{R}_\beta}^{\vec{r}}d\vec{r'}\dot{\vec{A}}(\vec{r'},t)\Bigr)
|\chi_\beta\rangle  
\nonumber\\
 \tilde{O}_{\alpha,\beta}(t)&:=& 
\langle\chi_{\alpha}| 
\e{\frac{i}{\hbar}\mathcal{F}_{\alpha,\beta}(\hat{\vec{r}},t)}|\chi_\beta\rangle
 \label{eq:a13}
 \end{eqnarray}  
In this form, the Peierls substitution
method\cite{peierls33_zp80_763} is formally exact.

In practice, $\mathcal{F}_{\alpha,\beta}(\vec{r},t)$ is neglected. For this to
be a good approximation, the basis set needs to be sufficiently
localized. 

Furthermore, the vector potential is approximated by a
constant. This is equivalent to the long-wavelength limit. It also
excludes dipole-forbidden, but quadrupole-allowed transitions. The
latter are not considered relevant in comparison with the strong
charge-transfer transitions in the present work.

With these approximations, by exploiting the orthonormality of our
basisset, and after ignoring off-site terms of the dipole matrix
elements, we obtain
\begin{eqnarray}
i\hbar\partial_tc_{\alpha,n}&=&
\sum_\beta
{\rm e}^
{\frac{-iq}{\hbar}\vec{A}(t)(\vec{R}_{\beta}-\vec{R}_{\alpha})}
\langle\chi_\alpha|\hat{H}[\vec{0}]|\chi_\beta\rangle
c_{\beta,n}
\nonumber\\
&+&\sum_\beta\delta_{R_\alpha,R_\beta}
\biggl(-q\dot{A}(\vec{t})
\langle\chi_\alpha|\vec{r}-\vec{R}_\beta|\chi_\beta\rangle\biggr)
c_{\beta,n}
\nonumber\\
 \label{eq:a14}
\end{eqnarray}  
The first term on the right-hand side describes charge-transfer
transitions, while the second term describes dipole-allowed onsite
transitions. The latter vanish in our model and are included here only
for the sake of completeness.

The Peierls phase only affects off-site matrix elements. In our case,
these are the hopping matrix elements. Thus, the only change required
to incorporate the excitation is to multiply the hopping matrix
elements with the time-dependent Peierls phase.

\subsection{Temperatures}
\label{sec:temperatures}

\subsubsection{Phonon temperature}
The temperature $T_{ph}$ of the phonon degrees of freedom has been
evaluated from the kinetic energy of Jahn-Teller-active and breathing
phonon modes of the oxygen ions. We use the relation
\begin{eqnarray}
\sum_{i=1}^{N_{\rm{O}}}\frac{1}{2}M_O\dot{\vec{R}}_i^2
= \frac{N_{\rm{O}}}{2}k_BT_{ph}
\end{eqnarray}
 where index $i$ runs over all $N_{\text{O}}$ oxygen ions in the unit
 cell and $M_O$ is the mass of the oxygen ion.  

There is only one degree of freedom per oxygen atom in our simulation,
because only three phonon modes per formula unit are considered. These
are the modes with strong electron-phonon coupling, which receive the
energy directly from the excited electrons and holes. Only later,
these ``hot'' modes dissipate their energy into the other phonon modes
and the spin system.

\subsubsection{Electron temperature}
\label{sec:electrontemperature}
The temperature of the electronic degrees of freedom are obtained from
the occupations of the Born-Oppenheimer wave functions. For that
purpose, we extract the one-particle wave functions
$|\psi_j(t)\rangle$ and the instantaneous one-particle Hamiltonian
$\hat{h}^{BO}(t)$ acting on the electrons. The Hamiltonian is the
Born-Oppenheimer Hamiltonian for the instantaneous spin distribution
and atomic positions.

Let $|\phi^{BO}_j(t)\rangle$ be the eigenstates and
$\epsilon^{BO}_j(t)$ the eigenvalues of the one-particle Hamiltonian
$\hat{h}^{BO}(t)$.  The occupations $\bar{f}_j$ of the
Born-Oppenheimer states $|\phi^{BO}_j(t)\rangle$ are obtained from
their projections onto the occupied wave functions
$|\psi_j(t)\rangle$ as
\begin{eqnarray}
\bar{f}_j(t):=\sum_{n=1}^{N_e}
\left|\langle\psi_n(t)|\phi^{BO}_j(t)\rangle\right|^2
\;.
\end{eqnarray}

Electron temperature $T_{\psi}$ and electron chemical potential $\mu$
are determined such that energy and particle number coincide with
those of a thermal distribution, i.e.
\begin{eqnarray}
\sum_j\bar{f}_j&=&\sum_j\left(\e{(\epsilon^{BO}_j-\mu)/(k_BT_{\psi})}+1\right)^{-1}
\nonumber\\
\sum_j\bar{f}_j\epsilon^{BO}_j&=&\sum_j
\left(\e{(\epsilon^{BO}_j-\mu)/(k_BT_{\psi})}+1\right)^{-1}\epsilon_j^{BO}
\end{eqnarray}

In order to compare the instantaneous distributions
$(\bar{f}_j,\epsilon_j)$ to the Fermi distribution
$f_{T,\mu}(\epsilon)=1/[1+\exp(\frac{1}{k_BT}(\epsilon-\mu))]$, we will
plot
\begin{eqnarray}
z(\bar{f})=\mathrm{arctanh}(1-2\bar{f})
\end{eqnarray}
because this transforms a Fermi distribution into a linear
function of the energy:
\begin{eqnarray}
z\big(f_{T,\mu}(\epsilon)\big)=\frac{1}{k_BT}(\epsilon-\mu)
\end{eqnarray}
As a result, we can read the quasi temperature from the slope and the
quasi Fermi level from the zero of the interpolated line through the
data points $\Big(z(\bar{f}_j),\epsilon^{BO}_j\Big)$.

\subsubsection{Temperature of the spin subsystem}
We extract the temperature of the spin subsystem, i.e. the spins of
the {\ttg}~electrons, analogously to that of the {\eg}~electrons. For
each time step, we extract a Born-Oppenheimer Hamiltonian
\begin{eqnarray}
\mathbf{h}^{BO,S}_R&:=&
m_S
\left( \begin{array}{cc} B_{z,R} &B_{x,R}-iB_{y,R}
\\B_{x,R}+iB_{y,R}& -B_{z,R}\end{array}\right)
\label{eq:boham}
\end{eqnarray}
with $\vec{B}_R$ defined in Eq.~\ref{eq:effmagfield} and the absolute
value $m_S$ of the {\ttg}~magnetic moment.

The projections of the instantaneous Pauli spinors $\vec{a}_R(t)$ onto
the eigenvectors $\vec{a}^{BO}_{j,R}(t)$ of $\mathbf{h}^{BO,S}_R(t)$
yield occupations
\begin{eqnarray}
\tilde{f}_{j,R}(t)=\biggl|\sum_{\sigma\in\{\uparrow,\downarrow\}}
a^*_{\sigma,R}(t)a^{BO}_{\sigma,j,R}(t)
\biggr|^2
\label{eq:boocc}
\end{eqnarray}
for ground state with $j=0$ and excited state with $j=1$.

The comparison with the internal energy for non-interacting spins in
an external magnetic field provides a relation 
\begin{eqnarray}
\sum_R\Bigl(\tilde{f}_{0,R}-\tilde{f}_{1,R}\Bigr)&=&
\sum_R\tanh\left(\frac{m_S|\vec{B}_R|}{k_BT_S}\right)
\label{eq:tempfromf}
\end{eqnarray}
which is resolved for the instantaneous temperature $T_{S}(t)$ of the
spin system.

A more detailed derivation of the expressions summarized here is
provided in appendix~\ref{app:spintemperature}.

\section{Temperature of the spin subsystem}
\label{app:spintemperature}
The temperature of the spin system is extracted analogously to that of
the electrons. We consider a system of uncoupled spins in a
magnetic-field distribution $\vec{B}_R$ defined by the local
Born-Oppenheimer Hamiltonian for the spin system according to
Eq.~\ref{eq:boham}.  The free energy of this system is
\begin{eqnarray}
F_T=-k_BT\ln\sum_{\vec{\sigma}} {\rm e}^{-\frac{1}{k_BT} E_{\vec{\sigma}}}
\end{eqnarray}
where the energy of a spin distribution  $\vec{\sigma}$ is
\begin{eqnarray}
E_{\vec{\sigma}}=-\sum_R m_S|\vec{B}_R| (-1)^{\sigma_R}\;.
\end{eqnarray}
$\sigma_R\in\{0,1\}$ characterizes the ground and excited state of the
local spin $\vec{S}_R$ of the {\ttg}~electrons at site $R$.  The
absolute value of the magnetic moment related to the spin $\vec{S}_R$ 
of the three {\ttg}~electrons at a
given site is
$m_S=|\gamma\frac{3\hbar}{2}|$.

This yields for the free energy 
\begin{eqnarray}
F_T=-k_BT\sum_R \ln\left[2\cosh\left(\frac{m_S|\vec{B}_R|}{k_BT}\right)\right]
\end{eqnarray}

The instantaneous temperature of the spin system is extracted by
comparing the energy obtained from the instantaneous spin distribution
\begin{eqnarray}
E(t)=\sum_R \vec{a}^*_R(t)\mathbf{h}^{BO,S}_R(t)\vec{a}_R(t)
\label{eq:instspinen}
\end{eqnarray}
with the energy $\langle E\rangle_T=\beta\partial_\beta F_T$ of the
thermal ensemble, where $\beta=1/(k_BT)$.

With the eigenstates $\vec{a}^{BO}_{j,R}$ and the eigenvalues
$\tilde{\epsilon}_{j,R}$ of the Born-Oppenheimer Hamiltonian
$\mathbf{h}^{BO,S}_R$, the instantaneous energy
Eq.~\ref{eq:instspinen} is
\begin{eqnarray}
E(t)=\sum_R \tilde{f}_{j,R}(t)\tilde{\epsilon}_{j,R}(t)
\end{eqnarray}
with occupations
\begin{eqnarray}
\tilde{f}_{j,R}(t)=\biggl|\vec{a}^*_R(t)\vec{a}^{BO}_{j,R}(t)\biggr|^2
\end{eqnarray}
for the local ground state with $j=0$ and the excited state with
$j=1$.

The requirement 
\begin{eqnarray}
\sum_R\sum_{j=0}^1 \tilde{f}_{j,R}\tilde{\epsilon}_{j,R}
&\stackrel{!}{=}&\beta\partial_\beta F_\beta=\left\langle E\right\rangle_T
\end{eqnarray}
provides an expression for the instantaneous temperature $T(t)$
\begin{eqnarray}
\sum_R
\big(\tilde{f}_{0,R}-\tilde{f}_{1,R}\big)
= \sum_R 
\tanh\left(\frac{m_S|\vec{B}_R|}{k_BT}\right)
\label{eq:spintempappendix}
\end{eqnarray}

\section{Frequencies of the coherent phonon modes}
\label{app:phmodefit}
\begin{figure}[hbtp]
\begin{center}
\includegraphics[width=0.9\linewidth]{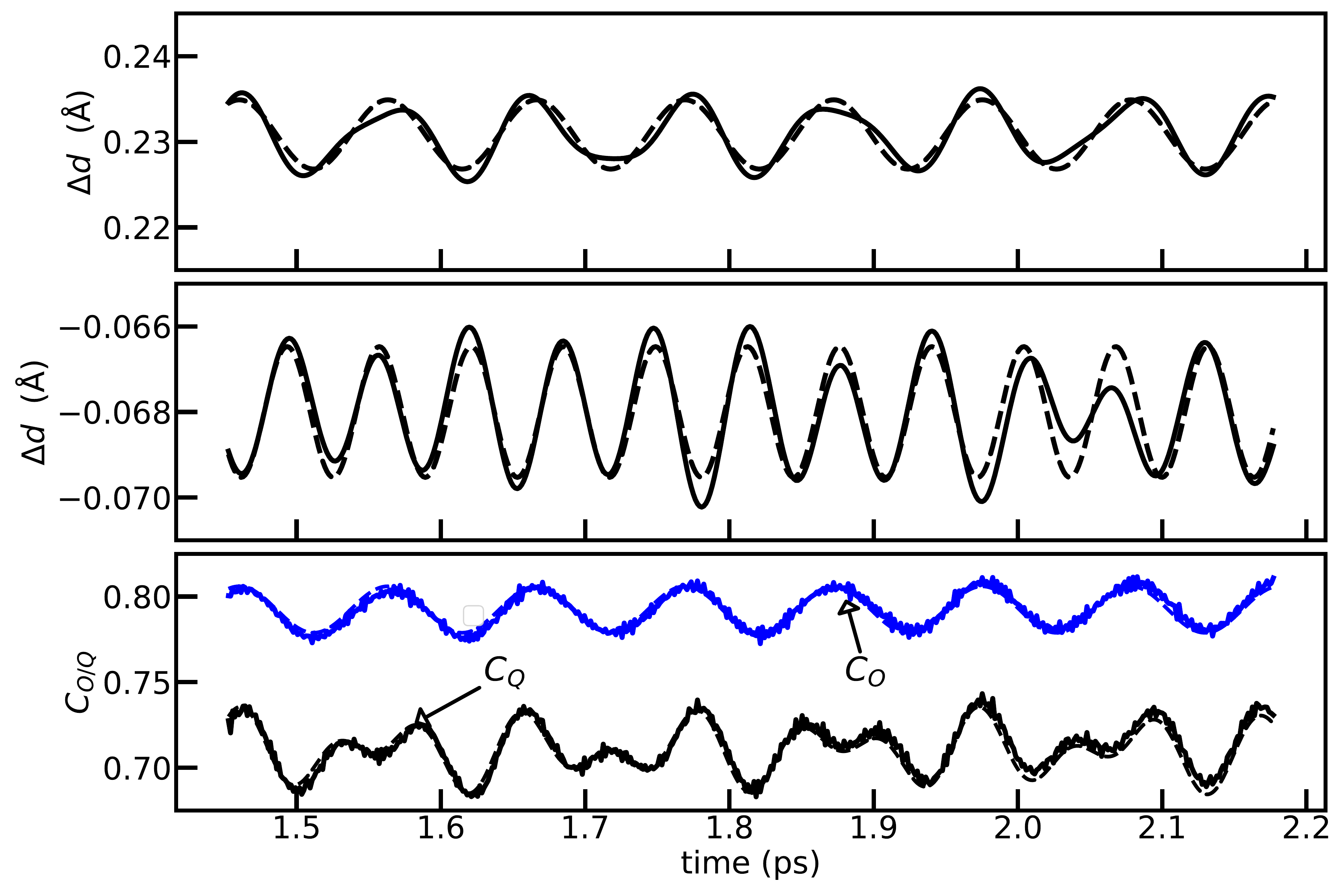}
\end{center}
\caption{\label{fig:coherentmodefreq}Fit of the phonon modes and
  diffraction intensities.  The top figure shows $\Delta d_{||}^{3+}$
  (red) and the fit (dashed). The middle graph shows $\Delta d_y^{4+}$
  and its fit. The bottom graph are the correlation functions $C_Q$
  for charge and $C_O$ for orbital order and their fits.  The derived
  frequencies are 9.7~THz and 15.7~THz. The trajectory has been
  performed for a light-amplitude of
  $A_0=0.20$~$\rm{\hbar/(ea_0)}$. (color online)}
\end{figure}

The frequencies of the coherent modes, present in regimes I and II,
have been extracted by non-linear curve fitting of a superposition of a
constant and two cosine functions with amplitude, frequency and phase
shift as variable parameters. The quality of the fit is shown in
\fig{fig:coherentmodefreq}. The fit gives a frequency of 9.7~THz and
15.7~THz. 

The vibration of 15.7~THz is dominant in the $\Delta
d^{4+}_y$ and can be attributed to the planar breathing mode on the
corner sites of the CE-type magnetic structure, which couples to the
charge transfer from the central to the corner sites.

The lower frequency with $\nu=9.7$~THz is a Jahn-Teller mode on the
central site of a trimer.

In order to confirm that the oscillations of the charge- and
orbital-order correlation functions are a direct consequence of the
coherent phonons, the correlation functions have been fitted with same
two frequencies. The perfect agreement shown in
\fig{fig:coherentmodefreq} supports our conjecture.

\end{document}